\documentclass[twoside,twocolumn,english,prc,amsmath,a4paper,myheadings]{article}
\usepackage{mathpazo}
\usepackage[T1]{fontenc}
\usepackage[latin9]{inputenc}
\usepackage{geometry}
\usepackage{color}
\usepackage{verbatim}
\usepackage{calc}
\usepackage{textcomp}
\usepackage{amsmath}
\usepackage{amssymb}
\usepackage{graphicx}
 
\makeatletter

\providecommand{\tabularnewline}{\\}

\newcommand{\lyxaddress}[1]{
\par {\raggedright #1
\vspace{1.4em}
\noindent\par}
}

\def\@oddhead{\rightmark \hfill Perturbative QCD concerning light and heavy 
flavor in the EPOS4 framework \hfill \thepage}
\def\@evenhead{\thepage \hfill K. Werner and B. Guiot \hfill}
\def\fnum@table{\tablename~{\bf\thetable}}
\def\fnum@figure{\figurename~{\bf\thefigure}}
\def\tablename{\footnotesize{\bf Table}}
\def\figurename{\footnotesize{\bf Figure}}

\usepackage{dcolumn}
\usepackage[font=small,labelfont=bf]{caption}

\def\citet{\cite}

\newcommand{\lineup}{ }
\newcommand{\linelow}{   }

\AtBeginDocument{
  
}

\makeatother

\usepackage{babel}
\begin{document}
\twocolumn[   \begin{@twocolumnfalse}  

\title{Perturbative QCD concerning light and heavy flavor in the EPOS4 framework }

\date{}
\author{K. Werner$^{(a)}$ and B. Guiot$^{(b)}$}
\maketitle

\lyxaddress{\begin{center}
$^{(a)}$SUBATECH, Nantes University \textendash{} IN2P3/CNRS \textendash{}
IMT Atlantique, Nantes, France\\
$^{(b)}$Universidad Tecnica Federico Santa Maria y Centro Cientifico-Tecnologico
de Valparaiso, Valparaiso, Chile
\par\end{center}}
\begin{quote}
We recently introduced new concepts, implemented in EPOS4, which allows us
to consistently accommodate factorization and saturation in high-energy
proton-proton and nucleus-nucleus collisions, in a rigorous parallel
scattering framework. EPOS4 has a modular structure and in this paper,
we present in detail how the ``single scattering module'' 
(the main EPOS4 building block) is related to perturbative QCD, 
and how these calculations are performed, with particular care
being devoted to heavy flavor contributions. We discuss similarities and 
differences compared to the usual pQCD approach based on factorization.\\
$\quad$
\end{quote}
\end{@twocolumnfalse}]

\section{Introduction\label{=======Introduction=======}}

Factorization \cite{Collins:1989,Ellis:1996} (in connection with
asymptotic freedom \cite{Gross:1973,Politzer:1973}) is a powerful
concept to reliably compute inclusive cross sections for high transverse
momentum ($p_{t}$) particle production in proton-proton (pp) scattering
at very high energies. However, there are very interesting cases not
falling into this category, like high-multiplicity events in proton-proton
scattering in the TeV energy range, where a very large number of parton-parton
scatterings contribute. Such events are particularly interesting,
since the CMS collaboration observed long-range near-side angular
correlations for the first time in high-multiplicity proton-proton
collisions \cite{CMS:2010ifv}, which was before considered to be
a strong signal for collectivity in heavy-ion collisions. Studying
such high-multiplicity events (and multiplicity dependencies of observables)
goes much beyond the frame covered by factorization, and we need an
appropriate tool, able to deal with multiple scatterings, which must
happen in parallel at high energies. 

Although everybody will agree that multiple scatterings should conserve
energy-momentum, the way it is implemented is fundamentally different
in EPOS compared to other models. 
Concerning momentum sharing, the EPOS4 method employs multiple
scattering laws
of the form $f_{1}(p_{1})\times f_{2}(p_{2})..\times\delta(p-\sum p_{i})$.
\textbf{The delta function here is crucial}, like the probability
law in a microcanonical ensemble. We refer to this method as ``rigorous
parallel scattering scenario'', for unbiased momentum sharing.
This is very different compared to a structure like 
$f_{1}(p_{1}|p)\times f_{2}(p_{2}|p-p_{1})\times f_{3}(p_{3}|p-p_{1}-p_{2})$
with conditional probabilities (as it is usually done), which may
perfectly conserve momentum but in a sequential manner. 

In \cite{werner:2023-epos4-overview}, we show that such a ``rigorous
parallel scattering scheme'' can be constructed based on S-matrix
theory (see also \cite{Gribov:1967vfb,Gribov:1968jf,GribovLipatov:1972,Abramovsky:1973fm,Drescher:2000ha}),
which we will briefly review  in the following. We show \textendash{}
and this is far from trivial \textendash{} how to accommodate factorization
and saturation in an energy conserving parallel scattering scenario.
The starting point is the elastic scattering T-matrix %
\begin{equation}
iT=\sum_{n=0}^{\infty}\int dX\,\frac{1}{n!}\,V\times\left\{ iT_{\mathrm{Pom}}\times...\times iT_{\mathrm{Pom}}\right\} \times V\times\delta,\label{multiple-tmatrix-pp}
\end{equation}
expressed in terms of ``elementary'' T-matrices\footnote{To be more precise: 
what we call $T$ is the Fourier transform of
the T-matrix with respect to the transverse momentum exchange, divided
by twice the invariant mass squared of the considered process. }%
{} $T_{\mathrm{Pom}}$, the latter ones representing parton-parton scattering,
with the ``vertex'' \emph{V} representing the connection to the
projectile and target remnants. The elementary T-matrices depend on
the light-cone momentum fractions $x_{i}^{\pm}$of the incoming partons,
in addition to the energy squared $s$, and the impact parameter $b$. The vertices depend on the light-cone momentum fractions of the remnants $x_{\mathrm{remn}}^{+}$ (projectile
side) or $x_{\mathrm{remn}}^{-}$ (target side). Most important: the
$\delta$ which stands for 
\begin{equation}
\delta(1-\sum x_{i}^{+}-x_{\mathrm{remn}}^{+})\delta(1-\sum x_{i}^{-}-x_{\mathrm{remn}}^{-}),
\end{equation}
to assure energy-momentum conservation in an unbiased fashion (this
is what we mean by a rigorous parallel scattering scenario). The
symbol $\int dX$ stands for integrating over all these momentum fractions.
The expression can be easily generalized for nucleus-nucleus scattering
\cite{werner:2023-epos4-overview}.

So far we discussed only elastic scattering, the connection with inelastic
scattering provides the optical theorem in impact parameter representation,
\begin{equation}
\sigma_{\mathrm{tot}}=\int d^{2}b\,\mathrm{cut}\,T,
\end{equation}
with $\mathrm{cut\,}T\equiv\mathrm{\frac{1}{i}disc}T$, where $\mathrm{disc}\,T$
is the s-channel discontinuity $T(s+i\epsilon)-T(s-i\epsilon),$ and
with $T$ being per definition the Fourier transformed T-matrix divided
by $2s$. So we need to compute the ``cut'' of the complete diagram,
$\mathrm{cut}\,T$, i.e., for pp we need to evaluate 
\begin{equation}
\mathrm{cut}\left\{ V\times iT_{\mathrm{Pom}}\times...\times iT_{\mathrm{Pom}}\times V\right\} ,\label{cut-multiple-pomeron}
\end{equation}
which corresponds to the sum of all possible cuts, considering, in
particular, all possibilities of cutting or not any of the parallel
Pomerons. We have finally a sum of products with some fraction of the
Pomerons being cut (cut Pomerons), the others not (uncut
Pomerons), referred to as cutting rules. We define $G$ to
be the cut of a single Pomeron,
\begin{equation}
G=\mathrm{cut}\,T_{\mathrm{Pom}}.\label{definition-G}
\end{equation}
Cut Pomerons represent inelastic scattering ( $a+b\to$ many partons)
and uncut Pomerons elastic scatterings ($a+b\to a+b$).

\begin{figure}[h]
\centering{}\includegraphics[scale=0.21]{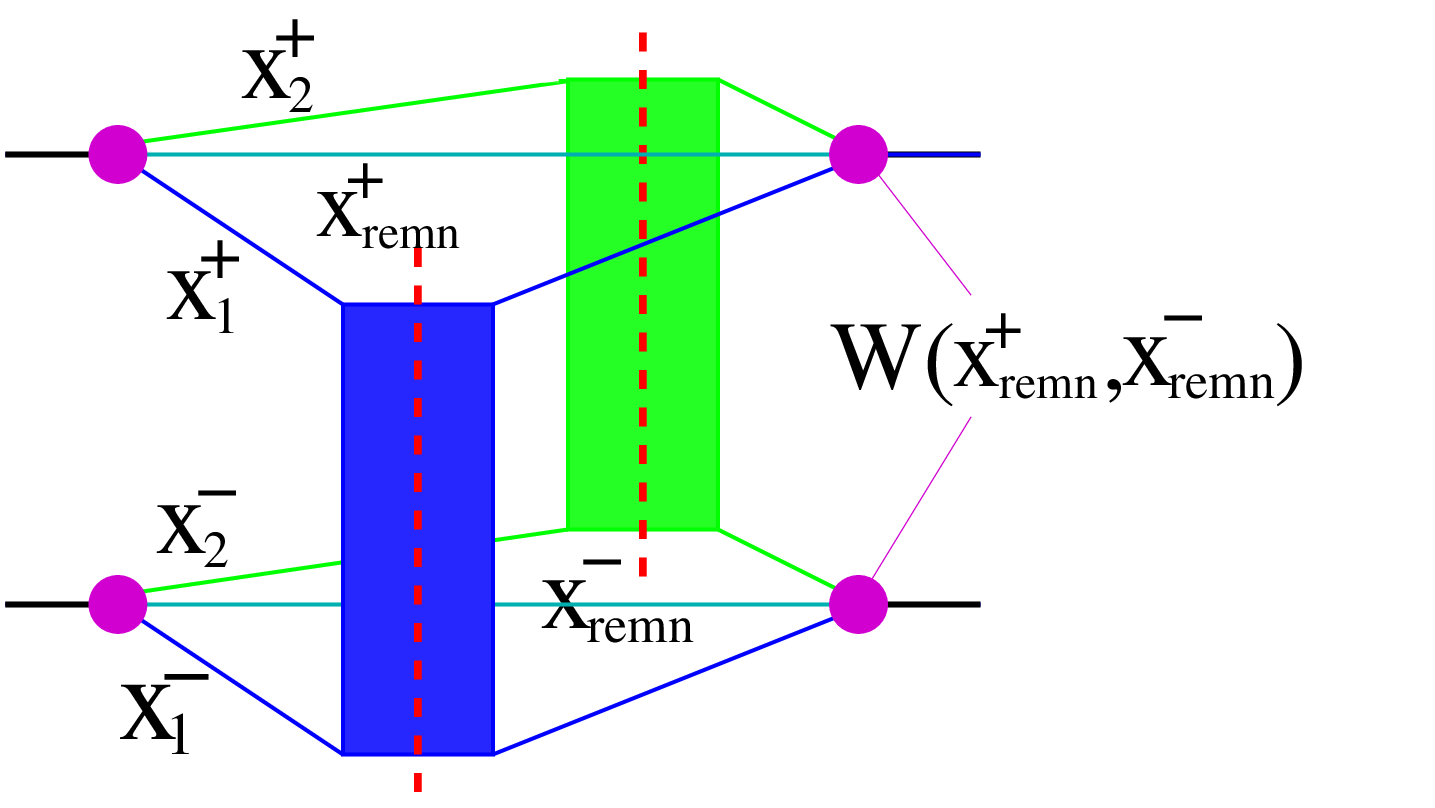}
\caption{Double scattering, each box representing a cut Pomeron (single inelastic
scattering). \label{two-boxes}}
\end{figure}

The uncut Pomerons are finally summed over and the corresponding variables 
integrated out, so eventually the total cross section is expressed as a sum 
over products of $G$, very similar to Eq. (\ref{multiple-tmatrix-pp})
with the vertices $V(x_{\mathrm{remn}}^{+})V(x_{\mathrm{remn}}^{-})$
replaced by some effective vertex function $W(x_{\mathrm{remn}}^{+},x_{\mathrm{remn}}^{-})$.
A great advantage of this T-matrix formalism is its modular structure.
Multiple scattering is expressed in terms of modules ($G$) representing 
a single scattering each, as shown in the case of two inelastic scatterings
in Fig. \ref{two-boxes}.
 The precise content of these modules will be discussed later - this is 
where all the perturbative QCD physics is hidden, and discussing that 
will be the main purpose of this paper.

\begin{figure}[h]
\centering{}%
\begin{minipage}[c]{0.2\columnwidth}%
\noindent \begin{center}
\includegraphics[scale=0.2]{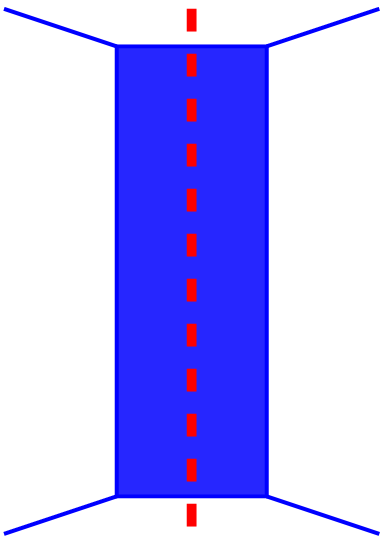} 
\par\end{center}%
\end{minipage}%
\begin{minipage}[c]{0.2\columnwidth}%
\noindent \begin{center}
\textcolor{blue}{\large{}{} 
\[
\boldsymbol{\boldsymbol{=G(x^{+},x^{-},s,b)}}
\]
}{\large{} }
\par\end{center}%
\end{minipage}$\qquad\qquad$\caption{The cut Pomeron $G$. \label{box-equal-G}}
\end{figure}

The important new issue in \cite{werner:2023-epos4-overview} is the
understanding of how energy conservation ruins factorization,
and how to solve this problem via an appropriate definition of $G$.
The cut Pomeron $G=\mathrm{cut\,}T_{\mathrm{Pom}}$ is the fundamental
quantity in the EPOS formalism. For the moment, we consider the Pomeron
as a ``box'', with two external legs representing two incoming particles
(nucleon constituents) carrying light-cone momentum fractions $x^{+}$ and
$x^{-}$, so $G=G(x^{+},x^{-},s,b)$, with the energy squared $s$,
and the impact parameter $b$, see Fig. \ref{box-equal-G}.
Let us define the ``Pomeron energy fraction'' $x_{\mathrm{PE}}=x^{+}x^{-}=M_{\mathrm{Pom}}^{2}/s,$
with $M_{\mathrm{Pom}}$ being the transverse mass of the Pomeron,
which is the crucial variable characterizing cut Pomerons: the bigger
$x_{\mathrm{PE}}$, the bigger the Pomeron's invariant mass and the
number of produced particles. Large invariant masses also favor high
$p_{t}$ jet production. We also define, for a given cut Pomeron connected
to projectile nucleon $i$ and target nucleon $j$, a ``connection
number'' $N_{\mathrm{conn}}=(N_{\mathrm{P}}+N_{\mathrm{T}})/2,$
with $N_{\mathrm{P}}$ being the number of Pomerons connected to \emph{i,
}and with $N_{\mathrm{T}}$ being the number of Pomerons connected
to \emph{j. }The case $N_{\mathrm{conn}}=1$ corresponds to an isolated
Pomeron, which may take all the energy of the initial nucleons, whereas
in the case of $N_{\mathrm{conn}}>1$ the energy for a given Pomeron will
be shared with others. To quantify the effect of energy sharing,
we define $f^{(N_{\mathrm{conn}})}(x_{\mathrm{PE}}$) to be the inclusive
$x_{\mathrm{PE}}$ distribution , i.e., the probability that a single
Pomeron carries an energy fraction $x_{\mathrm{PE}}$, for Pomerons
with given values of $N_{\mathrm{conn}}$. The distribution for $N_{\mathrm{conn}}>1$
will be deformed with respect to the $N_{\mathrm{conn}}=1$ case,
due to energy sharing, and we define the corresponding ``deformation
function'' $R_{\mathrm{deform}}(x_{\mathrm{PE}})$ as a ratio of
$f^{(N_{\mathrm{conn}})}(x_{\mathrm{PE}}$) over $f^{(1)}(x_{\mathrm{PE}}$).
As shown in \cite{werner:2023-epos4-overview}, this function can
be calculated and tabulated. As discussed in very much detail later,
we also calculate and tabulate some function $G_{\mathrm{QCD}}(Q^{2},\,x^{+},x^{-},s,b)$,
which contains as a basic element a cut parton ladder based on DGLAP
parton evolutions \cite{GribovLipatov:1972,AltarelliParisi:1977,Dokshitzer:1977}
from the projectile and target side, with an elementary QCD cross
section in the middle, $Q^{2}$ being the low virtuality cutoff in
the DGLAP evolution. The latter is usually taken to be constant and
of the order of $1\,$GeV, whereas we allow any value. With all this
preparation, we are now able to connect $G$ (used in the multi-Pomeron
diagrams) and $G_{\mathrm{QCD}}$ (making the link to QCD), as follows:\\
\vspace*{0.cm}

\noindent\fbox{\begin{minipage}[t]{1\columnwidth - 2\fboxsep - 2\fboxrule}%
For each cut Pomeron, for given $x^{\pm}$, $s$, $b$, and
$N_{\mathrm{conn}}$,\\ 
we postulate: 
\begin{equation}
\boldsymbol{G(\!\,x^{+}\!\!,x^{-}\!\!,s,b)\!=\frac{n}{R_{\mathrm{deform}}^{(N_{\mathrm{conn}})}(x_{\mathrm{PE}})}G_{\mathrm{QCD}}(Q_{\mathrm{sat}}^{2}\,,\,x^{+}\!\!,x^{-}\!\!,s,b)},\label{fundamental-epos4-equation}
\end{equation}
\vspace*{-0.3cm}%
\end{minipage}}\medskip{}
\\
with $n$ being  independent on $x^{\pm}$. 
Most importantly, $G$ does not depend on  $N_{\mathrm{conn}}$, but  $Q_{\mathrm{sat}}^{2}$ does, it ``compensates'' the  $N_{\mathrm{conn}}$ dependence of $R_{\mathrm{deform}}^{(N_{\mathrm{conn}})}$.
Considering  even a large value of $N_{\mathrm{conn}}$, we can prove
that the distribution $f^{(N_{\mathrm{conn}})}(x_{\mathrm{PE}}$)
is proportional to a product of vertex functions times $G_{\mathrm{QCD}}(Q_{\mathrm{sat}}^{2}\,,\,x^{+}\!\!,x^{-}\!\!,s,b)$,
and so the deformation cancels out! So, for any number of $N_{\mathrm{conn}}$,
the inclusive particle production is governed by a single Pomeron
represented by $G_{\mathrm{QCD}}$, which does not depend on $N_{\mathrm{conn}}$,
apart from the implicit $N_{\mathrm{conn}}$ dependence of $Q_{\mathrm{sat}}^{2}$.
Consequently, the $p_{t}$ dependence of outgoing partons will be independent
of $N_{\mathrm{conn}}$ in the hard domain (high $p_{t}$), 
see \cite{werner:2023-epos4-overview} for more details. 
Computing a
$p_{t}$ distribution according to $G_{\mathrm{QCD}}$ amounts precisely
to factorization or binary scaling in AA scattering, but - again -
only obtained for hard processes. In other words, computing  $R_{\mathrm{AA}}$ in a full AA simulation, we get unity at large $p_{t}.$

Let us come back to our earlier statement saying that ``energy conservation ruins factorization''. 
The statement actually depends on how one relates $G$  and $G_{\mathrm{QCD}}$.
In earlier versions, we adopted what we thought at the time to be ``the natural Pomeron definition'', 
but what we now call ``the naive Pomeron definition'', 
namely $G=G_{\mathrm{QCD}}$. As discussed in very much detail in \cite{werner:2023-epos4-overview},
this leads unavoidably to the following problem: An inclusive pp $p_t$ distribution will
be a superposition of contributions for different connection  numbers. 
With increasing connection numbers, these contributions get more and more deformed 
(suppressed at large $x_{{\rm PE}}$), which corresponds to a suppression of the yields at high $p_t$. 
Therefore the minimum bias $p_t$ distribution will be suppressed at large $p_t$ 
compared to the single Pomeron distribution. On the other hand, factorization requires inclusive cross
sections to be given by a single cut Pomeron, since based on this diagram one obtains formulas
as $f_{{\rm PDF}}\otimes\hat{\sigma}\otimes f_{{\rm PDF}}$, corresponding to factorization.

The fact that multiple Pomeron interactions reduce for inclusive cross sections to a single cut Pomeron 
(leading to factorization if one assumes ``the naive Pomeron definition''), refers to 
``AGK cancellations'' \cite{Abramovsky:1973fm}, shown to be valid in a scenario without energy conservation. 
As discussed above, including energy sharing (as it should be), ruins first ``AGK cancellations'' 
and, as a consequence, factorization. 
With a ``proper Pomeron definition'' as employed in EPOS4, we recover ``AGK cancellations'' in the sense 
that inclusive cross sections can be expressed in terms of a single Pomeron expression  $G_{\mathrm{QCD}}$, 
although in reality multiple scatterings contribute. 
But this statement is only true for $p_t$ values  bigger than the relevant 
saturation scales of the different multiple scattering contributions, 
where ``relevant'' refers to the relative weight of the contributions.

The new EPOS4 framework is able to recover factorization at large
$p_{t}$ (a difficult task in the parallel scattering formalism).
This  allows us to compute, tabulate, and employ ``EPOS PDFs'', and 
based on these, we may compute inclusive cross sections as a simple convolution of 
two PDFs and an elementary pQCD cross section, and the result will be   
identical -- at large $p_{t}$ -- to the simulation
results. But to make this work in practice, we need high precision
and appropriate methods for both simulation and cross section calculations.
In older versions, for example, we used in the simulations frequently
a ``redo'' whenever ``kinematic problems'' showed up, whereas
such situations should be avoided.

As a side remark, the EPOS4 multiple scattering scheme is quite different to 
what is usually called 'multiple parton interactions' (MPI): the former refers to multiple cut
Pomerons with external legs being soft partons, whereas the latter treats the scattering 
of two hard partons, based on multiple parton distribution functions, generalizing 
the factorization approach. The review \cite{MPI:2018} has actually two parts, namely ``hard MPI'' 
and ``soft MPI'', and the two have little in common.

The main purpose of this paper is to provide detailed information
about the calculation of $G_{\mathrm{QCD}}$, based on perturbative
QCD, with special care concerning heavy flavors. We discuss the implementation
(for the first time in the EPOS framework) of the ``backward parton
evolution method'', which allows much better control of the hard
processes. We discuss important ``kinematic'' issues connected
to processes involving charm and bottom, taking into account 12 different ``reaction classes'' for the cross section calculations, since the kinematics is quite different, e.g., for the Born processes $gg\to gg$ and $gg\to b\bar{b}$. A useful side-effect concerning the new strategies,
in particular the ``backward evolution'', is the fact that many
formulas are very similar to what is used in models based on ``factorization''.
We may compare our EPOS PDFs with ``standard'' PDFs, but there are also
fundamental differences: in EPOS, we have first to deal with evolutions
for each parton ladder, with an initial distribution of the corresponding
parton distribution of the type $\delta(1-x)$, and these singularities
need to be taken care of. Related to this, dealing with singularities
is the major challenge for cross section calculations as well as for
parton generation.

We mentioned already the modular structure of the approach, where
the multiple scattering is expressed in terms of the cut Pomerons
$G$, and the latter one corresponds to some function $G_{\mathrm{QCD}}$.
This function $G_{\mathrm{QCD}}$ itself has a modular structure,
the modules being vertex functions, a soft evolution function, and
most importantly the ``cut parton ladder'', which is (up to an impact
parameter dependent factor) equal to a parton-parton cross section.
In section \ref{========00003DEPOS4-building-block========00003D=00003D},
we will discuss how the EPOS4 building block $G_{\mathrm{QCD}}$ is
related to parton-parton cross sections, and in section \ref{========00003Dparton-parton-xsections========00003D=00003D},
we will discuss how the integrated and differential parton-parton  cross sections can be computed. The former are needed to compute $G_{\mathrm{QCD}}$, while the latter are necessary for the parton generation.

\section{Relating the EPOS4 building block $\boldsymbol{\boldsymbol{G_{\mathrm{QCD}}}}$
to parton-parton cross sections \label{========00003DEPOS4-building-block========00003D=00003D}}

As discussed in the last section (see also \cite{werner:2023-epos4-overview}),
the multiple scattering contributions to the total cross section are
expressed in terms of (products of) cut Pomeron expressions $G$,
and the latter ones are related the ``real QCD expressions'' $G_{\mathrm{QCD}}$
via Eq. (\ref{fundamental-epos4-equation}). In this sense, $G_{\mathrm{QCD}}$
is the basis of everything, the fundamental building block of EPOS4. 
This $G_{\mathrm{QCD}}$ depends on the saturation scale $Q_{\mathrm{sat}}^{2}$,
which is of fundamental importance in the   EPOS4 framework 
\cite{werner:2023-epos4-overview}, but here the focus is on 
the details for the calculation of $G_{\mathrm{QCD}}$, where $Q_{\mathrm{sat}}^{2}$
is ``only'' a constant, representing the low virtuality cutoff (and we use simply symbols
like $Q_1^2$ and $Q_2^2$ as arguments).

We will use different symbols, like $G$, $T$, $\boldsymbol{\mathrm{T}}$,
and $\sigma$, all being related to each other. For clarity, we recall
the definitions in Table \ref{the-four-symbols}.
\begin{table}[h]
\begin{tabular*}{1\columnwidth}{@{\extracolsep{\fill}}|c|l|}
\hline 
 $\boldsymbol{\mathrm{T}}$  & %
\begin{minipage}[t]{0.85\columnwidth}%
Diagonal element of the elastic scattering T-matrix as defined in
standard quantum mechanics textbooks, where the asymptotic state is
a system of two protons or two nuclei%
\end{minipage}\tabularnewline
\hline 
 $T$  & %
\begin{minipage}[t]{0.85\columnwidth}%
Fourier transform with respect to the transverse momentum exchange
of the elastic scattering T-matrix $\boldsymbol{\mathrm{T}}$, divided
by $2s$ (formulas are simpler using this representation)%
\end{minipage}\tabularnewline
\hline 
$G$  & %
\begin{minipage}[t]{0.85\columnwidth}%
Defined as $G=\mathrm{cut}\,T=2\mathrm{Im}T=\frac{1}{i}\mathrm{disc}T$
(where ``disc'' refers to the variable $s$), referring to the inelastic
process associated with the cut of the elastic diagram corresponding
to $T$%
\end{minipage}\tabularnewline
\hline 
 $\sigma$  & %
\begin{minipage}[t]{0.85\columnwidth}%
Integrated inclusive parton-parton scattering cross section, which
is useful because $\boldsymbol{\mathrm{T}}$, $T$, and $G$ may be
expressed in terms of $\sigma$%
\end{minipage}\tabularnewline
\hline 
\end{tabular*}\caption{The symbols $G$, $T$, $\boldsymbol{\mathrm{T}}$, and $\sigma$.
\label{the-four-symbols}}

\end{table}

Concerning the structure of $G_{\mathrm{QCD}}$, we expect on both
projectile and target side two possibilities, namely a valence quark
or a sea quark/gluon being the first perturbative parton, and in addition
soft contributions, correspondingly we have 
\begin{align}
G_{\mathrm{QCD}} & =G_{\mathrm{QCD}}^{\mathrm{sea-sea}}+G_{\mathrm{QCD}}^{\mathrm{sea-val}}+G_{\mathrm{QCD}}^{\mathrm{val-sea}}+G_{\mathrm{QCD}}^{\mathrm{val-val}}\nonumber \\
 & +G_{\mathrm{soft}}+G_{\mathrm{psoft}}.
\end{align}

In this section, we will discuss the different $G$'s one by one,
and in particular how the $G_{\mathrm{QCD}}^{K}$ are related to parton-parton
cross sections $\sigma$, the latter one discussed in great detail
in section \ref{========00003Dparton-parton-xsections========00003D=00003D}.

\subsection{The usual factorization approach for proton-proton scattering}

Certain elements of the EPOS4 scheme are similar to the usual factorization
approach, which we will briefly review in the following. Based on
the factorization hypothesis, the inclusive parton production cross
section in proton-proton scattering is given as \cite{Ellis:1996}
\begin{align}
 & E_{3}E_{4}\frac{d^{6}\sigma_{\mathrm{incl}}}{d^{3}p_{3}d^{3}p_{4}}=\sum_{klmn}\int\!\int\!\!dx_{1}dx_{2}\,f_{\mathrm{PDF}}^{k}(x_{1},\mu_{\mathrm{F}}^{2})f_{\mathrm{PDF}}^{l}(x_{2},\mu_{\mathrm{F}}^{2})\nonumber \\
 & \qquad\frac{1}{32s\pi^{2}}\bar{\sum}|\mathcal{M}^{kl\to mn}|^{2}\delta^{4}(p_{1}+p_{2}-p_{3}-p_{4})\frac{1}{1+\delta_{mn}},\label{eq:factoriz}
\end{align}
where the parton distribution function (PDF) $f_{\mathrm{PDF}}^{a}$
represents the number of partons of species $a$ entering the hard
scattering, and where $p_{1}$ and $p_{2}$ ($p_{3}$ and $p_{4}$)
are the momenta of the incoming (outgoing) partons. The PDFs can be
considered to be known from deep inelastic electron-proton scattering,
the amplitudes $\mathcal{M}$ can be computed, and thus one obtains
the inclusive cross section as a simple integral. The factorization
scale $\mu_{\mathrm{F}}^{2}$ represents the scale which allows to
separate ``long range (soft)'' and ``short range (hard)'' parts
of the interaction, such that the former ones are part of the
proton structure, whereas the latter ones show up in the perturbative
matrix elements $\mathcal{M}$. This procedure has been shown to successfully
describe experimental jet production data. In the EPOS4 framework,
a similar ``factorization formula'' will be used but in a different
context. %

\subsection{Cut parton ladder $\boldsymbol{\boldsymbol{G_{\mathrm{QCD}}^{\mathrm{hard}}}}$
and the parton-parton cross section $\boldsymbol{\boldsymbol{\sigma_{\mathrm{hard}}^{ij}}}$}

\begin{figure}[h]
\centering{}\includegraphics[scale=0.25]{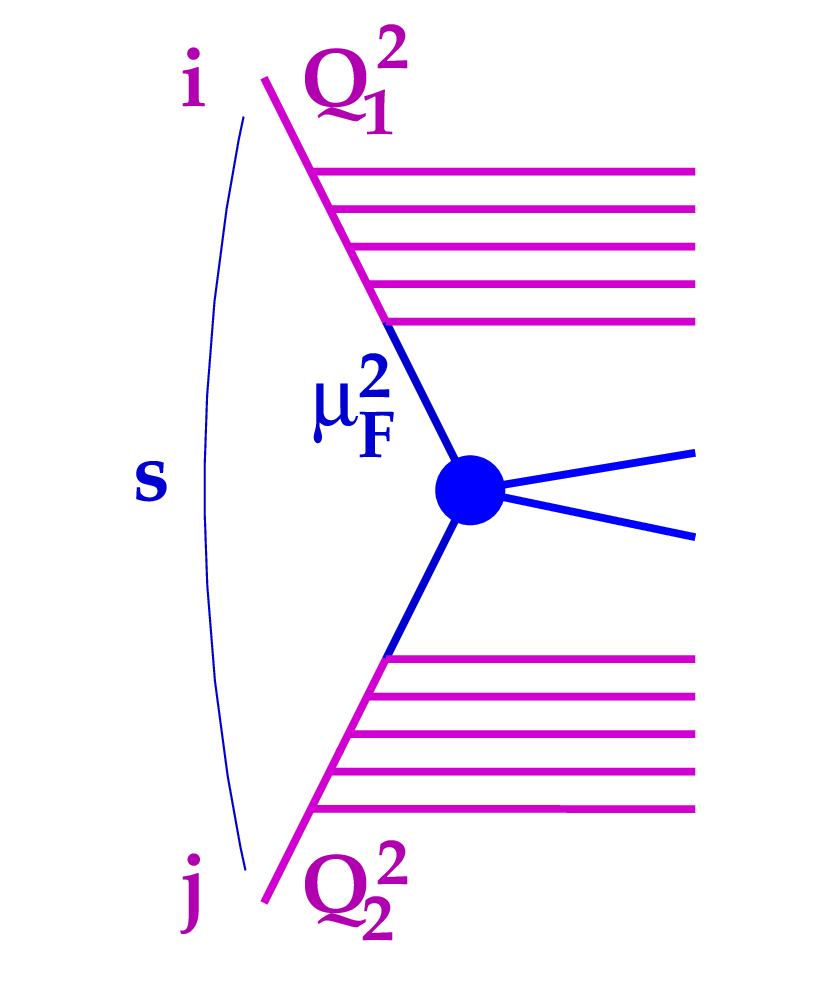}\caption{The parton-parton cross section $\sigma_{\mathrm{hard}}^{ij}(Q_{1}^{2},Q_{2}^{2},s)$.\label{ladder-3}}
\end{figure}

\begin{figure}[h]
\centering{}\includegraphics[scale=0.35]{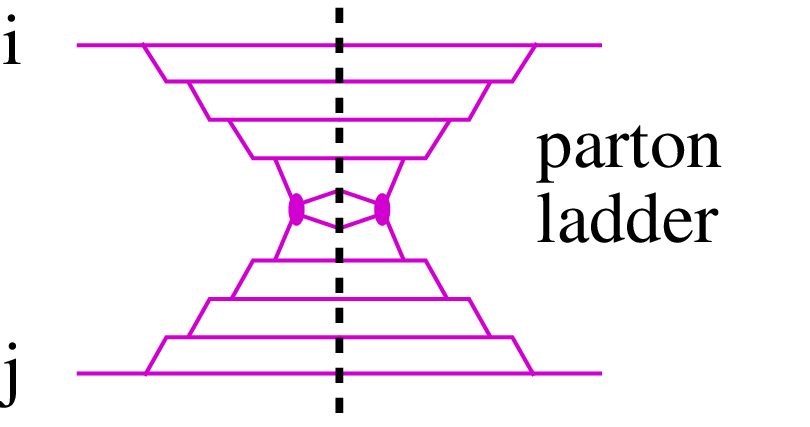}\caption{Cut parton ladder $G_{\mathrm{QCD}}^{\mathrm{hard,ij}}$.\label{single-pomeron-graph-1}}
\end{figure}

We first discuss a somewhat simplified situation, namely a simple
parton ladder. Let us consider the scattering of two partons with
flavors $i$ and $j$ carrying virtualities $Q_{1}^{2}$ and $Q_{2}^{2}$
, both first evolving via parton emissions with ordered virtualities
up to $\mu_{F}^{2}$ before interacting as an elementary $2\to2$
pQCD (hard) scattering (see Fig. \ref{ladder-3}).
The corresponding di-jet production
cross section can be written as 
\begin{align}
 & \sigma_{\mathrm{hard}}^{ij}(Q_{1}^{2},Q_{2}^{2},s)=\sum_{klmn}\int dx_{1}dx_{2}\int\frac{d^{3}p_{3}d^{3}p_{4}}{E_{3}E_{4}}\nonumber \\
 & \qquad E_{\mathrm{QCD}}^{ik}(x_{1},Q_{1}^{2},\mu_{\mathrm{F}}^{2})E_{\mathrm{QCD}}^{jl}(x_{2},Q_{2}^{2},\mu_{\mathrm{F}}^{2})\label{definition-sigma-hard}\\
 & \qquad\frac{1}{2s}\,\frac{1}{16\pi^{2}}\bar{\sum}|\mathcal{M}^{kl\to mn}|^{2}\delta^{4}(p_{1}+p_{2}-p_{3}-p_{4})\frac{1}{1+\delta_{mn}},\nonumber 
\end{align}
where the momenta of the outgoing partons (jets) are integrated out.
This expression is very similar to the usual factorization formula
Eq. (\ref{eq:factoriz}), but with the parton distribution functions
$f_{\mathrm{PDF}}^{k}(x_{i},\mu_{\mathrm{F}}^{2})$ replaced by $E_{\mathrm{QCD}}^{ik}(x_{K},Q_{K}^{2},\mu_{\mathrm{F}}^{2})$,
representing parton evolution starting at virtuality $Q_{K}^{2}$
with a distribution $\delta(x-1)\delta_{ki}$, but using the same
DGLAP evolution \cite{GribovLipatov:1972,AltarelliParisi:1977,Dokshitzer:1977}.
The corresponding cut diagram, referred to as ``cut parton ladder'',
is shown in Fig. \ref{single-pomeron-graph-1}.
Concerning the corresponding elastic scattering T-matrix, we assume
\begin{equation}
\boldsymbol{\boldsymbol{\mathrm{T}}}_{\mathrm{hard}}^{ij}(Q_{1}^{2},Q_{2}^{2},s,t)=i\,s\,\sigma_{\mathrm{hard}}^{ij}(Q_{1}^{2},Q_{2}^{2},s)\times\exp(R_{\mathrm{hard}}^{2}t),
\end{equation}
with $R_{\mathrm{hard}}^{2}\approx0$ \cite{GribovLipatov:1972},
which is compatible with the usual relation $\sigma_{\mathrm{hard}}^{ij}=2\mathrm{Im}\,\boldsymbol{\mathrm{T}}_{\mathrm{hard}}^{ij}(t=0)/(2s)$.
Furthermore assuming a purely transverse momentum exchange $t=-q_{\bot}^{2}$,
the Fourier transform and division by $2s$ gives %
\begin{align}
 & T_{\mathrm{hard}}^{ij}(Q_{1}^{2},Q_{2}^{2},s,b)=\frac{1}{8\pi^{2}s}\int d^{2}q_{\bot}e^{-iq_{\bot}b}\boldsymbol{\boldsymbol{\mathrm{T}}}_{\mathrm{hard}}^{ij}(s,t)\label{g-hard-0}\\
 & =\frac{i}{2}\sigma_{\mathrm{hard}}^{ij}(Q_{1}^{2},Q_{2}^{2},s)\,\frac{1}{4\pi R_{\mathrm{hard}}^{2}}\exp\left(-\frac{b^{2}}{4R_{\mathrm{hard}}^{2}}\right).\label{g-hard-1}
\end{align}
For the corresponding $G=\mathrm{cut}\,T_{\mathrm{hard}}=2\mathrm{Im}\,T_{\mathrm{hard}}$,
we get
\begin{align}
 & G_{\mathrm{QCD}}^{\mathrm{hard,ij}}(Q_{1}^{2},Q_{2}^{2},s,b)\label{g-hard}\\
 & =\sigma_{\mathrm{hard}}^{ij}(Q_{1}^{2},Q_{2}^{2},s)\,\frac{1}{4\pi R_{\mathrm{hard}}^{2}}\exp\left(-\frac{b^{2}}{4R_{\mathrm{hard}}^{2}}\right).\nonumber 
\end{align}
So the cut parton ladder expression $G$ is simply the product of
the dijet production cross section $\sigma_{\mathrm{hard}}^{ij}$
times a Gaussian impact parameter dependence. 

\subsection{Relating $\boldsymbol{\boldsymbol{G_{\mathrm{QCD}}^{\mathrm{val-val}}}}$
to the parton-parton cross section $\boldsymbol{\boldsymbol{\sigma_{\mathrm{hard}}^{ij}}}$}

Here we consider the ``val-val'' contribution, where on both sides
a valence quark is the first perturbative parton. We might imagine
in the multiple Pomeron formula (as Eq. (\ref{multiple-tmatrix-pp}),
but using $\boldsymbol{\mathrm{T}}$ and not $T$) for each ``val-val''
Pomeron an expression like 
\begin{equation}
\sum_{i,j}\int\frac{dz^{+}dz^{-}}{z^{+}z^{-}}\boldsymbol{\mathrm{T}}_{\mathrm{hard}}^{ij}(...,z^{+}z^{-}s,...)F^{i}(z^{+})F^{j}(z^{-}),\label{val-val-initial-formula}
\end{equation}
where the indices $i$ and $j$ refer to the flavors of the valence
quarks, with some ``vertex functions'' $F^{a}$, and with $s$
being the proton-proton squared cms energy. 

\begin{figure}[h]
\centering{}\includegraphics[scale=0.35]{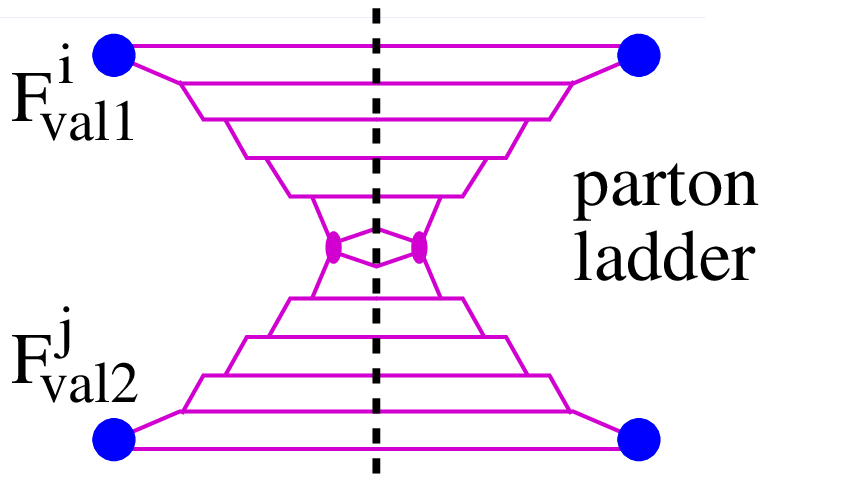}\caption{The contribution $G_{\mathrm{QCD}}^{\mathrm{val-val}}$ . \label{G-val-val-1}}
\end{figure}

However, the external legs of our Pomerons should always be colorless
objects, carrying light-cone momentum fractions $x^{+}$on the projectile
and $x^{-}$ on the target side. So each valence quark has a partner
(antiquark or diquark), with the valence quark carrying a light-cone
momentum fraction $z^{\pm}$ and its partner $x^{\pm}-z^{\pm}$, the
sum of both being $x^{\pm}.$ The partner is emitted (as a timelike
parton) immediately, and the valence quark starts the parton evolution.
The corresponding cut diagram is indicated in Fig. \ref{G-val-val-1}.
We imagine vertex functions $F_{\mathrm{val}\,1}^{i}$ and $F_{\mathrm{val}\,2}^{j}$
with three arguments each: the light-cone momentum fractions of  the
valence quark and of its partner, and the Mandelstam $t$. Let us
first look at the T-matrix. Instead of Eq. (\ref{val-val-initial-formula}),
we use for each ``val-val'' Pomeron an expression (including integration)
like 
\begin{align}
 & \sum_{i,j}\int\frac{dz^{+}dz^{-}}{z^{+}z^{-}}\int dx^{+}dx^{-}\boldsymbol{\mathrm{T}}_{\mathrm{hard}}^{ij}(Q_{0}^{2},Q_{0}^{2},z^{+}z^{-}s,t)\nonumber \\
 & \qquad\qquad F_{\mathrm{val}\,1}^{i}(z^{+},x^{+}-z^{+},t)F_{\mathrm{val}\,2}^{j}(z^{-},x^{-}-z^{-},t),
\end{align}
which can then be written as an integration $\int dx^{+}dx^{-}/(x^{+}x^{-})$
referring to the ``white'' legs, with an integrand that can be
interpreted as the corresponding T-matrix, i.e. 
\begin{equation}
\int\frac{dx^{+}dx^{-}}{x^{+}x^{-}}\mathrm{\mathbf{T}}_{\mathrm{val}-\mathrm{val}}\!\left(x^{+},x^{-},s,t\right),
\end{equation}
with the ``val-val'' T-matrix
\begin{eqnarray}
 &  & \mathrm{\mathbf{T}}_{\mathrm{val}-\mathrm{val}}\!\left(x^{+},x^{-},s,t\right)\label{x}\\
 &  & =\sum_{i,j}\int\!\!\!dz^{+}dz^{-}\frac{x^{+}x^{-}}{z^{+}z^{-}}\:\mathrm{\mathbf{T}}_{\mathrm{hard}}^{ij}\left(Q_{0}^{2},Q_{0}^{2},z^{+}z^{-}s,t\right)\nonumber \\
 &  & \qquad F_{\mathrm{val}\,1}^{i}(z^{+},x^{+}-z^{+},t)\;F_{\mathrm{val}\,2}^{j}(z^{-},x^{-}-z^{-},t).\nonumber 
\end{eqnarray}
For $F_{\mathrm{val}\,1}^{i}$, $F_{\mathrm{val}\,2}^{j}$, and $\mathrm{\mathbf{T}}_{\mathrm{hard}}^{ij}$,
the $t$ dependence can be factored out as $e^{R^{2}t},$ with parameters
$R_{\mathrm{val}\,1}^{2}$, $R_{\mathrm{val}\,2}^{2}$, and $R_{\mathrm{hard}}^{2}$.
We compute $G$ as usual as twice the imaginary part of the Fourier
transform of the T-matrix divided by $2sx^{+}x^{-}$, and we get (using
Eqs. (\ref{g-hard-0},\ref{g-hard-1}) with $z^{+}z^{-}s$ instead
of $s$)\\
\noindent\fbox{\begin{minipage}[t]{1\columnwidth - 2\fboxsep - 2\fboxrule}%
\vspace*{-0.3cm}
\begin{eqnarray}
 &  & G_{\mathrm{QCD}}^{\mathrm{val}-\mathrm{val}}(x^{+},x^{-},s,b)=\sum_{i,j}\int dz^{+}dz^{-}\label{final-G-val-val}\\
 &  & \quad F_{\mathrm{val}\,1}^{i}(z^{+},x^{+}-z^{+},0)\,F_{\mathrm{val}\,2}^{j}(z^{-},x^{-}-z^{-},0)\nonumber \\
 &  & \quad\sigma_{\mathrm{hard}}^{ij}\!\left(Q_{0}^{2},Q_{0}^{2},z^{+}z^{-}s\right)\,\frac{1}{4\pi R^{2}}\exp\!\left(-\frac{b^{2}}{4R^{2}}\right)\nonumber 
\end{eqnarray}
\vspace*{-0.3cm}%
\end{minipage}}\medskip{}
\\
with
\begin{equation}
R^{2}=R_{\mathrm{hard}}^{2}+R_{\mathrm{val}\,1}^{2}+R_{\mathrm{val}\,2}^{2}.
\end{equation}

\subsection{Relating $\boldsymbol{\boldsymbol{G_{\mathrm{QCD}}^{\mathrm{sea-sea}}}}$
to the parton-parton cross section $\boldsymbol{\boldsymbol{\sigma_{\mathrm{hard}}^{ij}}}$ }

\begin{figure}[h]
\centering{}\includegraphics[scale=0.35]{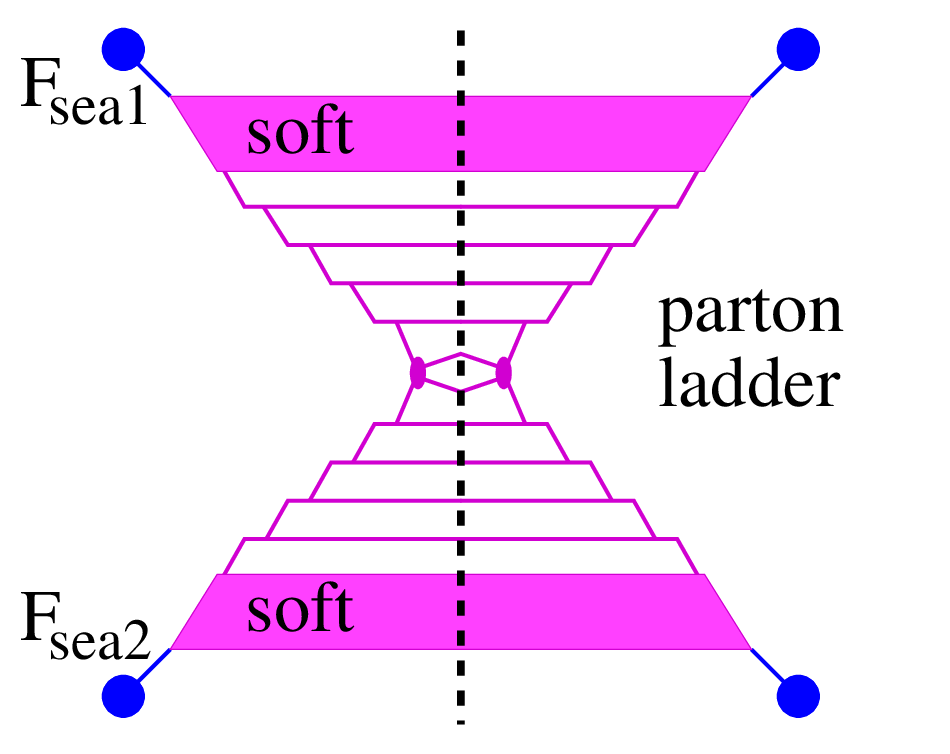}\caption{The contribution $G_{\mathrm{QCD}}^{\mathrm{sea-sea}}$ . \label{G-sea-sea}}
\end{figure}

For the ``sea-sea'' contribution, we expect a ``soft block'' preceding
the first pertubative parton, as indicated in Fig. \ref{G-sea-sea}.
The vertices  $F_{\text{sea}\,1}$ and $F_{\text{sea}\,2}$ couple the parton ladder to the projectile and target nucleons. In addition, we have
three blocks, the two soft blocks and in between the parton ladder
discussed earlier. The corresponding elastic scattering T-matrix for
the latter is $\boldsymbol{\boldsymbol{\mathrm{T}}}_{\mathrm{hard}}^{jk}$
and those for the soft blocks are per definition $\boldsymbol{\boldsymbol{\mathrm{T}}}_{\mathrm{soft}}^{j}$
and $\boldsymbol{\boldsymbol{\mathrm{T}}}_{\mathrm{soft}}^{k}$, with
$j$ and $k$ being the flavors of the two parton which connect the
soft blocks to the parton ladder. In principle we have for each of
these partons a four-dimensional loop integral, but based on the assumption
that transverse momenta and virtualities are negligible compared
to longitudinal momenta, they can be reduced to one-dimensional integrals
\cite{Drescher:2000ha},
\begin{align}
 & i\boldsymbol{\boldsymbol{\mathrm{T}}}_{\mathrm{sea-sea}}(Q_{1}^{2},Q_{2}^{2},x^{+},x^{-},s,t)\nonumber \\
 & =\sum_{ij}\int\!\frac{dz^{+}}{z^{+}}\frac{dz^{-}}{z^{-}}\mathrm{Im}\,\boldsymbol{\boldsymbol{\mathrm{T}}}_{\mathrm{soft}}^{i}\!\!\left(Q_{1}^{2},\frac{s_{0}}{z^{+}},t\right)\,\mathrm{Im}\,\boldsymbol{\boldsymbol{\mathrm{T}}}_{\mathrm{soft}}^{j}\!\!\left(Q_{2}^{2},\frac{s_{0}}{z^{-}},t\right)\nonumber \\
 & \qquad i\boldsymbol{\boldsymbol{\mathrm{T}}}_{\mathrm{hard}}^{ij}(Q_{1}^{2},Q_{2}^{2},z^{+}z^{-}x^{+}x^{-}s,t)F_{\mathrm{sea\,1}}(x^{+},t)\;F_{\mathrm{sea}\,2}(x^{-},t)\label{T-sea-sea}
\end{align}
What is the reason for getting the imaginary part of the soft T-matrices?
The loop integrals may be written as $\int dk^{+}dk^{-}d^{2}k_{t}...$,
and the $k^{-}$ variable can be related to the Mandelstam variables
$s$ and $u$ (for the soft block). Then the usual branch cuts translate
into cuts for $k^{-}$, and a rotation of the integration path, 
\begin{figure}[h]
\centering{}%
\begin{minipage}[c]{0.4\columnwidth}%
\includegraphics[scale=0.25]{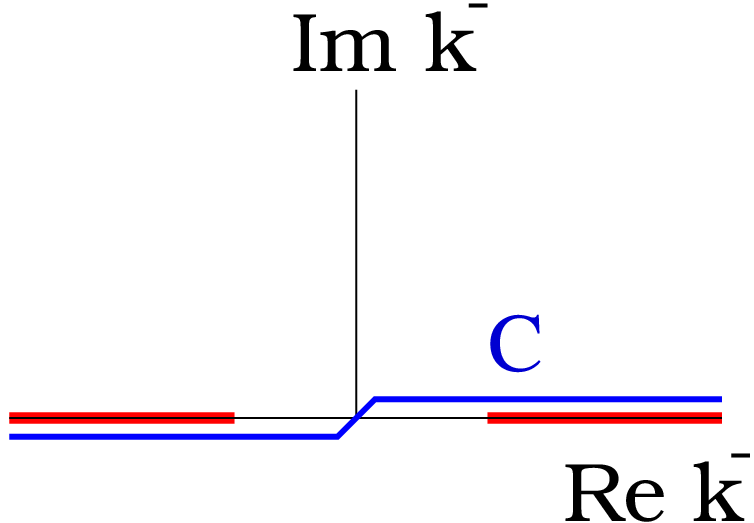}%
\end{minipage}%
\noindent\begin{minipage}[c]{0.1\columnwidth}%
\begin{center}
{\LARGE{}$\mathbf{\mathbf{\to}}$}
\par\end{center}%
\end{minipage}%
\begin{minipage}[c]{0.4\columnwidth}%
{}%
\begin{minipage}[c]{0.4\columnwidth}%
\includegraphics[scale=0.25]{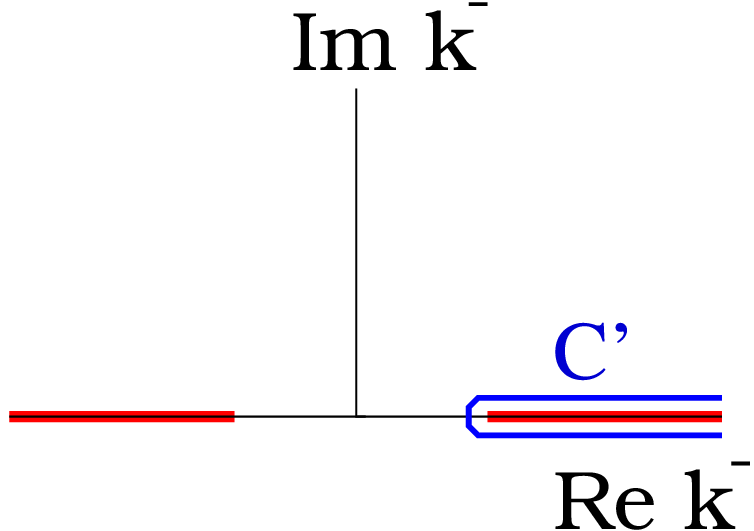}%
\end{minipage}%
\end{minipage}\caption{The  integration path in the $k^{-}$ plane. \label{path-kt-plane}}
\end{figure}
 as shown in Fig. \ref{path-kt-plane}, transforms $\int_{-\infty}^{\infty}dk^{-}...$
into $\int_{0}^{\infty}dk^{-}\mathrm{disc}...$ which is equal to
$\int_{0}^{\infty}dk^{-}\mathrm{Im}...$ . In order to compute
\begin{align}
 & G_{\mathrm{QCD}}^{\mathrm{sea-sea}}(Q_{1}^{2},Q_{2}^{2},x^{+},x^{-},s,b)\\
 & \qquad=2\mathrm{Im}\,T_{\mathrm{sea-sea}}(Q_{1}^{2},Q_{2}^{2},x^{+},x^{-},s,b),\nonumber 
\end{align}
with as usual $T$ being the Fourier transform divided by $2\hat{s}=2x^{+}x^{-}s$
of the T-matrix, 
\begin{eqnarray}
 &  & T_{\mathrm{sea-sea}}(Q_{1}^{2},Q_{2}^{2},x^{+},x^{-},s,b)=\frac{1}{8\pi^{2}x^{+}x^{-}s}\\
 &  & \quad\int d^{2}q_{\bot}e^{-iq_{\bot}b}\boldsymbol{\boldsymbol{\mathrm{T}}}_{\mathrm{sea-sea}}(Q_{1}^{2},Q_{2}^{2},x^{+},x^{-},s,t),\nonumber 
\end{eqnarray}
we note that for all T-matrices, the $t$ dependence can be factored
out as
\begin{equation}
\boldsymbol{\boldsymbol{\mathrm{T}}}_{\mathrm{soft}}^{i}\!\!\left(...,t\right)=\boldsymbol{\boldsymbol{\mathrm{T}}}_{\mathrm{soft}}^{i}\!\!\left(...,0\right)\times\exp\left(\left(R_{\mathrm{soft}}^{2}+\alpha'\ln(\hat{s})\right)\,t\right),
\end{equation}
\begin{equation}
\boldsymbol{\boldsymbol{\mathrm{T}}}_{\mathrm{hard}}^{ij}(...,t)=\boldsymbol{\boldsymbol{\mathrm{T}}}_{\mathrm{hard}}^{ij}(...,0)\times\exp\left(R_{\mathrm{hard}}^{2}\,t\right).
\end{equation}
The $t$ dependence of $F_{\mathrm{sea}\,1}$ and $F_{\mathrm{sea}\,2}$ can also be factored out as  $e^{R_{\mathrm{sea}\,K}^{2}t},$
with parameters $R_{\mathrm{sea}\,1}^{2}$, $R_{\mathrm{sea}\,2}^{2}$,
which gives a simple overall $t$ dependence of the r.h.s of Eq. (\ref{T-sea-sea})
as $e^{R^{2}t}$, so that we get easily 
\begin{eqnarray}
 &  & G_{\mathrm{sea-sea}}(Q_{1}^{2},Q_{2}^{2},x^{+},x^{-},s,b)=\sum_{ij}\int\!dz^{+}dz^{-}\nonumber \\
\nonumber \\
 &  & \qquad\mathrm{Im}\,\boldsymbol{\boldsymbol{\mathrm{T}}}_{\mathrm{soft}}^{i}\!\!\left(Q_{1}^{2},\frac{s_{0}}{z^{+}},0\right)\mathrm{Im}\,\boldsymbol{\boldsymbol{\mathrm{T}}}_{\mathrm{soft}}^{j}\!\!\left(Q_{2}^{2},\frac{s_{0}}{z^{-}},0\right)\\
 &  & \qquad\frac{1}{z^{+}z^{-}x^{+}x^{-}s}\mathrm{Im}\boldsymbol{\boldsymbol{\mathrm{T}}}_{\mathrm{hard}}^{ij}(Q_{1}^{2},Q_{2}^{2},z^{+}z^{-}x^{+}x^{-}s,0)\nonumber \\
 &  & \qquad F_{\mathrm{sea\,1}}(x^{+},0)\;F_{\mathrm{sea}\,2}(x^{-},0)\:\frac{1}{4\pi R^{2}}\exp\left(-\frac{b^{2}}{4R^{2}}\right).\nonumber 
\end{eqnarray}
Using
\begin{equation}
\frac{1}{\hat{s}}\mathrm{Im}\boldsymbol{\boldsymbol{\mathrm{T}}}_{\mathrm{hard}}^{ij}(Q_{1}^{2},Q_{2}^{2},\hat{s},0)=\sigma_{\mathrm{hard}}^{ij}(Q_{1}^{2},Q_{2}^{2},\hat{s}),
\end{equation}
and defining so-called ``soft evolution functions''
\begin{equation}
E_{\mathrm{soft}}^{i}(Q^{2},z)=\mathrm{Im}\,\boldsymbol{\boldsymbol{\mathrm{T}}}_{\mathrm{soft}}^{i}\!\!\left(Q^{2},\frac{s_{0}}{z},0\right),\label{definition-e-soft}
\end{equation}
we get finally\\
\noindent\fbox{\begin{minipage}[t]{1\columnwidth - 2\fboxsep - 2\fboxrule}%
\vspace*{-0.3cm}
\begin{align}
 & G_{\mathrm{QCD}}^{\mathrm{sea-sea}}(Q_{1}^{2},Q_{2}^{2},x^{+},x^{-},s,b)=\sum_{ij}\int\!dz^{+}dz^{-}\label{final-G-sea-sea}\\
 & \qquad\quad F_{\mathrm{sea\,1}}(x^{+},0)\,F_{\mathrm{sea}\,2}(x^{-},0)\,E_{\mathrm{soft}}^{i}(Q_{1}^{2},z^{+})E_{\mathrm{soft}}^{j}(Q_{2}^{2},z^{-})\nonumber \\
 & \qquad\quad\sigma_{\mathrm{hard}}^{ij}(Q_{1}^{2},Q_{2}^{2},z^{+}z^{-}x^{+}x^{-}s)\,\frac{1}{4\pi R^{2}}\exp\left(-\frac{b^{2}}{4R^{2}}\right),\nonumber 
\end{align}
\vspace*{-0.3cm}%
\end{minipage}}\medskip{}
\\
with $R$ being given explicitly as 
\begin{equation}
R^{2}=R_{\mathrm{sea}\,1}^{\,2}+R_{\mathrm{sea}\,2}^{\,2}+2R_{\mathrm{soft}}^{2}+\alpha'\ln(\frac{1}{z^{+}})+\alpha'\ln(\frac{1}{z^{-}})+R_{\mathrm{hard}}^{2}.
\end{equation}
The precise form of $E_{\mathrm{soft}}^{i}$ and $F_{\mathrm{sea}\,K}$ 
will be discussed later, the former is based on a parametrization
(in Regge-pole form) of the soft T-matrix, the latter has a simple
power law form. 

\subsection{Relating $\boldsymbol{\boldsymbol{G_{\mathrm{QCD}}^{\mathrm{val-sea}}}}$
and $\boldsymbol{\boldsymbol{G_{\mathrm{QCD}}^{\mathrm{sea-val}}}}$
to the parton-parton cross section $\boldsymbol{\boldsymbol{\sigma_{\mathrm{hard}}^{ij}}}$ }

Having discussed the two contributions ``val-val'' and ``sea-sea'',
referring to contributions with respectively two valence quarks and
two sea quarks as ``first perturbative partons'' entering the parton
ladder, we easily obtain the expressions for ``val-sea'', namely
\\
\noindent\fbox{\begin{minipage}[t]{1\columnwidth - 2\fboxsep - 2\fboxrule}%
\vspace*{-0.3cm}
\begin{eqnarray}
 &  & G_{\mathrm{QCD}}^{\mathrm{val}-\mathrm{sea}}(Q_{2}^{2},x^{+},x^{-},s,b)=\sum_{i,j}\int dz^{+}dz^{-}\label{final-G-val-sea}\\
 &  & \quad F_{\mathrm{val}\,1}^{i}(z^{+},x^{+}-z^{+},0)\,F_{\mathrm{sea}\,2}(x^{-},0)\,E_{\mathrm{soft}}^{j}(Q_{2}^{2},z^{-})\nonumber \\
 &  & \quad\sigma_{\mathrm{hard}}^{ij}\!\left(Q_{0}^{2},Q_{2}^{2},z^{+}z^{-}x^{-}s\right)\,\frac{1}{4\pi R^{2}}\exp\!\left(-\frac{b^{2}}{4R^{2}}\right),\nonumber 
\end{eqnarray}
\vspace*{-0.3cm}%
\end{minipage}}\medskip{}
\\
with $R$ being given explicitly as 
\begin{equation}
R^{2}=R_{\mathrm{val}\,1}^{2}+R_{\mathrm{sea}\,2}^{\,2}+R_{\mathrm{soft}}^{2}+\alpha'\ln(\frac{1}{z^{-}})+R_{\mathrm{hard}}^{2}.
\end{equation}
The expression for ``sea-val'' is \\
\noindent\fbox{\begin{minipage}[t]{1\columnwidth - 2\fboxsep - 2\fboxrule}%
\vspace*{-0.3cm}
\begin{eqnarray}
 &  & G_{\mathrm{QCD}}^{\mathrm{sea}-\mathrm{val}}(Q_{1}^{2},x^{+},x^{-},s,b)=\sum_{i,j}\int dz^{+}dz^{-}\label{final-G-sea-val}\\
 &  & \quad F_{\mathrm{sea\,1}}(x^{+},0)\,E_{\mathrm{soft}}^{i}(Q_{1}^{2},z^{+})\,F_{\mathrm{val}\,2}^{j}(z^{-},x^{-}-z^{-},0)\nonumber \\
 &  & \quad\sigma_{\mathrm{hard}}^{ij}\!\left(Q_{1}^{2},Q_{0}^{2},z^{+}z^{-}x^{+}s\right)\,\frac{1}{4\pi R^{2}}\exp\!\left(-\frac{b^{2}}{4R^{2}}\right),\nonumber 
\end{eqnarray}
\vspace*{-0.3cm}%
\end{minipage}}\medskip{}
\\
with $R$ being given explicitly as 
\begin{equation}
R^{2}=R_{\mathrm{sea}\,1}^{\,2}+R_{\mathrm{soft}}^{2}+\alpha'\ln(\frac{1}{z^{+}})+R_{\mathrm{val}\,2}^{2}+R_{\mathrm{hard}}^{2}.
\end{equation}

\subsection{The vertices \emph{F} }

The main formulas of the preceding sections, Eqs. (\ref{final-G-val-val},\ref{final-G-sea-sea},\ref{final-G-val-sea},\ref{final-G-sea-val}),
allow to express the different $G_{\mathrm{QCD}}^{J}$ in terms of
``modules'', among them the vertices $F_{\mathrm{sea}\,K}$ and
$F_{\mathrm{val}\,K}$ with $K\in\{1,2\}$, which are simple functions,
to be discussed in the following. In the case of pp scattering, the functions
for $K=1$ and $K=2$ are identical, whereas for $\pi$p or Kp scattering
they are in general different (more precisely, the form is the same
but not the parameters). The vertices $F_{\mathrm{sea}\,K}$ and $F_{\mathrm{val}\,K}^{i}$ are given as\\
\noindent\fbox{\begin{minipage}[t]{1\columnwidth - 2\fboxsep - 2\fboxrule}%
\vspace*{-0.3cm}
\begin{align}
F_{\mathrm{sea}\,K}(x) & =\gamma_{\mathrm{sea}\,K}\,\,x^{-\alpha_{\mathrm{sea}\,K}},\\
F_{\mathrm{val}\,K}^{i}(z,z') & =N^{-1}\,q_{\mathrm{val}}^{i}(z)(1-z)^{\alpha_{\mathrm{I}\!\mathrm{R}}-1-\alpha_{\mathrm{remn}}}z'{}^{-\alpha_{\mathrm{I}\!\mathrm{R}}},\label{f-val-1}
\end{align}
\vspace*{-0.3cm}%
\end{minipage}}\medskip{}
\\
with $N=\Gamma\!\left(1+\alpha_{\mathrm{remn}}\right)\,\Gamma\!\left(1-\alpha_{\mathrm{I}\!\mathrm{R}}\right)/\Gamma\!\left(2+\alpha_{\mathrm{remn}}-\alpha_{\mathrm{I}\!\mathrm{R}}\right)$,
with $q_{\mathrm{val}}^{i}$ being a standard valence quark distribution
function for a small value $Q_{0}^{2}$ of the virtuality (see \cite{Drescher:2000ha}
Appendix C.2). From Eq. (\ref{f-val-1}), and using $V(z)=z^{\alpha_{\mathrm{remn}}}$
as remnant vertex, we obtain as parton distribution $f^{i}$ at $Q_{0}^{2}$
\begin{equation}
f^{i}(z)=\int_{0}^{1-z}dz'\,F_{\mathrm{val}}^{i}(z,z')\,V(1-z-z')=q_{\mathrm{val}}^{i}(z),
\end{equation}
having used $\int_{0}^{1}dy\,y{}^{-\alpha_{\mathrm{I}\!\mathrm{R}}}\,(1-y)^{\alpha_{\mathrm{remn}}}=$$\frac{\Gamma(1-\alpha_{\mathrm{I}\!\mathrm{R}})\Gamma(1+\alpha_{\mathrm{remn}})}{\Gamma(2-\alpha_{\mathrm{I}\!\mathrm{R}}+\alpha_{\mathrm{remn}})}$.
So we get $f^{i}(z)=q_{\mathrm{val}}^{i}(z)$, as it should be, which
justifies our choice of $F_{\mathrm{val}\,K}^{i}$.

\subsection{The soft elements\emph{ E}$_{\boldsymbol{\boldsymbol{\mathrm{soft}}}}$
and \emph{G}$_{\boldsymbol{\boldsymbol{\mathrm{soft}}}}$}

Based on the asymptotic Regge-pole expression, the soft T-matrix $\boldsymbol{\mathrm{T}}_{\mathrm{soft}}^{i}(Q^{2},s,t)$
is (for the moment) assumed to not depend on $Q^{2}$ and to be proportional
to to $i\,s^{1+B_{\mathrm{soft}}}\exp\left((2R_{\mathrm{soft}}^{2}+\alpha'\ln\frac{s}{s_{0}})t\right)$,
with a ``Regge-pole intercept'' $1+B_{\mathrm{soft}}=\alpha_{\mathrm{soft}}(0)$,
where $B_{\mathrm{soft}}$ is a parameter close to zero. The ``soft
evolution functions'' $E_{\mathrm{soft}}^{i}(Q^{2},z)$ are defined
in Eq. (\ref{definition-e-soft}) as the imaginary part of $\boldsymbol{\mathrm{T}}_{\mathrm{soft}}^{i}$
for $t=0$ and $s=s_{0}/z$, so they are up to constants equal to
$(s_{0}/z)^{1+B_{\mathrm{soft}}}$. We add a splitting into a quark,
as well as a soft-Pomeron-ladder coupling of the form $(1-z)^{C_{\mathrm{soft}}}$, to get (we drop the $Q^{2}$ argument)\\
\noindent\fbox{\begin{minipage}[t]{1\columnwidth - 2\fboxsep - 2\fboxrule}%
\vspace*{-0.3cm}
\begin{eqnarray}
 &  & E_{\mathrm{soft}}^{g}\left(z\right)=\left(1-w_{\mathrm{split}}\right)\,E_{\mathrm{soft}}\left(z\right),\label{esoft-g-}\\
 &  & E_{\mathrm{soft}}^{q}\left(z\right)=w_{\mathrm{split}}\,\int_{z}^{1}\!d\xi\,f_{\mathrm{split}}(\xi)\,E_{\mathrm{soft}}\left(\frac{z}{\xi}\right),\label{esoft-q-}\\
 & \mathrm{} & \mathrm{with}\,E_{\mathrm{soft}}\left(z\right)=A_{\mathrm{soft}}\,z^{-1-B_{\mathrm{soft}}}\,(1-z)^{C_{\mathrm{soft}}},
\end{eqnarray}
\vspace*{-0.3cm}%
\end{minipage}}%
\medskip{}
\\
with a parameter $w_{\mathrm{split}}\in[0,1] $, and where we use $f_{\mathrm{split}}(\xi)=P_{g}^{q}(\xi)$,
since we are ``close to the perturbative domain''. 

Still based on the soft T-matrix expression $\propto i\,s^{1+B_{\mathrm{soft}}}\exp\left((2R_{\mathrm{soft}}^{2}+\alpha'\ln\frac{s}{s_{0}})t\right)$
and adding the same two vertices as for the ``sea-sea'' contribution,
we get the ``soft'' $G$ expressions as \\

\noindent %
\noindent\fbox{\begin{minipage}[t]{1\columnwidth - 2\fboxsep - 2\fboxrule}%
\vspace*{-0.3cm}
\begin{align}
 & G_{\mathrm{soft}}(x^{+},x^{-},s,b)=\\
 & \qquad\quad F_{\mathrm{sea\,1}}(x^{+},0)\,F_{\mathrm{sea}\,2}(x^{-},0)\nonumber \\
 & \qquad\quad D{}_{\mathrm{soft}}\times(x^{+}x^{-}s){}^{B_{\mathrm{soft}}}\,\frac{1}{4\pi R^{2}}\exp\left(-\frac{b^{2}}{4R^{2}}\right),\nonumber 
\end{align}
\vspace*{-0.3cm}%
\end{minipage}}\medskip{}
\\
with parameters $D_{\mathrm{soft}}$ and $B_{\mathrm{soft}}$, and
with $R$ being given explicitly as 
\begin{equation}
R^{2}=R_{\mathrm{sea}\,1}^{\,2}+R_{\mathrm{sea}\,2}^{\,2}+2R_{\mathrm{soft}}^{2}+\alpha'\ln(\frac{s}{s_{0}}).
\end{equation}

\subsection{The pseudosoft elements\emph{ E}$_{\boldsymbol{\boldsymbol{\mathrm{psoft}}}}$
and \emph{G}$_{\boldsymbol{\boldsymbol{\mathrm{psoft}}}}$}

The ``soft evolution functions'' $E_{\mathrm{soft}}^{i}$ are meant
to take care of the ``non-perturbative part'', before entering the
perturbative regime represented by $E_{\mathrm{QCD}}^{ik}(x,Q^{2},\mu_{\mathrm{F}}^{2})$.
Originally, we were using $Q^{2}$ being equal to some soft scale
$Q_{0}^{2}$ (typically  $1-2\,\mathrm{GeV}^{2}$).
In that case,  $E_{\mathrm{soft}}^{i}$ represents purely soft physics,
and we could actually drop $Q^{2}$ as an argument, $E_{\mathrm{soft}}^{i}=E_{\mathrm{soft}}^{i}(z)$. 
However, now we introduce evolution functions $E_{\mathrm{QCD}}^{ik}(x,Q_{\mathrm{sat}}^{2},\mu_{\mathrm{F}}^{2})$
where the virtuality cutoff is equal to the saturation scale, which
is in general bigger than $Q_{0}^{2}$, which means that the ``part
not taken care of'' by the QCD evolution, is not purely soft anymore. This is the domain of gluon fusions, but there should be emissions
as well and a narrowing of the momentum fraction distributions. To
take this into account we define ``pseudosoft evolution functions''
as\\

\noindent %
\noindent\fbox{\begin{minipage}[t]{1\columnwidth - 2\fboxsep - 2\fboxrule}%
\vspace*{-0.3cm}
\begin{align}
 & E_{\mathrm{psoft}}^{k}(Q_{\mathrm{sat}}^{2},\xi)\\
 & =\int dz\,dy\Big\{\sum_{i}E_{\mathrm{soft}}^{i}(z)E_{\mathrm{QCD}}^{ik}(y,Q_{0}^{2},Q_{\mathrm{sat}}^{2})\delta(\xi-zy)\Big\},\nonumber 
\end{align}
\vspace*{-0.3cm}%
\end{minipage}}\medskip{}
\\

\noindent which is meant to replace the soft evolution functions used
so far. This is actually needed to have the same
$p_{t}$ distribution of emitted partons at large $p_{t}$ for different
values of $Q_{\mathrm{sat}}^{2}$.

Having replaced the soft evolution functions with the pseudosoft ones,
we need a modification of $G_{\mathrm{soft}}$ as well. We have little
guidance from theory, so we use the same parametrization as for $G_{\mathrm{soft}}$
\\

\noindent %
\noindent\fbox{\begin{minipage}[t]{1\columnwidth - 2\fboxsep - 2\fboxrule}%
\vspace*{-0.3cm}
\begin{align}
 & G_{\mathrm{psoft}}(Q_{\mathrm{sat}}^{2},x^{+},x^{-},s,b)=\label{g-psoft}\\
 & \qquad\quad F_{\mathrm{sea\,1}}(x^{+},0)\,F_{\mathrm{sea}\,2}(x^{-},0)\nonumber \\
 & \qquad\quad D{}_{\mathrm{psoft}}\times(x^{+}x^{-}s){}^{B_{\mathrm{psoft}}}\,\frac{1}{4\pi R^{2}}\exp\left(-\frac{b^{2}}{4R^{2}}\right),\nonumber 
\end{align}
\vspace*{-0.3cm}%
\end{minipage}}\medskip{}
\\
just with different parameters $B_{\mathrm{psoft}}$ and $D_{\mathrm{psoft}}$,
which may depend on $Q_{\mathrm{sat}}^{2}$. This is enough to compute
$G_{\mathrm{QCD}}$. But because of $Q_{\mathrm{sat}}^{2}>Q_{0}^{2}$,
we expect that hard pQCD processes should occur, producing light and
heavy flavor hadrons. This will be discussed more in the next section. 

\section{Parton-parton cross sections in EPOS4 involving light and heavy flavors
\label{========00003Dparton-parton-xsections========00003D=00003D}}

As discussed in section \ref{=======Introduction=======}
(see also \cite{werner:2023-epos4-overview}), the multiple scattering
contributions to the total cross section are expressed in terms of
(products of) cut Pomeron  expressions $G$, each one representing
a single scattering.  They  are related to the ``real
QCD expressions'' $G_{\mathrm{QCD}}$ via Eq. (\ref{fundamental-epos4-equation}),
in other words, $G_{\mathrm{QCD}}$ is the fundamental building block
of the multiple scattering framework of EPOS4. We showed in section
\ref{========00003DEPOS4-building-block========00003D=00003D},
that $G_{\mathrm{QCD}}$ has several contributions, each one being
composed of ``modules'', with the ``key'' module being the integrated
inclusive parton-parton scattering cross section $\sigma_{\mathrm{hard}}^{ij}$,
where $i$ and $j$ refer to the flavors of the two partons. This
``module'' $\sigma_{\mathrm{hard}}^{ij}$ contains all the pQCD
calculations. We refer to Table \ref{the-four-symbols} in section
\ref{========00003DEPOS4-building-block========00003D=00003D}
to recall the definitions of the symbols $G$ and $\sigma$ and their
relation to the T-matrix. In this section, we discuss in detail how
to compute $\sigma_{\mathrm{hard}}^{ij}$.

Concerning notations: whereas in the previous section the symbol $s$
usually referred to the Mandestam $s$ of nucleon-nucleon scattering,
we will use in this section $s$ and $t$ referring to the Born process
of the hard scattering, which here plays a crucial role. Although
it is not always explicitly written, the pQCD matrix elements $\mathcal{M}$
are always considered to be given in term of $s$ and $t$ (that is how
the are tabulated).

\subsection{Integrated and differential partonic cross sections \label{-------integrated-differential-xs-------}}

\begin{figure}[h]
\centering{}\includegraphics[scale=0.25]{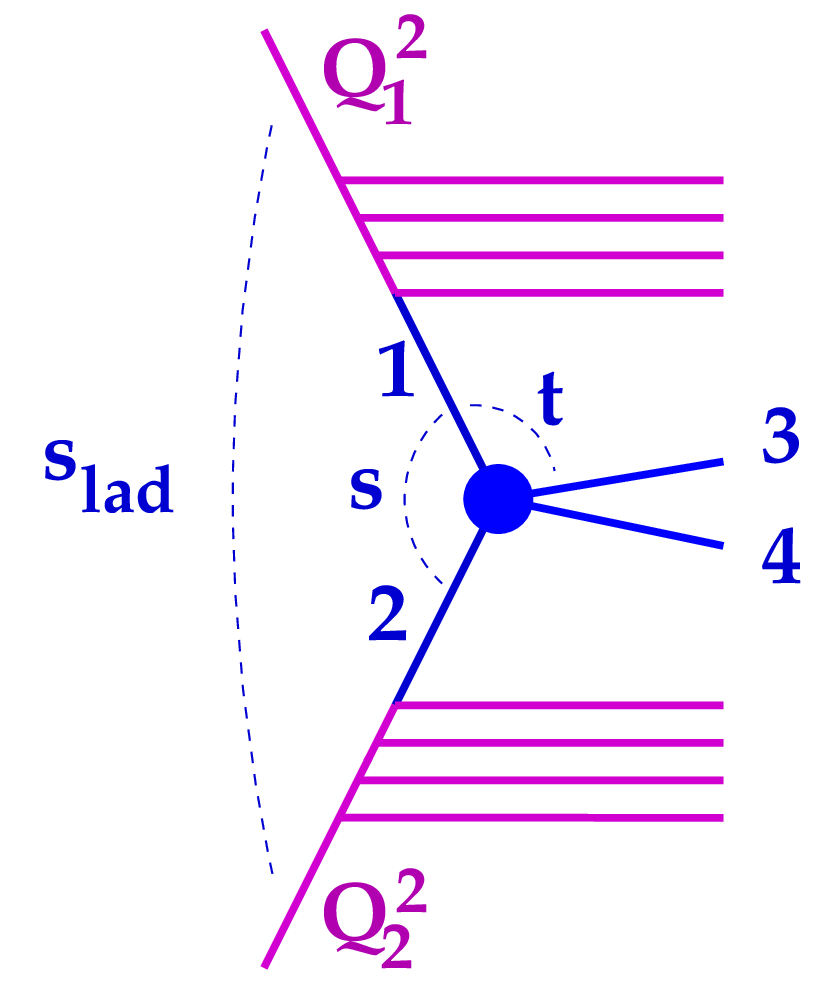}\caption{The differential parton-parton cross section $E_{3}E_{4}d^{6}\sigma_{\mathrm{hard}}^{ij}/d^{3}p_{3}d^{3}p_{4}$
considering the Born process $1+2\to3+4$. The solid lines represent
partons (quarks, antiquarks, or gluons).\label{ladder}}
\end{figure}

The main formula concerning the parton-parton scattering cross section
$\sigma_{\mathrm{hard}}^{ij}$ is Eq. (\ref{definition-sigma-hard}).
It is useful to split that formula, by first considering ``differential
cross sections''
\begin{align}
 & E_{3}E_{4}\frac{d^{6}\sigma_{\mathrm{hard}}^{ij}}{d^{3}p_{3}d^{3}p_{4}}=\sum_{klmn}\int\!\int\!\!dx_{1}dx_{2}\label{differential-cross-section}\\
 & \qquad\,E_{\mathrm{QCD}}^{ik}(x_{1},Q_{1}^{2},\mu_{\mathrm{F}}^{2})E_{\mathrm{QCD}}^{jl}(x_{2},Q_{2}^{2},\mu_{\mathrm{F}}^{2})\nonumber \\
 & \qquad\frac{1}{32s\pi^{2}}\bar{\sum}|\mathcal{M}^{kl\to mn}|^{2}\delta^{4}(p_{1}+p_{2}-p_{3}-p_{4})\frac{1}{1+\delta_{mn}},\nonumber 
\end{align}
representing the inelastic scattering of two partons with
virtualities $Q_{1}^{2}$ and $Q_{2}^{2}$ at a center-of-mass energy
squared $s_{\mathrm{lad}}$, with $x_{1/2}$ being the light-cone (LC)
momentum fractions (with respect to the LC momenta of the partons
before the evolution) of the partons entering the Born process of
the elementary parton-parton scattering $1+2\to3+4$, see Fig. \ref{ladder},
 with the corresponding flavors $k$, $l$, $m$ and $n$. The matrix
elements $\mathcal{M}$ is considered to be given in term of $s$ and
$t$, with $s=x_{1}x_{2}s_{\mathrm{lad}}$. Both partons first evolve
via parton emissions with ordered virtualities up to $\mu_{F}^{2}$
before interacting as an elementary $2\to2$ pQCD scattering. The
``integrated cross sections'' $\sigma_{\mathrm{hard}}^{ij}$ may
then be written as
\begin{equation}
\sigma_{\mathrm{hard}}^{ij}(s_{\mathrm{lad}},Q_{1}^{2},Q_{2}^{2})\!=\!\int\frac{d^{3}p_{3}d^{3}p_{4}}{E_{3}E_{4}}\left\{ E_{3}E_{4}\frac{d^{6}\sigma_{\mathrm{hard}}^{ij}}{d^{3}p_{3}d^{3}p_{4}}\right\} ,\label{integrated-cross-section}
\end{equation}
in terms of the ``differential cross section''. Both differential
and integrated cross sections are important in the EPOS4 framework:
\begin{itemize}
\item the integrated ones are needed to compute the weights for all possible
parallel scattering configurations, and to generate the corresponding configurations,
\item the differential cross sections are needed to compute parton distributions,
and generate partons in the Monte Carlo mode, for given configurations (generated in the first step).
\end{itemize}
Comparing Eq. (\ref{eq:factoriz}) and Eq. (\ref{differential-cross-section}),
we see that our differential cross section has the same structure
as the inclusive cross section based on factorization, but in our
case, we use evolution functions $E_{\mathrm{QCD}}$ rather than the
usual parton distribution functions $f_{\mathrm{PDF}}$, where $E_{\mathrm{QCD}}$
represents an evolution also according to DGLAP \cite{GribovLipatov:1972,AltarelliParisi:1977,Dokshitzer:1977},
but starting from a parton and not from a proton, with an initial
condition $E_{\mathrm{QCD}}^{ik}(z,Q_{1}^{2},Q_{1}^{2})=\delta_{ik}\;\delta(1-z)$.
Special care is needed to treat this ``singularity''.

We will eventually use dynamical saturation scales $Q_{\mathrm{sat\,1}}^{2}$
and $Q_{\mathrm{sat\,2}}^{2}$ as endpoint virtualities, so we compute
(via numerical integration) and then tabulate $\sigma_{\mathrm{hard}}^{ij}(s_{\mathrm{lad}},Q_{1}^{2},Q_{2}^{2})$
for a large range of possible values $s_{\mathrm{lad}}$, $Q_{1}^{2}$,
and $Q_{2}^{2}$ for all possible combinations of $i$ and $j$. During
the simulations, we use the predefined tables to compute $\sigma_{\mathrm{hard}}^{ij}$
via polynomial interpolation.

\begin{figure}[h]
\centering{}\includegraphics[scale=0.22]{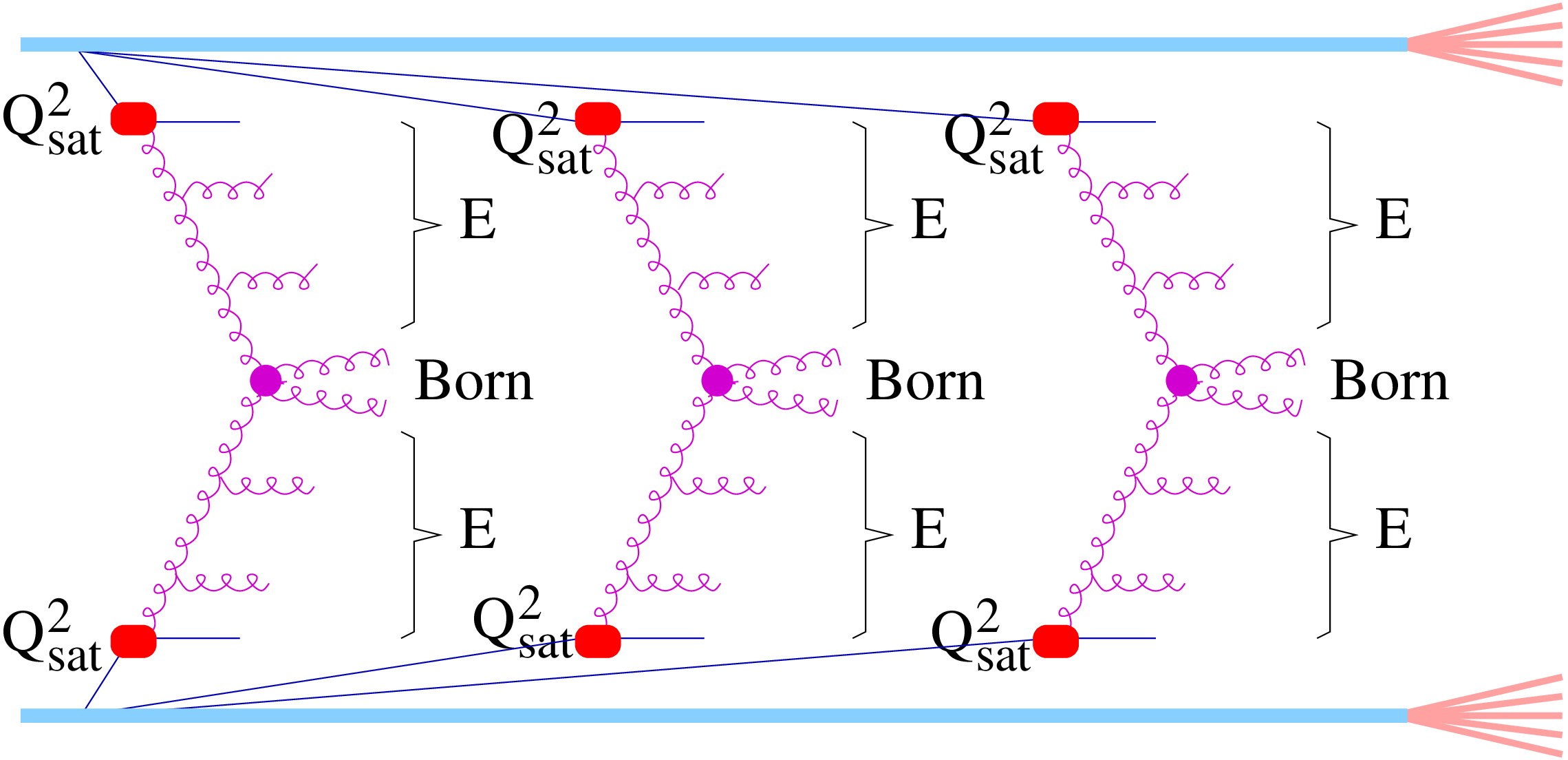}\caption{Multiple scattering configuration for three scatterings. \label{saturation-three-pomerons}}
\end{figure}

As discussed in detail later, all the kinematic variables shown in
fig. \ref{ladder} are related to each other: The scale $\mu_{\mathrm{F}}^{2}$ 
and the Mandelstam variable $t$ are related to $p_{t}^{2}$
(the transverse momentum of one of the
outgoing partons, say parton 3), 
the end virtualities $Q_{1}^{2}$ and $Q_{2}^{2}$
represent lower limits to $p_{t}^{2}$, but the precise relations
depend on the particular Born process. We deal with this problem by
introducing classes of cross sections, with the same Born kinematics
per class. The main reason that we need to be very careful concerning
parton kinematics is the fact that in our case, the differential cross
section formula Eq.~(\ref{differential-cross-section}) concerns one
single scattering in a multiple scattering configuration as shown
in Fig. \ref{saturation-three-pomerons}.
Here we show an example with three scatterings, but in AA collisions,
we have configurations with 1000 scatterings. So it is out of question
to use the standard method of Monte Carlo programming, namely rejection
in the case of forbidden kinematics, we simply have to avoid such cases.

\subsection{Parton evolution functions $\boldsymbol{\boldsymbol{E_{\mathrm{QCD}}}}$
\label{-------parton-evolution-------}}

Our parton evolution functions $E_{\mathrm{QCD}}^{ik}$ are similar
to the ``usual'' ones, obeying the same evolution equations, but
the initial conditions are different, and this requires different treatments.
We present here a new procedure to compute and tabulate these functions,
much faster than the techniques we used in EPOS3. The indices $i$
and $k$ refer to parton flavors in the form of integers with $0$
for a gluon, $1-6$ for quark flavors from $u$ to $t$, and the corresponding
negative numbers for the antiquarks. 

For a given end parton $i$, using $t=Q^{2}$, the evolution equation
for an evolution from $t_{a}$ to $t_{b}$ may be written as \cite{Ellis:1996}%
\begin{align}
 & E_{\mathrm{QCD}}^{ik}\left(x,t_{a},t_{b}\right)=\Delta^{i}(t_{a},t_{b})\delta(1-x)\delta_{ik}\label{evolution-equation-1}\\
 & \qquad+\sum_{j}\int\frac{dt}{t}\int\frac{dz}{z}\,E_{\mathrm{QCD}}^{ij}\left(\frac{x}{z},t_{a},t\right)\bar{P}_{j}^{k}\!(t,z)\,\Delta^{k}(t,t_{b})\nonumber 
\end{align}
with 
\begin{equation}
t_{a}\leq t\le t_{b},\;x\leq z\leq1-\epsilon\:.
\end{equation}
We use $\bar{P}=\frac{\alpha_{s}}{2\pi}P$, with $P$ being the splitting
functions without $()_{+}$ prescription, i.e., \cite{Ellis:1996}
\begin{align}
P_{q}^{q}(t,z) & =C_{F}\,(1+z^{2})/(1-z)\:,\\
P_{g}^{q}(t,z) & =\theta(|q|-N_{f}(t))\,\begin{array}{c}
\frac{1}{2}\end{array}(z^{2}+(1-z)^{2})\:,\\
P_{q}^{g}(t,z) & =C_{F}\,(1+(1-z)^{2})/z\:,\\
P_{g}^{g}(t,z) & =2C_{A}\,(\,z/(1-z)+(1-z)/z+z(1-z)\,)\:,
\end{align}
with quark indices $q\in\{-6,...,-1,1,...,6\}$ and $g=0$, with $C_{F}=4/3$,
$T_{R}=1/2$, $C_{A}=3$, and where 
\[
\theta(N_{f}-|q|)=\left\{ \begin{array}{cl}
1 & \mathrm{for}\:N_{f}(t)\ge|q|\\
0 & \mathrm{otherwise}
\end{array}\right.
\]
assures a possible emission of a quark of flavor $q$ if the number
of flavors $N_{f}(t)$ is at least as big as $|q|$. So for large
values of $t$, with $N_{f}(t)=5$, the weight for emitting a bottom
quark is identical to the one of a $u$ or $d$ quark. The symbols
$\Delta^{i}$ refer to the so-called Sudakov form factors, defined
as
\begin{equation}
\Delta^{q}(t_{1},t_{2})=\exp\left(-\int_{t_{1}\,^{2}}^{t_{2}}\frac{dt}{t}\frac{\alpha_{s}}{2\pi}\int_{\epsilon}^{1-\epsilon}dz\,P_{q}^{q}(z)\right),\label{Sudakov-q}
\end{equation}
and
\begin{align}
 & \;\Delta^{g}(t_{1},t_{2})=\nonumber \\
 & \exp\left(\,\!-\int_{t_{1}\,^{2}}^{t_{2}}\frac{dt}{t}\frac{\alpha_{s}}{2\pi}\int_{\epsilon}^{1-\epsilon}\,\,dz\,\left(\frac{1}{2}P_{g}^{g}(z)+N_{f}P_{g}^{q}(z)\right)\right),\label{Sudakov-g}
\end{align}
see Appendix \ref{-------Sudakov-factor-------}.%

\begin{figure}[b]
\begin{centering}
\includegraphics[scale=0.7]{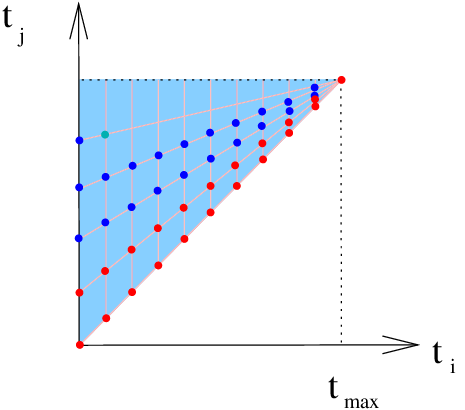}
\par\end{centering}
\caption{Tabulating the evolution function.\label{titj}}
\end{figure}

The fundamental difference with respect to the ``usual'' PDFs is
the fact that we have a singular initial condition, namely 
\begin{equation}
E_{\mathrm{QCD}}^{ab}\left(x,t_{a},t_{a}\right)=\delta(1-x)\delta_{ab},
\end{equation}
so we have to define the evolution function $\widetilde{E}_{\mathrm{QCD}}^{ab}\left(x,t_{a},t_{b}\right)$
for at least one emission, such that the evolution function can be
written as
\begin{equation}
E_{\mathrm{QCD}}^{ab}\left(x,t_{a},t_{b}\right)=\widetilde{E}_{\mathrm{QCD}}^{ab}\left(x,t_{a},t_{b}\right)+\Delta^{a}(t_{a},t_{b})\delta(1-x)\delta_{ab},\label{eq:eqcdtilde}
\end{equation}
with $\widetilde{E}_{\mathrm{QCD}}^{ab}\left(x,t_{a},t_{a}\right)=0$.
As proven in Appendix \ref{-------evolution-function-theorem-------},
we have for an arbitrary $t_{c}$ between $t_{a}$ and $t_{b}$ the
relation
\begin{equation}
E_{\mathrm{QCD}}^{ab}\left(x,t_{a},t_{b}\right)=\sum_{c}\int\frac{dz}{z}E_{\mathrm{QCD}}^{ac}\left(\frac{x}{z},t_{a},t_{c}\right)E_{\mathrm{QCD}}^{cb}\left(z,t_{c},t_{b}\right)\,,\label{evolution-abc-1}
\end{equation}
and using Eq. (\ref{eq:eqcdtilde}), we get%
{} 
\[
\widetilde{E}_{\mathrm{QCD}}^{ab}\left(x,t_{a},t_{b}\right)=\widetilde{E}_{\mathrm{QCD}}^{ab}\left(x,t_{a},t_{c}\right)\Delta^{b}(t_{c},t_{b})\qquad\qquad
\]
\begin{equation}
+\Delta^{a}(t_{a},t_{c})\widetilde{E}_{\mathrm{QCD}}^{ab}\left(x,t_{c},t_{b}\right)\label{eq:evo-abc}
\end{equation}
 
\[
\qquad\qquad\ +\sum_{c}\int\frac{dz}{z}\widetilde{E}_{\mathrm{QCD}}^{ac}\left(\frac{x}{z},t_{a},t_{c}\right)\widetilde{E}_{\mathrm{QCD}}^{cb}\left(z,t_{c},t_{b}\right).
\]
We tabulate $\widetilde{E}$, which means we compute $\widetilde{E}_{\mathrm{QCD}}^{ab}\left(x,t_{i},t_{j}\right)$
with $t_{\min}\leq t_{i}\leq t_{j}\leq t_{\max}$ (blue area in the
$t_{i}-t_{j}$ plane, in Fig. \ref{titj})
explicitly 
\begin{equation}
t_{i}=t_{\mathrm{min}}\left(\frac{t_{\mathrm{max}}}{t_{\min}}\right)^{(i-1)/(M-1)},
\end{equation}
 and 
\begin{equation}
t_{j}=t_{i}\left(\frac{t_{\mathrm{max}}}{t_{i}}\right)^{(j-1)/(M-1)},
\end{equation}
 for $i,j$ from $1$ to $M$. We first compute $\widetilde{E}_{\mathrm{QCD}}^{ab}\left(x,t_{i},t_{j}\right)$
for all $i$ and $j=1$ and $j=2$ in an iterative fashion. These
are the red points in Fig. \ref{titj}. We then compute the $\widetilde{E}$
values for $j=3,4,5...$, always for $i=1$ to $i=M$. We use Eq.
(\ref{eq:evo-abc}) as
\[
\widetilde{E}_{\mathrm{QCD}}^{ab}\left(x,t_{i},t_{j}\right)=\widetilde{E}_{\mathrm{QCD}}^{ab}\left(x,t_{i},t_{j-1}\right)\Delta^{b}(t_{j-1},t_{j})\qquad\qquad
\]
\begin{equation}
+\Delta^{a}(t_{i},t_{j-1})\widetilde{E}_{\mathrm{QCD}}^{ab}\left(x,t_{j-1},t_{j}\right)\label{eq:evo-abc-1}
\end{equation}
\[
\qquad\qquad\ +\sum_{j-1}\int\frac{dz}{z}\widetilde{E}_{\mathrm{QCD}}^{ac}\left(\frac{x}{z},t_{i},t_{j-1}\right)\widetilde{E}_{\mathrm{QCD}}^{cb}\left(z,t_{j-1},t_{j}\right).
\]
This works, since $\widetilde{E}_{\mathrm{QCD}}^{ab}\left(x,t_{i},t_{j-1}\right)$
and $\widetilde{E}_{\mathrm{QCD}}^{cb}\left(x,t_{j-1},t_{j}\right)$
are already known via interpolation from already tabulated values.
In this way, we are able to compute and tabulate the evolution functions
$\widetilde{E}_{\mathrm{QCD}}^{ab}$, and use them via polynomial
interpolation, and thus we know $E_{\mathrm{QCD}}^{ab}$ via Eq. (\ref{eq:eqcdtilde}).

\subsection{Computing integrated parton-parton cross sections with quark mass
dependent kinematics \label{-------integrated-parton-parton-xs-------}}

As discussed in the last section, the integrated parton-parton cross
section $\sigma_{\mathrm{hard}}^{ij}$ is a crucial element of the
EPOS4 framework. It may be written as (see Appendix \ref{subsec:LC-momentum-fraction}
Eq. (\ref{xsection-LC-momentum-integral}))
\begin{align}
 & \sigma_{\mathrm{hard}}^{ij}\!=\!\!\sum_{klmn}\int\!\!dx_{1}\,dx_{2}\,dt\,E_{\mathrm{QCD}}^{ik}(x_{1},Q_{1}^{2},\mu_{\mathrm{F}}^{2})E_{\mathrm{QCD}}^{jl}(x_{2},Q_{2}^{2},\mu_{\mathrm{F}}^{2})\nonumber \\
 & \qquad\qquad\qquad\times\frac{\pi\alpha_{s}^{2}}{s^{2}}\left\{ \frac{1}{g^{4}}\bar{\sum}|\mathcal{M}^{kl\to mn}|^{2}\right\} \frac{1}{1+\delta_{mn}},\label{eq:sigmahard}
\end{align}
where the indices $i$, $j$, $k$, $l$, $m$, $n$ refer to all
kinds of flavors (gluons and (anti)quarks up to bottom). The matrix
elements $\mathcal{M}$ is considered to be given in terms of $s$
and $t$, with $s=x_{1}x_{2}s_{\mathrm{lad}}$, with $s_{\mathrm{lad}}$
referring to the parton-parton scattering, see Fig. \ref{ladder}. EPOS was originally constructed
only for light flavors, simply considering two types of partons, massless
quarks and gluons. Very often, in the program, the two cases were
simply treated explicitly, which required major changes in the code
structure. For instance, the number of active flavors (which appears in $\alpha_{s}$ and the splitting function), which was a constant so far, now depends
on $Q^{2}$.

And there is a second challenge: In the ``massless case'' of earlier
versions, we could first do the sum over all Born processes and compute
``the'' Born cross section for a given pair of partons $k$ and
$l$, 
\begin{equation}
\sum_{mn}\left\{ \frac{1}{g^{4}}\bar{\sum}|\mathcal{M}^{kl\to mn}|^{2}\right\} 
\end{equation}
before doing the integrals, which considerably simplifies the calculations.
Now we need explicitly the information about the process $kl\to mn$,
since the masses involved in the process affect the kinematics (like
the relation between Mandelstam $t$ and $p_{t}$ and integration
limits). So the operations $\sum_{mn}$ and $\int dx_{1}dx_{2}dt$
can no longer be exchanged in general, but it can be within classes:
We introduce classes $K$ of elementary reactions $kl\to mn$ obeying
to the same kinematics and then compute all relevant quantities (to
be discussed in detail in the next section) depending on of $K$.
In this way we do the kinematics properly and still keep efficient
(fast) procedures, which is crucial in the EPOS4 framework. 

\subsubsection{Born kinematics}

In the last section, we were considering processes involving initial
partons $i$ and $j$ and two partons $k$ and $j$ entering the Born
process (let us note them $1$ and $2$), producing partons $m$ and
$n$ (let us note them $3$ and $4$), so altogether we have the Born
process $1+2\to3+4$. In the CMS, we write the parton momentum four-vectors
as 
\begin{equation}
p=(\sqrt{E\text{\texttwosuperior}+M\text{\texttwosuperior}},\vec{p}_{t},p_{z}),\quad\mathrm{with}\;M^{2}=p^{2},\label{fourvec}
\end{equation}
with $\vec{p}_{t}=0$ for the incoming ones. So per definition, $E$
is the modulus of the momentum, $E=|\vec{p}|$, in the CMS, and so we
have obviously for all four particles
\begin{equation}
p_{z}^{2}=E^{2}-p_{t}^{2}\:,
\end{equation}
and
\begin{equation}
E_{1}=E_{2}\,,\;E_{3}=E_{4}.
\end{equation}
In the following, $E$ (without index) refers to $E_{1}$ ($=E_{2}$)
and $E'$ to $E_{3}$ ($=E_{4}$). We will use the following definitions:
\begin{equation}
W=4E^{2},\;W'=4E'^{2}.
\end{equation}
From energy conservation, we find
\begin{eqnarray}
 & s=\left(\sqrt{E^{2}+m_{1}^{2}}+\sqrt{E^{2}+m_{2}^{2}}\right)^{2},\\
 & \,=\left(\sqrt{E'^{2}+m_{3}^{2}}+\sqrt{E'^{2}+m_{4}^{2}}\right)^{2}
\end{eqnarray}
which gives
\begin{eqnarray}
 & W=s-2\left(m_{1}^{2}+m_{2}^{2}\right)+\frac{1}{s}\left(m_{1}^{2}-m_{2}^{2}\right)^{2},\label{eq:variableW1}\\
 & W'=s-2\left(m_{3}^{2}+m_{4}^{2}\right)+\frac{1}{s}\left(m_{3}^{2}-m_{4}^{2}\right)^{2}.\label{eq:variableW2}
\end{eqnarray}
In case of $m_{1}=m_{2}=m,$ we have 
\begin{equation}
W=s-4m^{2}.
\end{equation}
In case of $m_{1}=0$, $m_{2}=m$ or $m_{1}=m$, $m_{2}=0$, we have
\begin{equation}
W=\frac{\left(s-m^{2}\right)^{2}}{s}\,,
\end{equation}
and in case of $m_{1}=m_{2}=0$, we get
\begin{equation}
W=s.
\end{equation}
Corresponding formulas apply for $W'$. Essentially all kinematic
relations will be expressed in terms of $W$ and $W'$.

\subsubsection{Constraints for \emph{p$\boldsymbol{_{t}}$} }

Crucial for the following discussions is the relation between the
factorization scale and the transverse momentum of the outgoing parton,
\begin{equation}
\mu_{\mathrm{F}}^{2}=\frac{p_{t}^{2}+\lambda M^{2}}{\kappa},
\end{equation}
or it's inverse
\begin{equation}
p_{t}^{2}=\kappa\mu_{\mathrm{F}}^{2}-\lambda M^{2}\equiv\Pi_{\mathrm{F}}\left(\mu_{\mathrm{F}}^{2}\right),
\end{equation}
with $M$ being the (maximum) mass of the partons involved in the
Born process, and where the two coefficients $\kappa$ and $\lambda$
represent the freedom in defining $\mu_{\mathrm{F}}^{2}\left(p_{t}^{2}\right)$.
In EPOS4.0.0, we use $\kappa=1$ and $\lambda=0$.
A choice  $\kappa=1$ and $\lambda=1$ makes little difference (and only at very small  $p_t$) 
for charm production,
but for bottom one should better use this choice, as we will do in future releases.

The $p_{t}^{2}$ values in the integration in the cross section formulas
of the last section are restricted, since we have 
\begin{equation}
\mu_{\mathrm{F}}^{2}\ge\max\!\left[Q_{1}^{2},Q_{2}^{2}\right],
\end{equation}
which amounts to 
\begin{equation}
p_{t}^{2}\geq p_{t\,\min}^{2}=\Pi_{\mathrm{F}}(\max\!\left[Q_{1}^{2},Q_{2}^{2}\right]).
\end{equation}
In reality, we have to use the Mandelstam $t$ rather than $p_{\bot}^{2}$
in the integration, so we need to find the limits for this
variable. 

\subsubsection{Constraints for \emph{s} }

The quantity $E'$ is precisely the upper limit for the transverse
momentum of particle $3$, i.e., 
\begin{equation}
p_{t}^{2}\leq p_{t\,\mathrm{max}}^{2}=E'^{2}=\frac{W'}{4}.
\end{equation}
To get non-zero results, we need 
\begin{equation}
p_{t\,\mathrm{min}}^{2}<p_{t\,\mathrm{max}}^{2}=E'^{2}=\frac{W'}{4}.
\end{equation}
which gives
\begin{equation}
W'=s-2\left(m_{3}^{2}+m_{4}^{2}\right)+\frac{1}{s}\left(m_{3}^{2}-m_{4}^{2}\right)^{2}>4p_{t\,\mathrm{min}}^{2},
\end{equation}
and solving the quadratic inequality equation, we get
\begin{equation}
s_{\min}=d\Big(1+\sqrt{1-\left((m_{3}^{2}-m_{4}^{2})/d\right)^{2}}\Big)\;\label{smin}
\end{equation}
\[
\mathrm{with}\quad d=m_{3}^{2}+m_{4}^{2}+2\,p_{t\,\mathrm{min}}^{2}.
\]
This is first of all a limit for the energy squared of the Born process,
but of course as well a limit for the ladder.

\subsubsection{Constraints for \emph{t} }

Concerning $t$, we have 
\begin{eqnarray}
 & t=\left(\sqrt{E'^{2}+m_{3}^{2}}-\sqrt{E^{2}+m_{1}^{2}}\right)^{2}\nonumber \\
 & \qquad\qquad-\left(\vec{p}_{t}-\vec{0}\right)^{2}-\left(p_{z}-E\right)^{2},
\end{eqnarray}
which leads to
\begin{eqnarray}
 & |t|=\frac{1}{2}\,\Big(\,\sqrt{W+4m_{1}^{2}}\sqrt{W'+4m_{3}^{2}}-2(m_{1}^{2}+m_{3}^{2})\qquad\nonumber \\
 & \qquad\qquad\qquad\mp\sqrt{W}\sqrt{W'-4p_{t}^{2}}\,\Big),\label{pt2t}
\end{eqnarray}
the ``$\mp$'' referring to respectively $p_{z}\ge0$ and $p_{z}\le0$.
We may invert Eq. (\ref{pt2t}), to obtain
\begin{eqnarray}
 & p_{t}^{2}=\frac{W'}{4}-\frac{1}{W}\,\Big(\,|t|-\frac{1}{2}\sqrt{W+4m_{1}^{2}}\sqrt{W'+4m_{3}^{2}}\nonumber \\
 & \qquad\qquad\qquad+m_{1}^{2}+m_{3}^{2}\,\Big)^{2},\label{eq:t2pt}
\end{eqnarray}
which allows us to compute $p_{t}^{2}$ for a given $t$.

From Eq. (\ref{pt2t}), we get
\[
|t|_{\mathrm{min/max}}\!=\!\frac{1}{2}\!\Big(\!\!\sqrt{W+4m_{1}^{2}}\sqrt{W'+4m_{3}^{2}}-2(m_{1}^{2}+m_{3}^{2})
\]
\begin{equation}
\qquad\mp\sqrt{W}\sqrt{W'-4p_{t\,\min}^{2}}\,\Big).\label{tminmax}
\end{equation}
Instead of integrating from $|t|_{\mathrm{min}}$ to $|t|_{\mathrm{max}}$,
one may define the maximum value $|t|_{\mathrm{max+}}$ of $|t|$
with $p_{z}\ge0$, i.e.,
\begin{equation}
|t|_{\mathrm{max+}}=\frac{1}{2}\,\Big(\,\sqrt{W+4m_{1}^{2}}\sqrt{W'+4m_{3}^{2}}-2(m_{1}^{2}+m_{3}^{2})\Big).\label{eq:tmaxplus}
\end{equation}
This quantity $|t|_{\mathrm{max+}}$ is the upper limit of $|t|$
actually used, since we may write
\begin{equation}
I=\int_{|t|_{\min}}^{|t|_{\max}}f(t)dt=\int_{|t|_{\min}}^{|t|_{\max+}}f(t)dt+\int_{|t|_{\max+}}^{|t|_{\max}}f(t)dt,
\end{equation}
and a variable transformation $t=2|t|_{\max+}-t'$ for the second
integral (and then replacing $t'$ by $t$) leads to 
\begin{equation}
I=\int_{|t|_{\min}}^{|t|_{\max+}}\left\{ \,f(t)\,+\,f(2|t|_{\max+}-t\,\right\} dt,
\end{equation}
since the upper limit of the transformed variable in the second integral
is $2|t|_{\max+}-|t|_{\max+}=|t|_{\max+}$ and the lower limit is
(using $|t|_{\min}+|t|_{\max}=2|t|_{\max+}$) given as $2|t|_{\max+}-|t|_{\max}=|t|_{\min}$.

\subsubsection{Cross section classes \label{cross-section-classes}}

From the discussion in the last sections, it is clear that when computing
cross sections, we need to identify explicitly contributions from
particular classes of elementary scatterings, obeying different
kinematics, due to the different masses being involved. We will use
the notation 
\begin{equation}
case\;m_{1}m_{2}m_{3}m_{4},
\end{equation}
representing the reaction $1+2\to3+4$ with the four masses being
$m_{1}$, $m_{2}$, $m_{3}$, and $m_{4}$. The masses for charm and
bottom quarks used in EPOS4.0.0 are $m_{c}=1.27\,\mathrm{GeV/c^{2}}$
and $m_{b}=4.18\,\mathrm{GeV/c^{2}}$. We will distinguish 12 different
classes (``light'' refers to massless partons) as follows:
\begin{enumerate}
\item $ll\to ll$ : scattering of light partons (case $0000$)
\item $cl\to cl$ : scattering of charmed quarks or antiquarks with light
partons (case $m0m0$ with $m=m_{c}$) 
\item $bl\to bl$ : scattering of bottom quarks or antiquarks with light
partons (case $m0m0$ with $m=m_{b}$) 
\item $l\bar{l}\to c\bar{c}$ : annihilation of light pairs into charmed
pairs (case $00mm$ with $m=m_{c}$) 
\item $l\bar{l}\to b\bar{b}$ : annihilation of light pairs into bottom
pairs (case $00mm$ with $m=m_{c}$) 
\item $c\bar{c}\to l\bar{l}$ : annihilation of charmed pairs into light
pairs (case $mm00$ with $m=m_{c}$) 
\item $b\bar{b}\to l\bar{l}$ : annihilation of bottom pairs into light
pairs (case $mm00$ with $m=m_{b}$) 
\item $c\bar{c}\to c\bar{c}$ : scattering of charmed pairs into charmed
pairs (case $mm\tilde{m}\tilde{m}$ with $m=\tilde{m}=m_{c}$) 
\item $c\bar{c}\to b\bar{b}$ : scattering of charmed pairs into bottom
pairs (case $mm\tilde{m}\tilde{m}$ with $m=m_{c}$, $\tilde{m}=m_{b}$) 
\item $b\bar{b}\to c\bar{c}$ : annihilation of bottom pairs into charm
pairs (case $mm\tilde{m}\tilde{m}$ with $m=m_{b}$, $\tilde{m}=m_{c}$) 
\item $b\bar{b}\to b\bar{b}$ : annihilation of bottom pairs into bottom
pairs (case $mm\tilde{m}\tilde{m}$ with $m=\tilde{m}=m_{b}$) 
\item $cb\to cb$ : scattering of charm and bottom partons (case $m\tilde{m}m\tilde{m}$
with $m=m_{c},$ $\tilde{m}=m_{b}$) 
\end{enumerate}
The cross section calculation [see Eq. (\ref{eq:sigmahard})] is then
given as
\begin{align}
 & \sigma_{\mathrm{hard}}^{ij}=\sum_{K=1}^{12}\:\sum_{klmn\sim K}\int\!\!dx_{1}\,dx_{2}\,dt\,\nonumber \\
 & \qquad\qquad\qquad E_{\mathrm{QCD}}^{ik}(x_{1},Q_{1}^{2},\mu_{\mathrm{F}}^{2})E_{\mathrm{QCD}}^{jl}(x_{2},Q_{2}^{2},\mu_{\mathrm{F}}^{2})\nonumber \\
 & \qquad\qquad\qquad\times\frac{\pi\alpha_{s}^{2}}{s^{2}}\left\{ \frac{1}{g^{4}}\bar{\sum}|\mathcal{M}^{kl\to mn}|^{2}\right\} \frac{1}{1+\delta_{mn}},
\end{align}
where ``$klmn\sim K$'' refers to indices corresponding to the class
$K$, where one then uses the integration limits and kinematic relation
according to $K$. For numerical efficiency, we use the explicit formulas
from Appendix \ref{=======appendix:born-kinematics=======}.
This is still not yet the final formula, as discussed in the next
section. %
{} 

\textbf{}%
\textbf{}%

\subsubsection{Cross sections for both-sided, single-sided, and no emissions}

Our evolution functions are based on the same equation as the usual
PDFs, but we have a singular initial condition, i.e., $E_{\mathrm{QCD}}^{ik}\left(x,t_{a},t_{a}\right)=\delta(1-x)\delta_{ik}.$
Therefore, as discussed in section \ref{-------parton-evolution-------},
we compute and tabulate $\tilde{E}_{\mathrm{QCD}}$, where $\tilde{E}_{\mathrm{QCD}}$
refers to the case of at least one emission, such that
\begin{align}
E_{\mathrm{QCD}}^{ab}\left(x,Q_{a}^{2},Q_{b}^{2}\right) & =\widetilde{E}_{\mathrm{QCD}}^{ab}\left(x,Q_{a}^{2},Q_{b}^{2}\right)\label{eq:eqcdtilde-1}\\
 & +\Delta^{a}(Q_{a}^{2},Q_{b}^{2})\delta(1-x)\delta_{ab}.\nonumber 
\end{align}
This takes explicitly care of the singular initial condition. Correspondingly,
we have to deal with four different integrated parton parton cross
sections. First, we have 
\begin{align}
 & \sigma_{\mathrm{hard}}^{\mathrm{both}\,ij}=\sum_{K=1}^{12}\:\sum_{klmn\sim K}\int\!\!dx_{1}\,dx_{2}\,dt\,\nonumber \\
 & \qquad\qquad\qquad\tilde{E}_{\mathrm{QCD}}^{ik}(x_{1},Q_{1}^{2},\mu_{\mathrm{F}}^{2})\tilde{E}_{\mathrm{QCD}}^{jl}(x_{2},Q_{2}^{2},\mu_{\mathrm{F}}^{2})\nonumber \\
 & \qquad\qquad\qquad\times\frac{\pi\alpha_{s}^{2}}{s^{2}}\left\{ \frac{1}{g^{4}}\bar{\sum}|\mathcal{M}^{kl\to mn}|^{2}\right\} \frac{1}{1+\delta_{mn}},\label{sigma-hard-both}
\end{align}
which represents the integrated parton-parton cross section with at
least one emission on each side, referred to as both-sided emissions.
Then we have single-sided emissions, i.e., 
\begin{align}
 & \sigma_{\mathrm{hard}}^{\mathrm{upper}\,ij}=\sum_{K=1}^{12}\:\underset{kjmn\sim K}{\sum_{kmn}}\int\!\!dx_{1}\,dt\,\nonumber \\
 & \qquad\qquad\qquad\tilde{E}_{\mathrm{QCD}}^{ik}(x_{1},Q_{1}^{2},\mu_{\mathrm{F}}^{2})\,\Delta^{j}(Q_{2}^{2},\mu_{\mathrm{F}}^{2})\nonumber \\
 & \qquad\qquad\qquad\times\frac{\pi\alpha_{s}^{2}}{s^{2}}\left\{ \frac{1}{g^{4}}\bar{\sum}|\mathcal{M}^{kj\to mn}|^{2}\right\} \frac{1}{1+\delta_{mn}},\label{sigma-hard-upper}
\end{align}
representing the integrated parton-parton cross section with at least
one emission on the upper side and no emission on the opposite side, and
\begin{align}
 & \sigma_{\mathrm{hard}}^{\mathrm{lower}\,ij}=\sum_{K=1}^{12}\:\underset{ilmn\sim K}{\sum_{lmn}}\int\!\!dx_{2}\,dt\,\nonumber \\
 & \qquad\qquad\qquad\Delta^{i}(Q_{1}^{2},\mu_{\mathrm{F}}^{2})\tilde{E}_{\mathrm{QCD}}^{jl}(x_{2},Q_{2}^{2},\mu_{\mathrm{F}}^{2})\nonumber \\
 & \qquad\qquad\qquad\times\frac{\pi\alpha_{s}^{2}}{s^{2}}\left\{ \frac{1}{g^{4}}\bar{\sum}|\mathcal{M}^{il\to mn}|^{2}\right\} \frac{1}{1+\delta_{mn}}.\label{sigma-hard-lower}
\end{align}
representing the integrated parton-parton cross section with at least
one emission on the lower side and no emission on the opposite side.
Finally, we have 
\begin{align}
 & \sigma_{\mathrm{hard}}^{\mathrm{none\,}ij}=\sum_{K=1}^{12}\:\underset{ijmn\sim K}{\sum_{mn}}\int\!\,dt\,\nonumber \\
 & \qquad\qquad\qquad\Delta^{i}(Q_{1}^{2},\mu_{\mathrm{F}}^{2})\,\Delta^{j}(Q_{2}^{2},\mu_{\mathrm{F}}^{2})\nonumber \\
 & \qquad\qquad\qquad\times\frac{\pi\alpha_{s}^{2}}{s^{2}}\left\{ \frac{1}{g^{4}}\bar{\sum}|\mathcal{M}^{ij\to mn}|^{2}\right\} \frac{1}{1+\delta_{mn}},\label{sigma-hard-none}
\end{align}
representing the integrated parton-parton cross section with no emission
on either side. The four
formulas are very similar, just in case of no emission, one replaces
$\tilde{E}_{\mathrm{QCD}}^{jl}$ by $\delta_{jl}\Delta^{j}$ and one
has therefore one summation less. 

In principle there is the possibility, for all these cross sections,
to add a so-called K-factor to compensate higher order effects, but
presently we use K-factor = 1 (in EPOS4.0.0), since we use already
a ``variable flavor-number scheme'', i.e., the number of active flavors
(in $\alpha_{s}$, in $\tilde{E}$ etc) depends on the virtuality:
$N_{f}=N_{f}(Q^{2})$.
In that case, LO calculations 
with a K-factor being unity
already give a fair description
of the data, as  has been demonstrated independently in
\cite{Guiot:2018kfy} (figure 4) and \cite{Guiot:2021vnp}. 

We calculate and tabulate $\sigma_{\mathrm{hard}}^{\mathrm{both\,}ij}$~,
$\sigma_{\mathrm{hard}}^{\mathrm{upper\,}ij}$~, and $\sigma_{\mathrm{hard}}^{\mathrm{none\,}ij}$~,
keeping in mind that $\sigma_{\mathrm{hard}}^{\mathrm{lower\,}ij}$
can be obtained from $\sigma_{\mathrm{hard}}^{\mathrm{upper\,}ij}$
by exchanging arguments. In the simulations, we use polynomial interpolation
based on these tables to get the cross sections $\sigma_{\mathrm{hard}}^{...\,ij}$.
The complete cross section is then the sum
\begin{equation}
\sigma_{\mathrm{hard}}^{\mathrm{\,}ij}=\sigma_{\mathrm{hard}}^{\mathrm{both\,}ij}+\sigma_{\mathrm{hard}}^{\mathrm{upper\,}ij}+\sigma_{\mathrm{hard}}^{\mathrm{lower\,}ij}+\sigma_{\mathrm{hard}}^{\mathrm{none\,}ij}.
\end{equation}
As discussed in section \ref{========00003DEPOS4-building-block========00003D=00003D},
the cross section $\sigma_{\mathrm{hard}}^{\mathrm{\,}ij}$ may then
be used to compute ``the EPOS4 building block'' $G_{\mathrm{QCD}}$,
the basic quantity in our parallel scattering scheme.

\subsubsection{Taming singularities in the cross section integrals \label{technical-issues-concerning}}

For all integrations in the last section, we need to be aware of the
fact that the integrands contain ``singularities'' like $1/(x-a)^{\lambda}$.
Even when the integration domains avoid such singularities, as $\int_{a+\epsilon}^{b}f(x)dx$
with $\epsilon\ll1$, the numerical integration (Gaussian integration
in our case) may give wrong results, if it is employed in a naive
fashion. One actually needs to properly transform the integration
variables to have smooth integrands. Let us consider an integral $\int_{a}^{b}g(x)dx$,
which may be transformed to $\int_{u(a)}^{u(b)}g(x(u))\frac{dx}{du}du$,
where $u$ is defined in $[-1,1]$. Let us define the coordinate transformation
$u\leftrightarrow x$ via 
\begin{equation}
\frac{\int_{a}^{x}g(x')dx'}{\int_{a}^{b}g(x')dx'}=\frac{u+1}{2},\label{trafo-x-to-u}
\end{equation}
which may be written as 
\begin{equation}
\frac{G(x)-G(a)}{G(b)-G(a)}=\frac{u+1}{2},\label{trafo-x-to-u-2}
\end{equation}
if the primitive $G$ of $g$ is known. Then we get $\int_{u(a)}^{u}g(x(u'))\frac{dx}{du}du'$
$\propto(u+1)/2$, which means the integrand is constant. If we have
to compute $\int_{a}^{b}f(x)dx$, we are looking for some ``simple
function'' $g(x)$, being close to $f(x)$, in order to use Eq. (\ref{trafo-x-to-u-2})
to define the coordinate transformation $u\leftrightarrow x$. We
then compute 
\begin{equation}
\int_{a}^{b}f(x)dx=\int_{u(a)}^{u(b)}f(x(u))\frac{dx}{du}du,
\end{equation}
where we expect the integrand $f(x(u))\frac{dx}{du}$ to vary slowly
with $u$ (since for $f=g$ it would be constant). Then based on the
$u$-integration variable, we use 14-point Gaussian integration. It
is not necessary to find a $g$ very close to $f$, it should be ``sufficiently
close'' to give a well-behaved transformed function, in particular
taking care of singularities. In the case of several singularities, one
needs to split the integration domain and use an appropriate coordinate
transformation for each piece. In the case of singularities, there are
always cutoffs which make the integrals mathematical well defined,
it is a purely numerical issue, without transformations, the integration
results could be completely wrong. 

To compute $\sigma_{\mathrm{hard}}^{\mathrm{both\,}ij}$ according
to Eq. (\ref{sigma-hard-both}), we first note $dx_{1}dx_{2}dt=dz\,dtdx_{1}/x_{1}$
with $z=x_{1}x_{2}$. Concerning the $z$ integration, we make a coordinate
transformation Eq. (\ref{trafo-x-to-u-2}) with 
\begin{equation}
G(z)=-z^{-\delta_{\mathrm{hard}}},
\end{equation}
because this gives (for $f\approx g$)
\begin{align}
\sigma_{\mathrm{hard}}^{\mathrm{both\,}ij} & =\int_{a/s}^{1}f(z)dz\\
 & \approx\int_{a/s}^{1}g(z)dz=-G(\frac{a}{s})+G(1)\propto s^{\delta_{\mathrm{hard}}},\nonumber 
\end{align}
which is what we expect. Actually $\sigma_{\mathrm{hard}}^{\mathrm{both\,}ij}\propto s^{\delta_{\mathrm{hard}}}$
defines the empirical technical parameter $\delta_{\mathrm{hard}}$
(the numerical value is 0.25). This case shows the importance of the
transformation to take care of a singular integrand, here of the form
$\delta_{\mathrm{hard}}\,z^{-\delta_{\mathrm{hard}}-1}$, which allows
nevertheless numerical integration with great precision. As mentioned
before, it is not important to have an ``approximation'' $g(z)$
everywhere close to $f(z)$, but it must take care of the singularity.
Next, we consider (inside the $z$ integral) the $t$ integration.
Here we make a coordinate transformation Eq. (\ref{trafo-x-to-u-2})
with 
\begin{equation}
G(t)=-1/t\,,
\end{equation}
which corresponds to $g(t)=1/t^{2}$, the expected $t$ singularity
from the Born process. Finally (inside the $z$ and the $t$ integral),
we consider the $x_{1}$ integration. Here we split the integration
domain, say $a\to b$, into $a\to c$ and $c\to b$, since we expect
singularities $1/(1-x_{1})$ and $1/x_{1}$, from $\tilde{E}_{\mathrm{QCD}}^{ij}(x_{1})\sim P_{i}^{j}(x_{1})$
close to $x_{1}=1$ and $x_{1}=0$. Concerning the integration $c\to b$,
we make a coordinate transformation Eq. (\ref{trafo-x-to-u-2}) with
\begin{equation}
G(x_{1})=-\ln(1-x_{1}),
\end{equation}
which corresponds to $g(x_{1})=1/(1-x_{1})$, corresponding to the
expected $x_{1}$ singularity for $x_{1}$ close to unity. For the
integration $a\to c$, we make a coordinate transformation Eq. (\ref{trafo-x-to-u-2})
with 
\begin{equation}
G(x_{1})=\ln(x_{1}),
\end{equation}
which corresponds to $g(x_{1})=1/x_{1}$, being the expected $x_{1}$
singularity for $x_{1}$close to zero. 

To compute $\sigma_{\mathrm{hard}}^{\mathrm{upper\,}ij}$ according
to Eq. (\ref{sigma-hard-upper}), we have only a $x_{1}$ and a $t$
integration, which are treated as the corresponding integration for
$\sigma_{\mathrm{hard}}^{\mathrm{both\,}ij}$. The cross section $\sigma_{\mathrm{hard}}^{\mathrm{lower\,}ij}$
is not computed but derived from $\sigma_{\mathrm{hard}}^{\mathrm{upper\,}ij}$
as 
\begin{equation}
\sigma_{\mathrm{hard}}^{\mathrm{lower\,}ij}(s_{\mathrm{lad}},Q_{1}^{2},Q_{2}^{2})=\sigma_{\mathrm{hard}}^{\mathrm{upper\,}ji}(s_{\mathrm{lad}},Q_{2}^{2},Q_{1}^{2}),
\end{equation}
i.e., by exchanging $i$ and $j$ as well as $Q_{1}^{2}$ and $Q_{2}^{2}$.
Finally for $\sigma_{\mathrm{hard}}^{\mathrm{none\,}ij}$ according
to (\ref{sigma-hard-none}), one needs only a $t$ integration, also done
in the same way as in the other cases.

\subsection{Differential parton-parton cross sections and backward evolution
\label{-------Differential-parton-parton -------}}

One of the (new) key elements in EPOS4 is the fact that even though we have in general to deal with multiple parallel ladders (see Fig. \ref{saturation-three-pomerons}),
which energy-momentum shared between them, it is possible \textendash{}
for each ladder \textendash{} to first generate the Born process (magenta
dot in Fig. \ref{saturation-three-pomerons}) and determine the corresponding outgoing partons, and then via backward evolution generate the parton emissions. This
has many advantages (compared to the forward evolution technique in
EPOS3): Not only does it correspond to the ``usual'' method employed
by factorization models, it also allows much better control of the
hard processes. 

\begin{figure}[h]
\centering{}\includegraphics[scale=0.25]{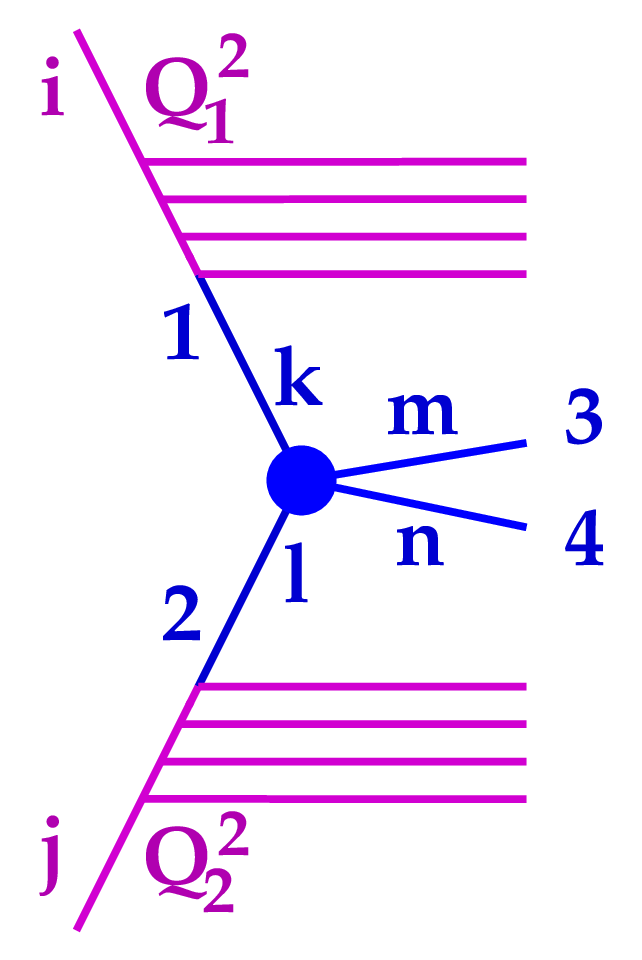}\caption{Parton-parton scattering, considering the Born process $1+2\to3+4$.
The solid lines represent partons (quarks, antiquarks, or gluons).\label{ladder-1}}
\end{figure}

We consider parton-parton scattering as shown in Fig. \ref{ladder-1}.
Starting from the differential parton-parton cross section Eq. (\ref{differential-cross-section}),
as shown in Appendix \ref{=======appendix:xsection-formulas=======}
Eq. (\ref{eq:LC-momentum-integral-1}), we may write
\[
\frac{d^{3}\sigma_{\mathrm{hard}}^{ij}}{dx_{1}\,dx_{2}\,dt}\!=\!\!\sum_{K=1}^{12}\:\sum_{klmn\sim K}E_{\mathrm{QCD}}^{ik}(x_{1},Q_{1}^{2},\mu_{\mathrm{F}}^{2})E_{\mathrm{QCD}}^{jl}(x_{2},Q_{2}^{2},\mu_{\mathrm{F}}^{2})
\]
\begin{equation}
\times\frac{\pi\alpha_{s}^{2}}{s^{2}}\left\{ \frac{1}{g^{4}}\bar{\sum}|\mathcal{M}^{kl\to mn}|^{2}\right\} \frac{1}{1+\delta_{mn}}.\label{dsigma-dx1-dx2-dt}
\end{equation}
Compared to Eq. (\ref{eq:LC-momentum-integral-1}), we added a ``$\sum_{K=1}^{12}$''
term and the ``$klmn\sim K$'' condition to account for 12 different
cross section classes, as discussed in section \ref{cross-section-classes}.
The variables $x_{1}$ and $x_{2}$ are the momentum fractions of
partons 1 and 2. This formula is very useful as a basis to generate
the hard process in the parton ladder, in the Monte Carlo procedure,
when employing the backward evolution. 

However, we should not forget that we always need to distinguish four
cases, namely both-sided, one-sided(lower), one-sided(upper), or no
emissions. Correspondingly, we have four equations. For both-sided
emissions, we use Eq. (\ref{dsigma-dx1-dx2-dt}), with $\tilde{E}$
instead of $E$. In case of one-sided emission, we replace one of
the $\tilde{E}_{\mathrm{QCD}}^{jl}$ by $\delta_{jl}\Delta^{j}$,
and in case of no emissions, we replace both $\tilde{E}_{\mathrm{QCD}}^{jl}$
by $\delta_{jl}\Delta^{j}$.

\subsubsection{Generating the hard process $\mathbf{1+2 \to 3+4}$ \label{Generating-the-hard-process }}

We first have to know if we have both-sided, one-sided, or no emissions.
So we first use the integrated cross sections $\sigma_{\mathrm{hard}}^{\mathrm{both\,}ij}$~,
$\sigma_{\mathrm{hard}}^{\mathrm{upper\,}ij}$~, $\sigma_{\mathrm{hard}}^{\mathrm{lower\,}ij}\,$,
and $\sigma_{\mathrm{hard}}^{\mathrm{none\,}ij}$~as weights to determine randomly what kind of emission we have. 

In the case of both-sided emissions, we define, based on Eq. (\ref{dsigma-dx1-dx2-dt})
with $\tilde{E}$ instead of $E$, the function $X_{klmn}^{ij}$ as
\begin{eqnarray}
 &  & X_{klmn}^{ij}(z,t,x_{1})=\left\{ \sigma_{\mathrm{hard}}^{\mathrm{both}\,ij}\right\} ^{-1}\\
 &  & \;\times\tilde{E}_{\mathrm{QCD}}^{ik}(x_{1},Q_{1}^{2},\mu_{\mathrm{F}}^{2})\tilde{E}_{\mathrm{QCD}}^{jl}(x_{2},Q_{2}^{2},\mu_{\mathrm{F}}^{2})\nonumber \\
 &  & \;\times\frac{\pi\alpha_{s}^{2}}{s^{2}}\left\{ \frac{1}{g^{4}}\bar{\sum}|\mathcal{M}^{kl\to mn}|^{2}\right\} \frac{1}{1+\delta_{mn}},\nonumber 
\end{eqnarray}
with $z=x_{1}x_{2}$. We want to generate for given $i$ and $j$
the variables $z$, $t$ and $x_{1}$, according to the probability
distribution
\begin{equation}
\mathrm{prob}(z,t,x_{1})=\sum_{K=1}^{12}\:\sum_{klmn\sim K}X_{klmn}^{ij}(z,t,x_{1}),
\end{equation}
which is not trivial due to the fact that the expressions $X_{klmn}^{ij}$
contain singularities, as discussed in section \ref{technical-issues-concerning},
where we discussed the calculation of $\sigma_{\mathrm{hard}}^{\mathrm{both}\,ij}$
as a sum of integrals over $X_{klmn}^{ij}$. We also showed how to
solve this problem. We had integrals of the form $\int X_{klmn}^{ij}dz\,dt\,dx_{1}$,
which could be done after three coordinate transformations. Let us
name the three integration variables $z_{1}$, $z_{2}$, and $z_{3}$,
i.e.,
\begin{align}
z_{1} & =z=x_{1}x_{2}\,,\\
z_{2} & =t\:,\\
z_{3} & =x_{1}\,.
\end{align}
 The three coordinate transformations $z_{i}\to u_{i}$ were defined
as 
\begin{equation}
\frac{\int_{a_{i}}^{z_{i}}g_{i}(x')dx'}{\int_{a_{i}}^{b_{i}}g_{i}(x')dx'}=\frac{u_{i}+1}{2},\label{trafo-x-to-u-1}
\end{equation}
for $i\in\{1,2,3\}$, with 
\begin{align}
g_{1}(z_{1}) & \propto\qquad z_{1}^{-\delta_{\mathrm{hard}}-1},\\
g_{2}(z_{2}) & \propto\qquad z_{2}^{-2},\\
g_{3}(z_{3}) & \propto\left\{ \begin{array}{cc}
(1-z_{3})^{-1} & \mathrm{for}\,z_{3}>0.8\\
z_{3}\,^{-1} & \mathrm{for}\,z_{3}<0.8
\end{array}\right..
\end{align}
There are two transformations for $z_{3}$, due to two singularities
(at $0$ and $1$), which required a splitting of the integral. The
three functions $g_{i}$ are useful also for the Monte Carlo
generation of the variables $z_{i}$: One may first generate the $z_{i}$
according to $g_{i}$, by first generating uniform random numbers
$u_{i}$ between -1 and 1, and then computing $z_{i}=z_{i}(u_{i})$
by inverting Eq. (\ref{trafo-x-to-u-1}) (which is easy). Then, one
accepts these proposals with the probability 
\begin{equation}
p_{\mathrm{accept}}=\frac{\mathrm{prob}(z_{1},z_{2},z_{3})}{M\,g_{1}(z_{1})g_{2}(z_{2})g_{3}(z_{3})},
\end{equation}
where $M$ has to be chosen such that $p_{\mathrm{accept}}\le1$.
In this ratio, the singular parts of $\mathrm{prob}(z_{1},z_{2},z_{3})$
are canceled out by the  $g_{i}$, and $p_{\mathrm{accept}}$ is
therefore a smooth function of the variables, with an acceptable rate
of rejections.

Then, for given $i$, $j$, $z_{1}$, $z_{2}$ and $z_{3}$, we generate
$K$ and the flavors $k$ and $l$ according to
\begin{equation}
\mathrm{prob}(K,k,l)=\frac{\underset{klmn\sim K}{\underset{mn}{\sum}}X_{klmn}^{ij}(z_{1},z_{2},z_{3})}{\underset{K,k,l}{\sum}\underset{klmn\sim K}{\underset{mn}{\sum}}X_{klmn}^{ij}(z_{1},z_{2},z_{3})},
\end{equation}
and for given $i$, $j$, $z_{1}$, $z_{2}$, $z_{3}$, $K$,
$k$, and $l$, we generate $m$ and $n$ according to
\begin{equation}
prob(m,n)=\frac{X_{klmn}^{ij}(z_{1},z_{2},z_{3})|_{klmn\sim K}}{\underset{mn}{\sum}X_{klmn}^{ij}(z_{1},z_{2},z_{3})|_{klmn\sim K}}.
\end{equation}
The two factors $\tilde{E}_{\mathrm{QCD}}^{ik}(...)$ and $\tilde{E}_{\mathrm{QCD}}^{jl}(...)$
do not depend on $m$ or $n$ and could be dropped but for the numerical
procedures it is easier to keep them, since we may call the same functions
as in the steps before. 

In the case of one-sided emissions (on the upper side), we define
\begin{eqnarray}
 &  & X_{kmn}^{ij}(x_{1},t)=\left\{ \sigma_{\mathrm{hard}}^{\mathrm{upper}\,ij}\right\} ^{-1}\\
 &  & \;\times\tilde{E}_{\mathrm{QCD}}^{ik}(x_{1},Q_{1}^{2},\mu_{\mathrm{F}}^{2})\Delta^{j}(Q_{2}^{2},\mu_{\mathrm{F}}^{2})\nonumber \\
 &  & \;\times\frac{\pi\alpha_{s}^{2}}{s^{2}}\left\{ \frac{1}{g^{4}}\bar{\sum}|\mathcal{M}^{kj\to mn}|^{2}\right\} \frac{1}{1+\delta_{mn}}.\nonumber 
\end{eqnarray}
We want to generate for given $i$ and $j$ the variables $x_{1}$
and $t$, according to the probability distribution
\begin{equation}
\mathrm{prob}(x_{1},t)=\sum_{K=1}^{12}\:\underset{kjmn\sim K}{\sum_{kmn}}X_{kmn}^{ij}(x_{1},t).
\end{equation}
Again, we use the same coordinate transformations as already used
for the integrations to compute $\sigma_{\mathrm{hard}}^{\mathrm{upper}\,ij}$.
Here, with 
\begin{align}
z_{1} & =x_{1},\\
z_{2} & =t,
\end{align}
the transformations $z_{i}\to u_{i}$ were defined via Eq. (\ref{trafo-x-to-u-1})
for $i\in\{1,2\}$, with 
\begin{align}
g_{1}(z_{1}) & \propto\left\{ \begin{array}{cc}
(1-z_{1})^{-1} & \mathrm{for}\,z_{1}>0.8\\
z_{1}\,^{-1} & \mathrm{for}\,z_{1}<0.8
\end{array}\right..\\
g_{2}(z_{2}) & \propto\qquad z_{2}^{-2}.
\end{align}
We first generate the $z_{i}$ according to $g_{i}$ (first generating
$u_{i}$, then inverting Eq. (\ref{trafo-x-to-u-1})) and then accept
these proposals with the probability 
\begin{equation}
p_{\mathrm{accept}}=\frac{\mathrm{prob}(z_{1},z_{2})}{M\,g_{1}(z_{1})g_{2}(z_{2})}.
\end{equation}
Then for given $i$, $j$, $z_{1}$, and $z_{2}$, we generate $K$
and the flavor $k$ according to
\begin{equation}
\mathrm{prob}(K,k)=\frac{\underset{kjmn\sim K}{\underset{mn}{\sum}}X_{kmn}^{ij}(z_{1},z_{2})}{\underset{K,k}{\sum}\underset{kjmn\sim K}{\underset{mn}{\sum}}X_{kmn}^{ij}(z_{1},z_{2})},
\end{equation}
and finally, for given $i$, $j$, $z_{1}$, $z_{2}$, $K$, and $k$,
we generate $m$ and $n$ according to
\begin{equation}
\mathrm{prob}(m,n)=\frac{X_{kmn}^{ij}(z_{1},z_{2})|_{kjmn\sim K}}{\underset{mn}{\sum}X_{kmn}^{ij}(z_{1},z_{2})|_{kjmn\sim K}}.
\end{equation}
In case of one-sided emissions on the lower side, we use the same
algorithm as for the upper side, just exchanging the upper side and lower
side variables and indices.

In case of no emissions, we define
\begin{eqnarray}
 &  & X_{mn}^{ij}(t)=\left\{ \sigma_{\mathrm{hard}}^{\mathrm{none}\,ij}\right\} ^{-1}\\
 &  & \;\times\Delta^{i}(Q_{1}^{2},\mu_{\mathrm{F}}^{2})\Delta^{j}(Q_{2}^{2},\mu_{\mathrm{F}}^{2})\nonumber \\
 &  & \;\times\frac{\pi\alpha_{s}^{2}}{s^{2}}\left\{ \frac{1}{g^{4}}\bar{\sum}|\mathcal{M}^{ij\to mn}|^{2}\right\} \frac{1}{1+\delta_{mn}}.\nonumber 
\end{eqnarray}
We want to generate for given $i$ and $j$ the variable $t$, according
to the probability distribution
\begin{equation}
\mathrm{prob}(t)=\sum_{K=1}^{12}\:\underset{ijmn\sim K}{\sum_{mn}}X_{mn}^{ij}(t),
\end{equation}
which can be done by defining $z_{1}=t,$ and then defining the transformation
$z_{1}\to u_{1}$ via Eq. (\ref{trafo-x-to-u-1}) for $i=1$, with
$g_{1}(z_{1})=z_{1}^{-2}$. We first generate $z_{1}$ according to
$g_{1}$ (first generating $u_{1}$, then inverting Eq. (\ref{trafo-x-to-u-1}))
and then accept these proposals with the probability 
\begin{equation}
p_{\mathrm{accept}}=\frac{\mathrm{prob}(z_{1})}{M\,g_{1}(z_{1})}.
\end{equation}
Then for given $i$, $j$, and $z_{1}$, we generate $K$ according
to
\begin{equation}
\mathrm{prob}(K)=\frac{\underset{ijmn\sim K}{\underset{mn}{\sum}}X_{mn}^{ij}(z_{1})}{\underset{K}{\sum}\underset{ijmn\sim K}{\underset{mn}{\sum}}X_{mn}^{ij}(z_{1})},
\end{equation}
and finally, for given $i$, $j$, $z_{1}$, and $K$, we generate
$m$ and $n$ according to
\begin{equation}
\mathrm{prob}(m,n)=\frac{X_{mn}^{ij}(z_{1},z_{2})|_{ijmn\sim K}}{\underset{mn}{\sum}X_{mn}^{ij}(z_{1},z_{2})|_{ijmn\sim K}}.
\end{equation}

\subsubsection{Backward evolution}

Knowing for given end flavours $i$ and $j$, the Born process variables
$x_{1}$, $x_{2}$ and $t$, and the flavors $k$, $l$ (ingoing)
and $m$, $n$ (outgoing), we generate the parton emission via backward
evolution. The variable $t$ (Mandelstam variable) allows computing
$p_{t}^{2}$ [see Eq. (\ref{eq:t2pt})], which corresponds in a unique
fashion to some factorization scale $\mu_{\mathrm{F}}^{2}$, which
is the starting value of the virtuality for the backward evolution.
We therefore define $Q_{0}^{2}=\mu_{\mathrm{F}}^{2}$. Be $k$ the
flavor of the corresponding parton. 

In the following, we will use the symbol ``$t$'' as virtuality,
to treat the parton evolution, it {\it does not} refer to the Mandelstam
variable. We define $t_{0}=Q_{0}^{2}$.

We will in the following consider the evolution from $t_{1}$ to $t_{0}$
with $t_{1}<t_{0}$. We write the evolution equation [see Eq.(\ref{evolution-equation-1})],
using $\Delta^{k}(t_{1},t_{0})=\Delta^{k}(t_{0})/\Delta^{k}(t_{1})$,
as
\begin{align}
 & E_{\mathrm{QCD}}^{ik}\left(x,t_{1},t_{0}\right)=\frac{\Delta^{k}(t_{0})}{\Delta^{k}(t_{1})}\,\delta_{ik}\,\delta(1-x)\label{evol-1-2}\\
 & \qquad+\int_{t_{1}}^{t_{0}}\frac{dt}{t}\frac{\Delta^{k}(t_{0})}{\Delta^{k}(t)}\int\frac{dz}{z}\,\sum_{a}\bar{P}_{a}^{k}\!(t,z)\,E_{\mathrm{QCD}}^{ia}\left(\frac{x}{z},t_{1},t\right).\nonumber 
\end{align}
The integrand of the second term corresponds to a last branching in
$[t,t+dt]$ , so the corresponding probability is
\begin{equation}
g(t)=\frac{1}{t}\,\frac{\Delta^{k}(t_{0})}{\Delta^{k}(t)}\int\frac{dz}{z}\sum_{a}\bar{P}_{a}^{k}(t,z)\frac{E_{\mathrm{QCD}}^{ia}(\frac{x}{z},t_{1},t)}{E_{\mathrm{QCD}}^{ik}(x,t_{1},t_{0})}.\label{eq: Proba-t}
\end{equation}
Using Eq. (\ref{evolution-equation-3}), we get
\begin{equation}
g(t)=\frac{\partial}{\partial t}\left\{ \frac{\Delta^{k}(t_{0})}{E_{\mathrm{QCD}}^{ik}(x,t_{1},t_{0})}\,\frac{E_{\mathrm{QCD}}^{ik}\left(x,t_{1},t\right)}{\Delta^{k}(t)}\right\} .
\end{equation}
The probability of a last branching after some $t<t_{0}$ is $\int_{t}^{t_{0}}g(t')dt'$,
which means that $t$ is generated via
\begin{equation}
\int_{t}^{t_{0}}g(t')dt'=r,
\end{equation}
with $r$ being a uniform random number. The integral can be easily
done, and we get
\begin{equation}
\int_{t}^{t_{0}}g(t')dt'=\left[\frac{\Delta^{k}(t_{0})\,E_{\mathrm{QCD}}^{ik}(x,t_{1},t')}{\Delta^{k}(t')\,E_{\mathrm{QCD}}^{ik}(x,t_{1},t_{0})}\right]_{t}^{t_{0}}=r,
\end{equation}
which leads to
\begin{equation}
1-\frac{\Delta^{k}(t_{0})\,E_{\mathrm{QCD}}^{ik}(x,t_{1},t)}{\Delta^{k}(t)\,E_{\mathrm{QCD}}^{ik}(x,t_{1},t_{0})}=r,
\end{equation}
Using $1-r$ instead of $r$, we obtain finally 
\begin{equation}
R\equiv\frac{\Delta^{k}(t_{0})\,E_{\mathrm{QCD}}^{ik}(x,t_{1},t)}{\Delta^{k}(t)\,E_{\mathrm{QCD}}^{ik}(x,t_{1},t_{0})}=r,\label{eq: Ratio R}
\end{equation}
for the generation of $t$ in the interval $[t_{1,}t_{0}]$. 

So far we essentially followed the standard procedures, as explained
in \cite{Ellis:1996}. What makes things more difficult in the EPOS4
framework is the fact that we do evolutions for each of the parallel
parton ladders, see Fig. \ref{saturation-three-pomerons}, and there
we consider the evolution starting from a parton, not from a proton.
The initial condition for the evolution is 
\begin{equation}
E_{\mathrm{QCD}}^{ik}\left(x,t_{\mathrm{ini}},t_{\mathrm{ini}}\right)=\delta(1-x)\delta_{ik},
\end{equation}
and this singularity $\delta(1-x)$ needs some special attention.
In Eq. (\ref{eq: Ratio R}), we need the full evolution function,
which may be written as
\begin{equation}
E_{\mathrm{QCD}}^{ik}(x,t_{1},t)=\tilde{E}_{\mathrm{QCD}}^{ik}\left(x,t_{1},t\right)+\frac{\Delta^{k}(t)}{\Delta^{k}(t_{1})}\delta_{ik}\delta(1-x),
\end{equation}
which separates the ``smooth'' and the ``singular'' part. For
the generation of $t$ based on $R=r$, we consider the two cases
$x=1$ and $x\neq1$ separately.\medskip{}

For the case $x=1$, we have $R=1,$ which simply reflects the fact
that there is no emission between $t_{1}$ and $t_{0}$, so we recover
the initial condition $x=1$, and we are done for this case, the emission
process is finished. \medskip{}

For the case $x\ne1$, we have
\begin{equation}
R\equiv\frac{\Delta^{k}(t_{0})\,\tilde{E}_{\mathrm{QCD}}^{ik}(x,t_{1},t)}{\Delta^{k}(t)\,\tilde{E}_{\mathrm{QCD}}^{ik}(x,t_{1},t_{0})}=r,
\end{equation}
which needs to be solved to get $t$. Some elementary root finder
(like the bisection method) will do the job. Once this is done, for this
value of $t$, the probability distribution for the momentum fraction
$z$ of the last branching is obtained from Eq. (\ref{eq: Proba-t})
as 
\begin{equation}
h(z)\propto\frac{1}{z}\sum_{a}\bar{P}_{a}^{k}(t,z)E_{\mathrm{QCD}}^{ia}(\frac{x}{z},t_{1},t),\label{eq: Proba-z}
\end{equation}
which gives 
\begin{equation}
h(z)\propto\frac{1}{z}\!\sum_{a}\bar{P}_{a}^{k}(t,z)\!\left\{ \!\tilde{E}_{\mathrm{QCD}}^{ia}\!\left(\frac{x}{z},t_{1},t\right)\!+\!\frac{\Delta^{a}(t)}{\Delta^{a}(t_{1})}\delta_{ia}\delta(1-\frac{x}{z})\!\right\} ,\label{eq: Proba-z-2}
\end{equation}
or
\begin{equation}
h(z)\propto\frac{1}{z}\!\sum_{a}\bar{P}_{a}^{k}(t,z)\tilde{E}_{\mathrm{QCD}}^{ia}\!\left(\frac{x}{z},t_{1},t\right)\!+\!\bar{P}_{i}^{k}(t,z)\frac{\Delta^{i}(t)}{\Delta^{i}(t_{1})}\delta(x-z),\label{eq: Proba-z-2-1}
\end{equation}
for $z\geq x$. We need to distinguish again two cases, $z>x$ and
$z=x$. The latter corresponds to a parton with a momentum fraction
unity before the splitting (which is the initial value), so the emission
process will be finished at the next iteration step. The probability
for the $z=x$ case is 
\begin{equation}
W_{1}=\frac{\bar{P}_{i}^{k}(t,x)\frac{\Delta^{i}(t)}{\Delta^{i}(t_{1})}}{\left\{ \int_{x}^{1}\frac{dz}{z}\sum_{a}\bar{P}_{a}^{k}(t,z)\tilde{E}_{\mathrm{QCD}}^{ia}\left(\frac{x}{z},t_{1},t\right)\right\} +\bar{P}_{i}^{k}(t,x)\frac{\Delta^{i}(t)}{\Delta^{i}(t_{1})}},
\end{equation}
where the integral can be done using the Gauss-Legendre method after an
appropriate coordinate transform, splitting the integration domain
into two parts, to treat separately the $z\to0$ (for small $x$)
and $z\to1$ regions. This probability increases when (during the
iteration) the virtuality approaches $t_{1}$ and the momentum fraction
approaches unity. With the probability $W_{2}=1-W_{1}$, we have the
case $z>x$, and here the probability distribution to generate $z$
is 
\begin{equation}
h_{2}(z)\propto\frac{1}{z}\sum_{a}\bar{P}_{a}^{k}(t,z)\tilde{E}_{\mathrm{QCD}}^{ia}\left(\frac{x}{z},t_{1},t\right).\label{eq: Proba-z-2-1-1}
\end{equation}
Finally, we generate the flavor $a$ according to 
\begin{equation}
\bar{P}_{a}^{k}(t,z)\tilde{E}_{\mathrm{QCD}}^{ia}\left(\frac{x}{z},t_{1},t\right),\label{eq: Proba-z-2-1-1-1}
\end{equation}
for the given values of $t$ and $z$ found earlier. \medskip{}

At each iteration step, we have these two cases, $z=x$ and $z>x$, where in one case the iteration is finished in the next step, with
the initial value $x=1$, whereas in the other case, the iteration
continues. So in any case, we eventually end up with the initial values.

\subsubsection{Monte Carlo results versus cross section formulas for a single Pomeron}

\begin{figure}[h]
\centering{}\includegraphics[bb=20bp 20bp 542bp 595bp,clip,scale=0.47]{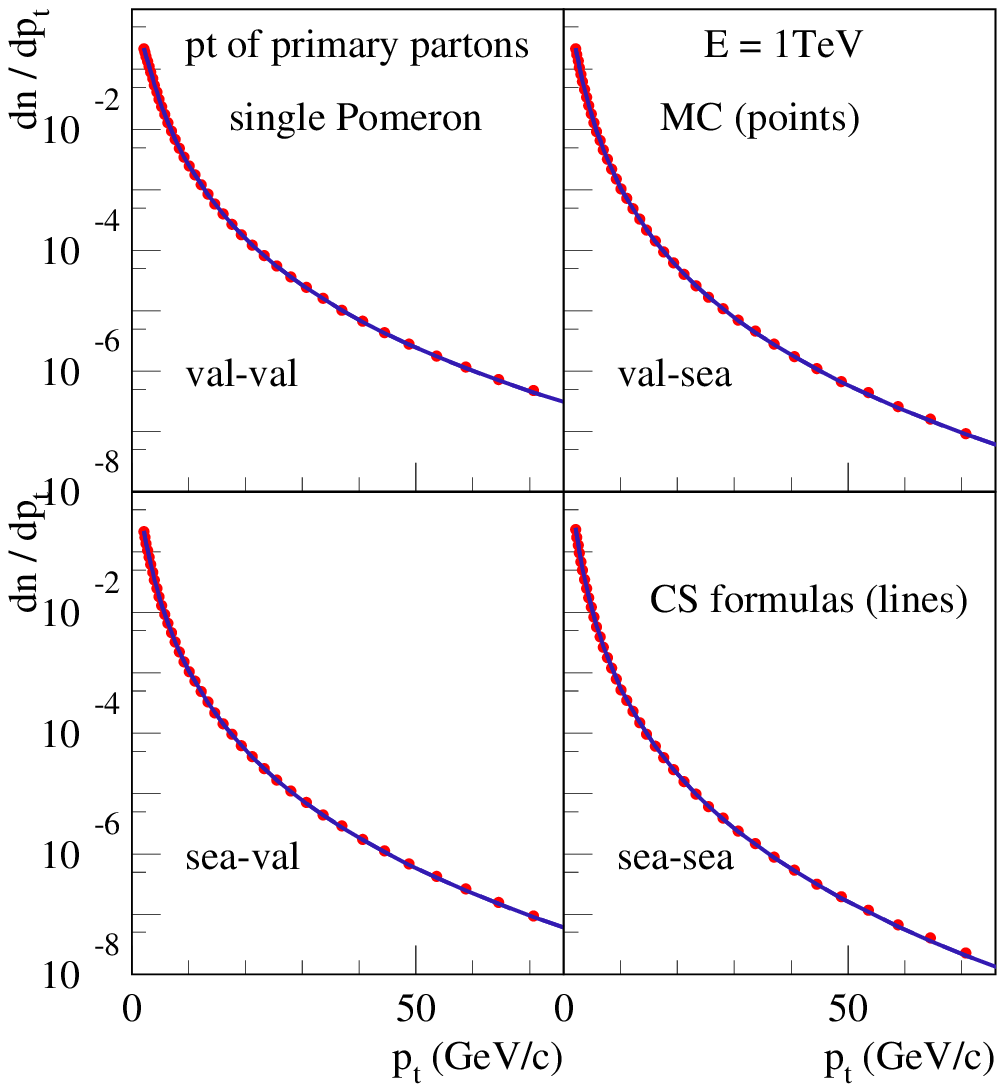}\caption{Transverse momentum distributions of primary partons for Pomerons
of type ``val-val'', ``val-sea'', ``sea-val'', and ``sea-sea''.
We compare the Monte Carlo results (points) to the corresponding curves
obtained with the help of explicit cross section formulas (see text).\label{fig: pt of primary partons} }
\end{figure}

An elementary distribution (for tests) is the transverse momentum
distribution $dn/dy\,dp_{t}$ for a given rapidity of primary partons, directly emitted in the Born process. Let us consider a single Pomeron,
carrying the full energy, i.e., $x^{+}=x^{-}=1$, for an energy $E=\sqrt{s}=1\,$TeV,
and let us consider $Q_{1}^{2}=Q_{2}^{2}=Q_{0}^{2}=O(1)$. We first
investigate the ``sea-sea'' contribution. Combining Eq. (\ref{final-G-sea-sea})
and Eq. (\ref{jet-differential-xsection}), after integrating out
the impact parameter $b$, we get (with a normalization constant $N$)
\begin{align}
 & \frac{d^{2}n^{\mathrm{sea-sea}}}{dydp_{t}^{2}}=N\sum_{ij}\int\!dz^{+}dz^{-}\\
 & \quad E_{\mathrm{soft}}^{i}(Q_{0}^{2},z^{+})E_{\mathrm{soft}}^{j}(Q_{0}^{2},z^{-})\frac{d^{2}\sigma_{\mathrm{hard}}^{ij}}{dydp_{t}^{2}}(Q_{0}^{2},Q_{0}^{2},z^{+}z^{-}s),\nonumber 
\end{align}
\vspace{-0.5cm}

\noindent with
\begin{align}
 & \frac{d^{2}\sigma_{\mathrm{hard}}^{ij}}{dydp_{t}^{2}}(Q_{0}^{2},Q_{0}^{2},s_{\mathrm{lad}})\\
 & =\sum_{klmn}\int dx\,E_{\mathrm{QCD}}^{ik}(x_{1},Q_{0}^{2},\mu_{\mathrm{F}}^{2})E_{\mathrm{QCD}}^{jl}(x_{2},Q_{0}^{2},\mu_{\mathrm{F}}^{2})\nonumber \\
 & \qquad\times\frac{\pi\alpha_{s}^{2}}{s^{2}}\left\{ \begin{array}{c}
\frac{1}{g^{4}}\end{array}\bar{\sum}|\mathcal{M}^{kl\to mn}|^{2}\right\} \frac{1}{1\!+\!\delta_{mn}}\,x_{1}x_{2}\frac{1}{x}\nonumber 
\end{align}
with $x_{1}=x+\begin{array}{c}
\frac{p_{t}}{\sqrt{s_{\mathrm{lad}}}}\end{array}e^{y}$ and $x_{2}=\begin{array}{c}
\frac{x_{1}}{x}\,\frac{p_{t}}{\sqrt{s_{\mathrm{lad}}}}\end{array}e^{-y}$. To compute $\mathcal{M}$, we need $s=x_{1}x_{2}s_{\mathrm{lad}}$
and $t=-p_{t}x_{1}\sqrt{s_{\mathrm{lad}}}e^{-y}$. In a similar way
we get explicit formulas to compute the ``val-val'', the ``val-sea'',
and the ``sea-val'' contribution. In Fig. \ref{fig: pt of primary partons},
we show transverse momentum distributions (for $y=0$) of primary
partons for Pomerons of type ``val-val'', ``val-sea'', ``sea-val'',
and ``sea-sea'', based on these explicit formulas (lines), compared
to the Monte Carlo results (points), using for the latter the methods
explained in section \ref{Generating-the-hard-process }. The Monte
Carlo results agree with those based on cross section formulas. This
may sound trivial, but in older EPOS versions with forward parton
evolution, this was not the case. A great advantage of the new method
is simply the fact that one can make rigorous tests, since the Monte
Carlo uses the same ``modules'' as the cross section formulas ($E_{\mathrm{soft}}$,
$E_{\mathrm{QCD}}$, and $\mathcal{M}$, all of them tabulated and available
via interpolation). In both cases one uses evolution functions $E_{\mathrm{QCD}}$
representing all parton emissions, whereas in the old method the emissions
in the Monte Carlo have been done one by one, and one needs to worry
about additional elements like the reconstruction of the momentum
four-vectors of the emitted partons.%

\subsubsection{EPOS PDFs \label{...... epos-pdfs.......}}

We may go one step further and provide formulas for inclusive momentum
distributions for pp scattering with one single Pomeron 
(identical to the full multiple scattering results at high $p_t$, where factorization applies), 
again first
considering ``sea-sea''. Combining Eqs. (\ref{cut-multiple-pomeron},\ref{definition-G},\ref{final-G-sea-sea},\ref{differential-cross-section}),
after integrating out the impact parameter $b$, we get
\begin{align}
 & E_{3}E_{4}\frac{d^{6}\sigma^{\mathrm{sea-sea}}}{d^{3}p_{3}d^{3}p_{4}}\\
 & =\int dx^{+}dx^{-}V(1-x^{+})V(1-x^{-})\sum_{ij}\int\!dz^{+}dz^{-}\nonumber \\
 & \quad F_{\mathrm{sea\,1}}(x^{+},0)\,F_{\mathrm{sea}\,2}(x^{-},0)\,E_{\mathrm{soft}}^{i}(Q_{0}^{2},z^{+})E_{\mathrm{soft}}^{j}(Q_{0}^{2},z^{-})\nonumber \\
 & \quad\sum_{klmn}\int\!\int\!\!dx_{1}dx_{2}E_{\mathrm{QCD}}^{ik}(x_{1},Q_{0}^{2},\mu_{\mathrm{F}}^{2})E_{\mathrm{QCD}}^{jl}(x_{2},Q_{0}^{2},\mu_{\mathrm{F}}^{2})\nonumber \\
 & \quad\frac{1}{32s\pi^{2}}\bar{\sum}|\mathcal{M}^{kl\to mn}|^{2}\delta^{4}(p_{1}+p_{2}-p_{3}-p_{4})\frac{1}{1+\delta_{mn}},\nonumber 
\end{align}

\noindent with $s=\xi^{+}\xi^{-}s_{\mathrm{pp}}$ being the squared
energy of the Born process, with $\xi_{1}=x^{+}z^{+}x_{1}$ and $\xi_{2}=x^{-}z^{-}x_{2}$.
For ``val-val'', combining Eqs. (\ref{cut-multiple-pomeron},\ref{definition-G},\ref{final-G-val-val},\ref{differential-cross-section}),
we get
\begin{align}
 & E_{3}E_{4}\frac{d^{6}\sigma^{\mathrm{val-val}}}{d^{3}p_{3}d^{3}p_{4}}\\
 & =\int dx^{+}dx^{-}V(1-x^{+})V(1-x^{-})\sum_{ij}\int\!dz^{+}dz^{-}\nonumber \\
 & \quad F_{\mathrm{val}\,1}^{i}(z^{+},x^{+}-z^{+},0)\,F_{\mathrm{val}\,2}^{j}(z^{-},x^{-}-z^{-},0)\nonumber \\
 & \quad\sum_{klmn}\int\!\int\!\!dx_{1}dx_{2}E_{\mathrm{QCD}}^{ik}(x_{1},Q_{0}^{2},\mu_{\mathrm{F}}^{2})E_{\mathrm{QCD}}^{jl}(x_{2},Q_{0}^{2},\mu_{\mathrm{F}}^{2})\nonumber \\
 & \quad\frac{1}{32s\pi^{2}}\bar{\sum}|\mathcal{M}^{kl\to mn}|^{2}\delta^{4}(p_{1}+p_{2}-p_{3}-p_{4})\frac{1}{1+\delta_{mn}},\nonumber 
\end{align}
with $s=\xi^{+}\xi^{-}s_{\mathrm{pp}}$ being the squared energy of
the Born process, with $\xi_{1}=z^{+}x_{1}$ and $\xi_{2}=z^{-}x_{2}$.
Corresponding formulas can be found for the ``val-sea'', and
the ``sea-val'' contribution 
(we do not consider ``psoft'' here, since it only contribute at low $p_t$). 
As a consequence, the complete momentum
distribution (the sum of the four contributions) may be written as
\begin{align}
 & E_{3}E_{4}\frac{d^{6}\sigma}{d^{3}p_{3}d^{3}p_{4}}\label{factorization-formula}\\
 & =\sum_{klmn}\int\!\!\int\!\!d\xi_{1}d\xi_{2}\,f_{\mathrm{PDF}}^{k}(\xi_{1},\mu_{\mathrm{F}}^{2})f_{\mathrm{PDF}}^{l}(\xi_{2},\mu_{\mathrm{F}}^{2})\nonumber \\
 & \quad\frac{1}{32s\pi^{2}}\bar{\sum}|\mathcal{M}^{kl\to mn}|^{2}\delta^{4}(p_{1}+p_{2}-p_{3}-p_{4})\frac{1}{1+\delta_{mn}},\nonumber 
\end{align}
with
\begin{align}
 & f_{\mathrm{PDF}}^{k}(\xi,\mu_{\mathrm{F}}^{2})\nonumber \\
 & =\int dx\,dz\,dy\Big\{\sum_{i}V(1-x)\,F_{\mathrm{sea}}(x,0)\,E_{\mathrm{soft}}^{i}(Q_{0}^{2},z)\nonumber \\
 & \qquad\qquad\qquad E_{\mathrm{QCD}}^{ik}(y,Q_{0}^{2},\mu_{\mathrm{F}}^{2})\delta(\xi-xzy)\Big\}\\
 & \quad+\int dx\,dz\,dy\Big\{\sum_{i}V(1-x)F_{\mathrm{val}}^{i}(z,x-z,0)\nonumber \\
 & \qquad\qquad\qquad E_{\mathrm{QCD}}^{ik}(y,Q_{0}^{2},\mu_{\mathrm{F}}^{2})\delta(\xi-zy)\Big\}.\nonumber 
\end{align}
These are the ``EPOS4 PDFs'' (parton distribution functions), composed
of a ``sea'' contribution (the first integral) and a ``valence''
contribution (the second integral).

Eq. (\ref{factorization-formula}) is identical to the ``usual''
factorization formula, which everybody uses. However, in our case,
it represents the \textbf{single Pomeron case}. The full simulation,
i.e., the \textbf{full multiple scattering case}, provides the same
result, but only at large transverse momenta, as we will show later. 

To make it very clear: In the EPOS4 formalism, we cannot use PDFs
as an input, we need several ``modules'' such as vertex functions
($V$, $F_{\mathrm{sea}}$, $F_{\mathrm{val}}^{i}$) and evolution
functions ($E_{\mathrm{soft}}^{i}$, $E_{\mathrm{QCD}}^{ik}$) to
construct the ``building blocks'' $G_{QCD}$, and then $G$ (cut
single Pomeron), being the basis of our multiple scattering scheme.
But we can use these ``modules'' to construct EPOS4 PDFs. 

\begin{figure}[h]
\centering{}\includegraphics[bb=20bp 20bp 590bp 780bp,clip,scale=0.45]{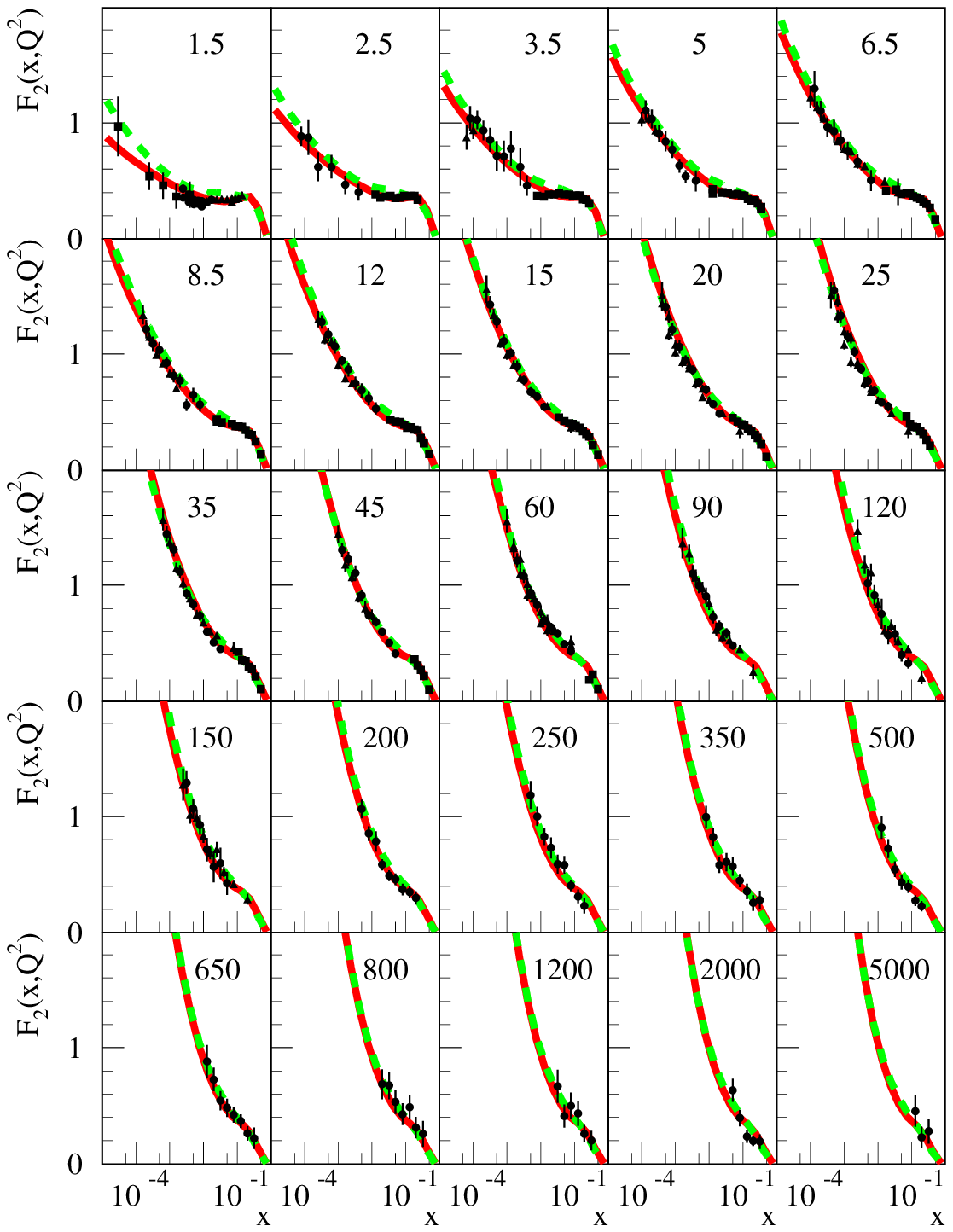}\vspace{-0.5cm}
\\
\caption{$F_{2}$ as a function of $x$ for different values of $Q^{2}$, the
latter one indicated (in units of $\mathrm{GeV}^{2}$) in the upper
right corners of each sub-plot . The red curves refer to EPOS PDFs,
the green ones to CTEQ PDFs, and the black points are data from ZEUS
and H1. \label{f2-epos}}
\end{figure}

Let us first compare the EPOS4 PDFs with other choices and with data.
At least the quark parton distribution functions can be tested and
compared with experimental data from deep inelastic electron-proton
scattering. The leading order structure function $F_{2}$ is given as 
\begin{equation}
F_{2}=\sum_{k}e_{k}^{2}\,x\,f_{\mathrm{PDF}}^{k}(x,Q^{2}),
\end{equation}
\vspace{-0.5cm}
\begin{align}
\mathrm{with}\qquad & x=x_{B}=Q^{2}/(2p\cdot q),\\
 & Q^{2}=-q^{2},
\end{align}
where $p$ is the momentum of the proton and $q$ the momentum of the
exchanged photon. In Fig. \ref{f2-epos},
we plot $F_{2}$ as a function of $x$ for different values of $Q^{2}$,
the latter one indicated (in units of $\mathrm{GeV}^{2}$) in the
upper right corners of each sub-plot . The red curve refers to EPOS
PDFs, the green one to CTEQ PDFs \cite{Dulat_2016-CTEQ-PDF}, and
the black points are data from ZEUS \cite{ZEUS96} and H1 \cite{H1-94,H1-96a,H1-96b}.
The two PDFs give very similar results, and both are close to the
experimental data.

\begin{figure}
\centering{}\includegraphics[bb=20bp 20bp 590bp 810bp,clip,scale=0.35]{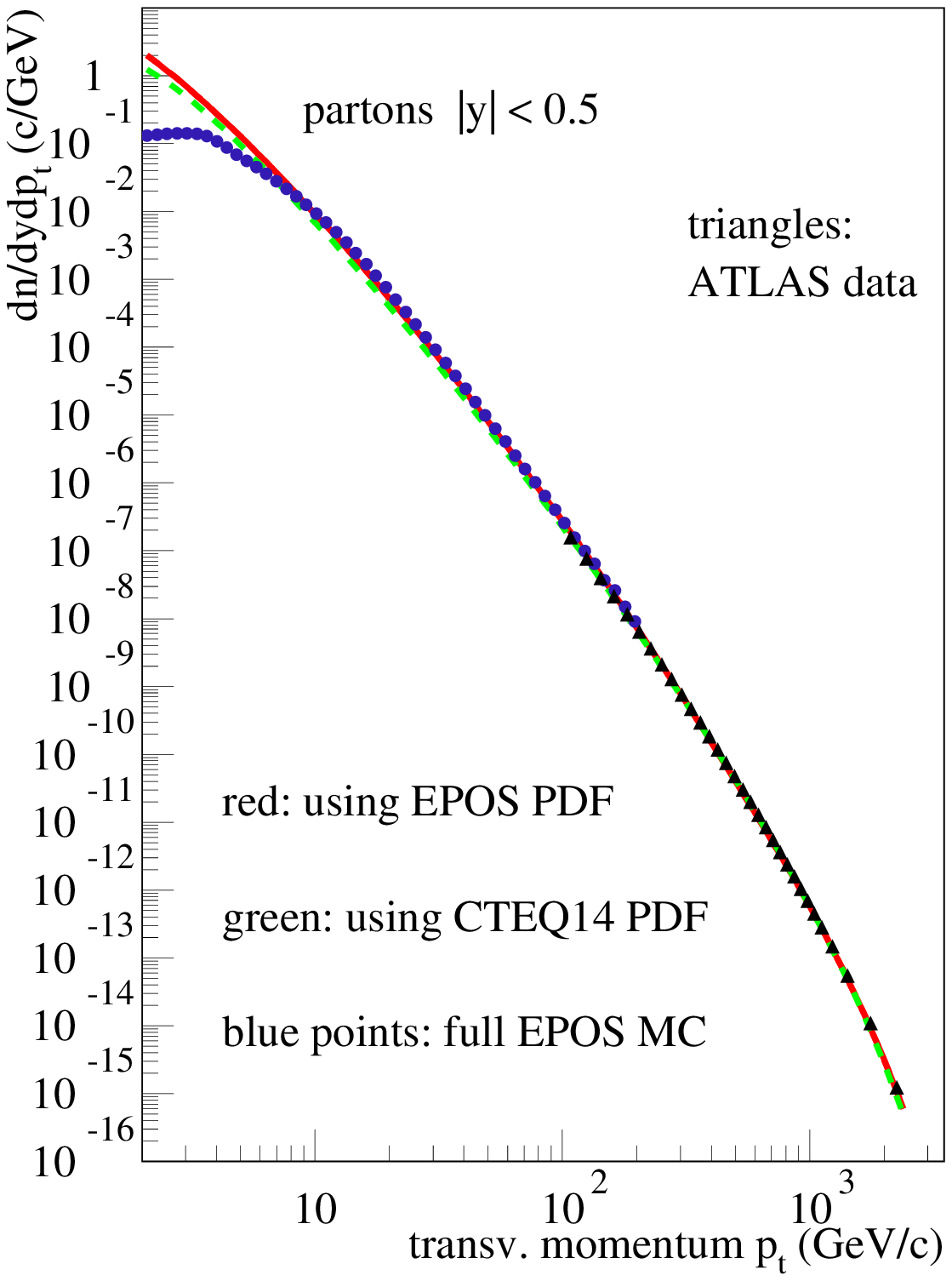}\\
\caption{Parton yield $dn/dp_{t}dy$ for pp at 13 TeV. We show results based
on EPOS PDFs (red full line), CTEQ PDFs (green dashed line), the full
EPOS simulation (blue circles), and experimental data from ATLAS (black
triangles). \label{parton-pt-spectra}}
\end{figure}

Having checked the EPOS PDFs, we will use these functions to compute
the jet (parton) cross section for pp at 13 TeV, using Eq. (\ref{factorization-formula}),
integrating out the momentum of the second parton and the azimuthal
angle of the first parton, which finally gives (see Eq. (\ref{jet-differential-xsection}),
with $E_{\mathrm{QCD}}^{ik}$ replaced by $f_{\mathrm{PDF}}^{k}$,
and $s_{\mathrm{lad}}$ by $s_{\mathrm{pp}}$)
\begin{align}
 & \frac{d^{2}\sigma}{dy\,dp_{t}^{2}}=\sum_{klmn}\int dx\,f_{\mathrm{PDF}}^{k}(x_{1},\mu_{\mathrm{F}}^{2})f_{\mathrm{PDF}}^{l}(x_{2},\mu_{\mathrm{F}}^{2})\nonumber \\
 & \qquad\times\frac{\pi\alpha_{s}^{2}}{s^{2}}\left\{ \begin{array}{c}
\frac{1}{g^{4}}\end{array}\bar{\sum}|\mathcal{M}^{kl\to mn}(s,t)|^{2}\right\} \frac{1}{1\!+\!\delta_{mn}}\,x_{1}x_{2}\frac{1}{x}\,,\label{jet-differential-xsection-1}\\
 & \mathrm{\quad with}\;x_{1}=x+\begin{array}{c}
\frac{p_{t}}{\sqrt{s_{\mathrm{pp}}}}\end{array}e^{y},\,x_{2}=\begin{array}{c}
\frac{x_{1}}{x}\,\frac{p_{t}}{\sqrt{s_{\mathrm{pp}}}}\end{array}e^{-y},\nonumber \\
 & \qquad\qquad s=x_{1}x_{2}\sqrt{s_{\mathrm{pp}}},\,t=-p_{t}x_{1}\sqrt{s_{\mathrm{pp}}}e^{-y},
\end{align}
with $\left\{ ...\right\} $being the form the squared matrix elements
are usually tabulated, with $\alpha_{s}=g^{2}/4\pi$. We define the
parton yield $dn/dp_{t}dy$ as the cross section $d\sigma/dy\,dp_{t}^{2}$,
divided by the inelastic pp cross section, times $2\,p_{t}$, showing
the result in Fig. \ref{parton-pt-spectra}.
We show results based on EPOS PDFs (red full line), CTEQ PDFs \cite{Dulat_2016-CTEQ-PDF}
(green dashed line), the full EPOS simulation (blue circles), and
experimental data from ATLAS \cite{ATLAS:2017ble} (black triangles).
At large values of $p_{t}$, all the different distribution agree,
whereas at low $p_{t}$ the EPOS Monte Carlo simulation results (using
the full multiple scattering scenario) are significantly below the
PDF results, as expected due to screening effects.

\begin{figure}[h]
\centering{}
\includegraphics[bb=20bp 40bp 590bp 600bp,clip,scale=0.35]{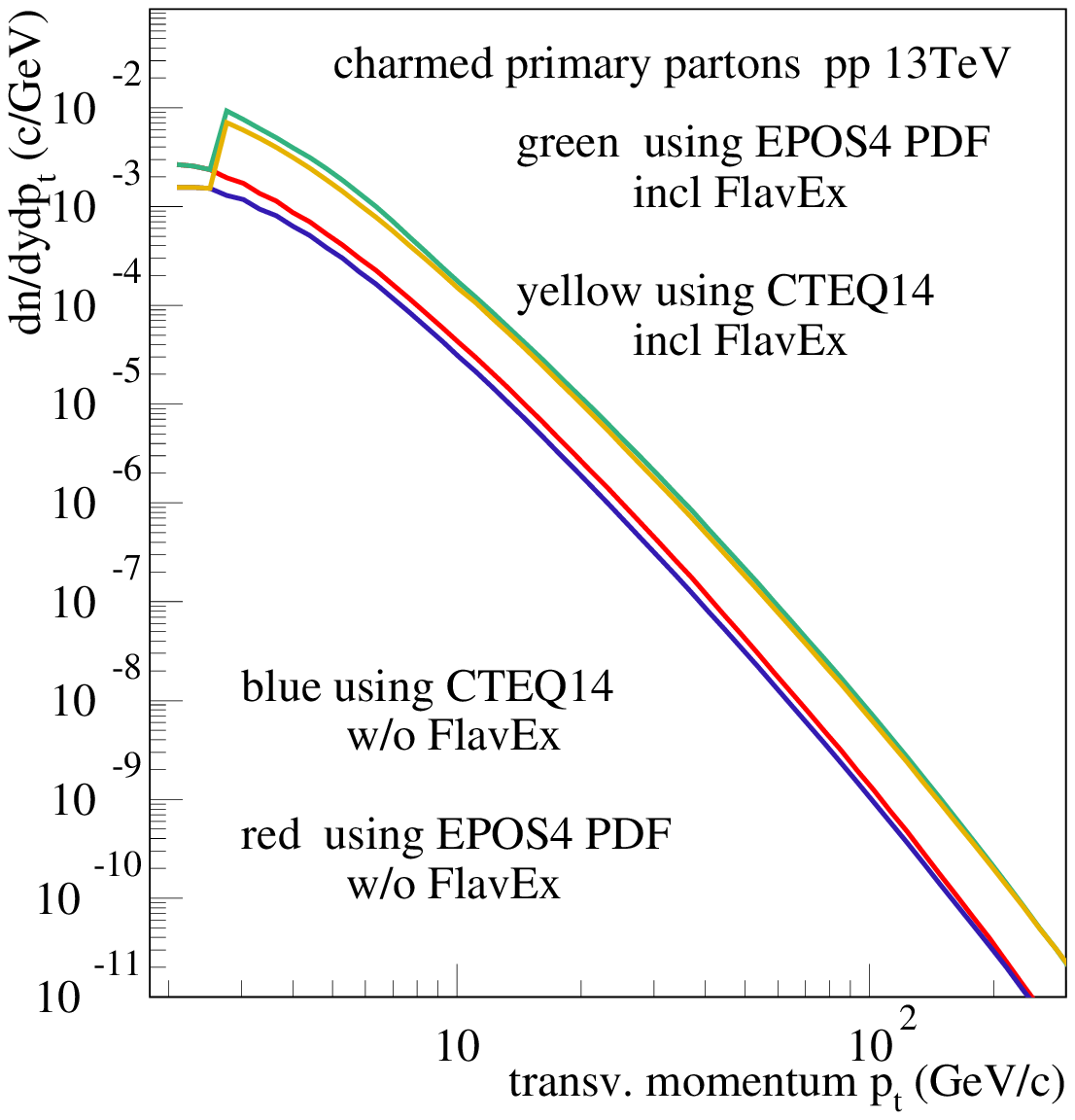}\caption{Transverse momentum distribution of charm quarks ($c$ and $\bar{c}$), based on EPOS PDFs and CTEQ PDFs. In both cases, we show results including
flavor excitation (incl FlavEx) or not (w/o FlavEx). 
The ``jumps'' mark the activation of charm flavor excitation for $\mu_{\mathrm{F}}>2\,m_c$.
\label{charmed-primary-partons}}
\end{figure}

\begin{figure}[h]
\centering{}\includegraphics[bb=20bp 40bp 590bp 600bp,clip,scale=0.35]{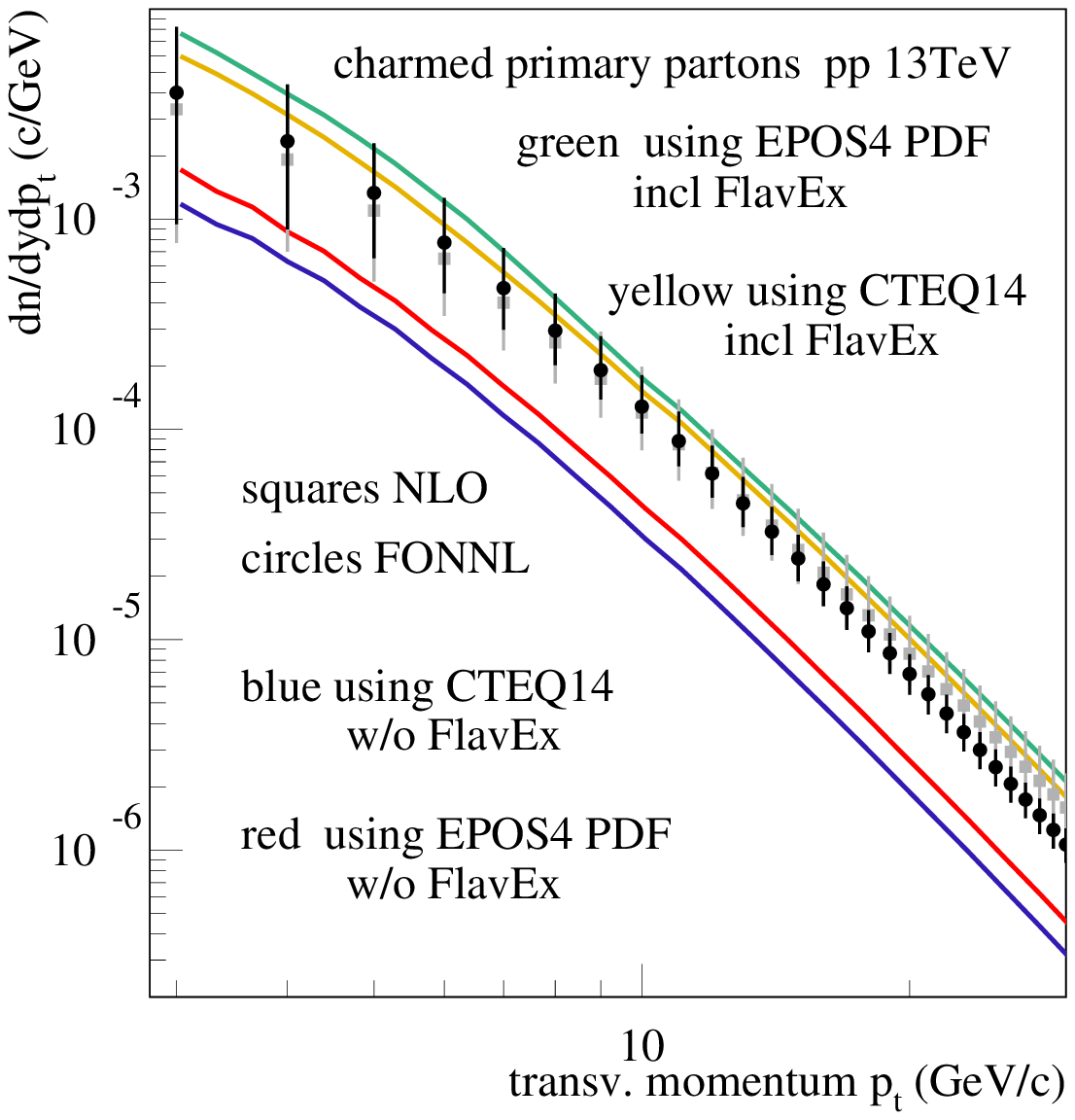}\caption{Transverse momentum distribution of charm quarks ($c$ and $\bar{c}$),
based on EPOS PDFs and CTEQ PDFs. In both cases, we show results including
flavor excitation (incl FlavEx) or not (w/o FlavEx). The black circles
represent FONLL results, and the gray squares NLO calculations.\label{charmed-primary-partons-1}}
\end{figure}

Let us consider the production of charmed primary partons ($c$ and
$\bar{c}$) for pp at 13 TeV, still based on EPOS4 PDFs, according
to Eq. (\ref{jet-differential-xsection-1}). The term ``primary''
means that here we do not consider charm created in the timelike
cascade (in the full simulation we do, of course). We have the elementary
Born scattering $kl\to mn$, where $k,l$ are the incoming and $m,n$
the outgoing parton flavors. Let us note light flavor partons as ``$L$''
and heavy flavor ones (here charm) as ``$H$''. We may produce charm
via $HL\to HL$ (``flavor excitation'') or via $LL\to H\bar{H}$.
We will in the following consider two cases: including flavor excitation
(incl FlavEx) or not (w/o FlavEx), for two calculations: based on
EPOS4 PDFs and based on CTEQ PDFs \cite{Dulat_2016-CTEQ-PDF}, see
Fig. \ref{charmed-primary-partons}.
The EPOS4 results are shown as green (incl FlavEx) and red lines (w/o
FlavEx), the CTEQ based results are shown as yellow (incl FlavEx)
and blue lines (w/o FlavEx). The EPOS4 and the CTEQ results are similar,
at large $p_{t}$ they are even identical. In both cases, the flavor
excitation contribution is largely dominant, over the whole $p_{t}$
range, above some threshold. In Fig. \ref{charmed-primary-partons-1},
\begin{figure}[h]
\centering{}\includegraphics[bb=20bp 40bp 590bp 600bp,clip,scale=0.35]{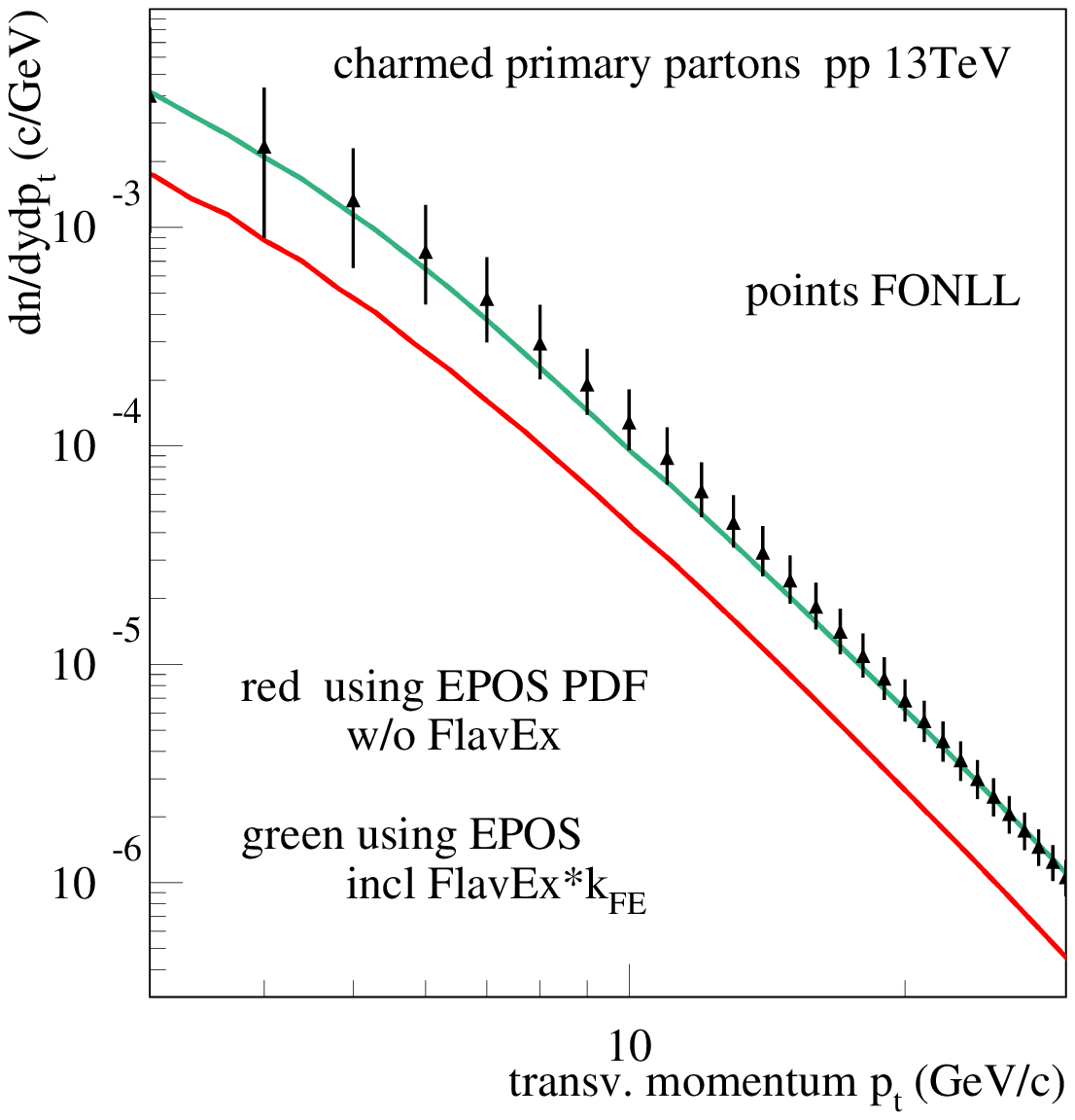}\caption{Transverse momentum distribution of charm quarks ($c$ and $\bar{c}$),
based on EPOS PDFs including flavor excitation fully (incl FlavEx)
or with a reduction factor $k_{\mathrm{FE}}$ (FlavEx{*}$k_{\mathrm{FE}}$).
The black circles represent FONLL results.\label{charmed-primary-partons-2}}
\end{figure}
we plot the same four curves (EPOS4 and the CTEQ based, with and without
flavor excitation), in a reduced $p_{t}$ range, together with FONLL
and NLO calculations \cite{Cacciari:2012ny}, the latter represented
by black circles (FONLL) and gray squares (NLO). The NLO results (at
least at large $p_{t})$ are quite close to the ones based on EPOS4
and CTEQ PDFs. In the EPOS4 framework, we use a k-factor equal unity;
however, we use a ``variable flavor number scheme'', i.e., the number
of allowed flavors depends on the virtuality, which means at large
$p_{t},$ which is correlated with large virtualities, we easily produce
heavy flavor. Concerning the FONLL results, they drop with increasing
$p_{t}$ considerably below the EPOS4 and the CTEQ based results.
In order to be close to FONLL, we multiply the flavor excitation contribution
by a factor $k_{\mathrm{FE}}<1,$ the result for a numerical value
of 0.4 is shown in Fig. \ref{charmed-primary-partons-2}.

\subsubsection{The pseudosoft contribution}

In the previous examples, only the contributions ``sea-sea'', ``val-val'',
``sea-val'', and ``val-sea'' were considered. The PDFs are actually
a sum of ``sea'' and ``val''. There is, however, also a ``pseudosoft''
contribution, see Eq. (\ref{g-psoft}). We expect the ``pseudosoft''
Pomeron to have an internal structure, allowing hard processes. We
assume some probability distribution $P(X)$ with $\left\langle X\right\rangle =O(1)$
of $X=Q^{2}/Q_{\mathrm{sat}}^{2}$, with $Q^{2}$ being the scale
associated with the hard process, and then for given $Q^{2}$, a probability
distribution of the form
\begin{equation}
E_{\mathrm{psoft}}^{i}(Q^{2},z^{+})E_{\mathrm{psoft}}^{j}(Q^{2},z^{-})\,\frac{1}{16\pi s^{2}}\bar{\sum}|\mathcal{M}^{ij\to mn}|^{2}\begin{array}{c}
\frac{1}{1+\delta_{mn}}\end{array}
\end{equation}
to generate a particular hard process.%
{} We expect the production of partons with transverse momenta of the
order of $Q_{\mathrm{sat}}^{2}$, in the range 1-10 GeV/c.

\subsubsection{A result for full EPOS }

\begin{figure}[h]
\centering{}\includegraphics[bb=20bp 20bp 590bp 810bp,clip,scale=0.45]{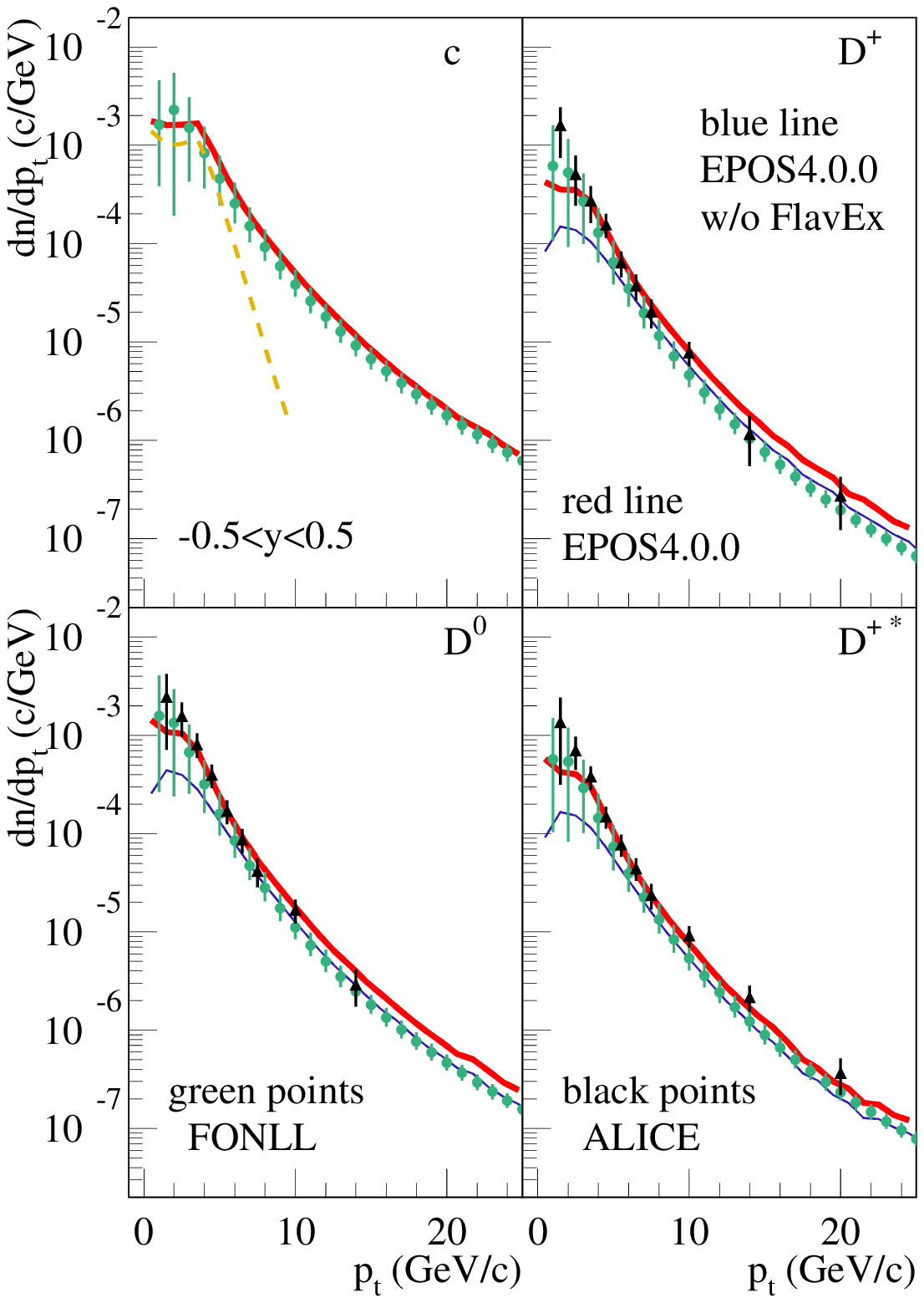}\\
\caption{Charmed partons and hadrons in pp collisions at 7 TeV. We show EPOS4
simulations, compared to FONLL results and ALICE data. \label{charmed-partons-mesons}}
\end{figure}
 In the examples discussed in \ref{...... epos-pdfs.......}, based
on EPOS PDFs, only primary partons were considered (directly produced
in the Born process). To get the complete picture (but without considering 
secondary interactions as hydro evolution and hadronic rescattering), 
we employ the full EPOS4 approach for primary scatterings, 
including multiple scattering, and also including
the timelike cascade, which includes Born processes like $g+g\to g+g$,
where each of the outgoing gluons may split into a $c\bar{c}$ pair.
In Fig. \ref{charmed-partons-mesons} (upper left panel),
we show the corresponding transverse momentum distribution of charmed
quarks (including antiquarks) in pp scattering at 7 TeV, compared
to FONLL results \cite{Cacciari:2012ny}. The yellow dashed line represents
the ``pseudosoft'' contribution, which visibly contributes at low
transverse momentum, as expected. 

In the full (primary scattering) approach, we consider not only the full 
partonic evolution
(initial-state and final-state cascade) but also hadronization.
This will be discussed in detail in section
\ref{=======color-connections=======},
but let us anticipate here some basic features:
cut parton ladders correspond in general to two chains of partons
$q-g-...-g-\bar{q}$ identified as kinky strings, with $q$ referring
to light flavor partons, and $g$ to gluons. The Born process or branchings
in the spacelike or the timelike cascade may lead to $Q\bar{Q}$
production, where $Q$ refers to ``heavy flavor\textquotedblright{}
(HF) quarks, i.e., charm or bottom. In this case, 
we end up with parton chains of the type $q-g-...-g-\bar{Q}$ and
$Q-g-...-g-q$, which will decay (among others) into HF hadrons. In
Fig. \ref{charmed-partons-mesons}, we show transverse momentum spectra
of $D^{+}$, $D^{-}$ mesons (upper right), $D^{0}$ mesons (lower
left), and $D^{+*}$, $D^{-*}$ mesons (lower right) in pp collisions
at 7 TeV. The red lines refer to EPOS4 simulations, the green points
to FONLL calculations \cite{Cacciari:2012ny}, and the black points
to ALICE data \cite{ALICE:2012-pp-charm-spectra}. We also show as
blue lines the EPOS4 results without flavor excitation. Many more
results can be found in \cite{werner:2023-epos4-overview}.

\section{From partons to strings: color flow \label{=======color-connections=======}}

In the previous section, we were discussing in detail the partonic
structure of the EPOS4 building blocks, the Pomerons. This allows
us to compute the weights of multiple Pomeron configurations and 
generate these. In the present section, we will discuss how to transform
these partonic structures into strings. The strings are then the basis
of hadron production via string decay or of the formation of a ``core''
(in case of high densities), serving as an initial condition of a
hydrodynamical evolution. The formation of strings for a particular
partonic configuration is based on the associated color flow. 

\subsection{An example}

\begin{figure}[h]
\noindent \centering{}\includegraphics[scale=0.25]{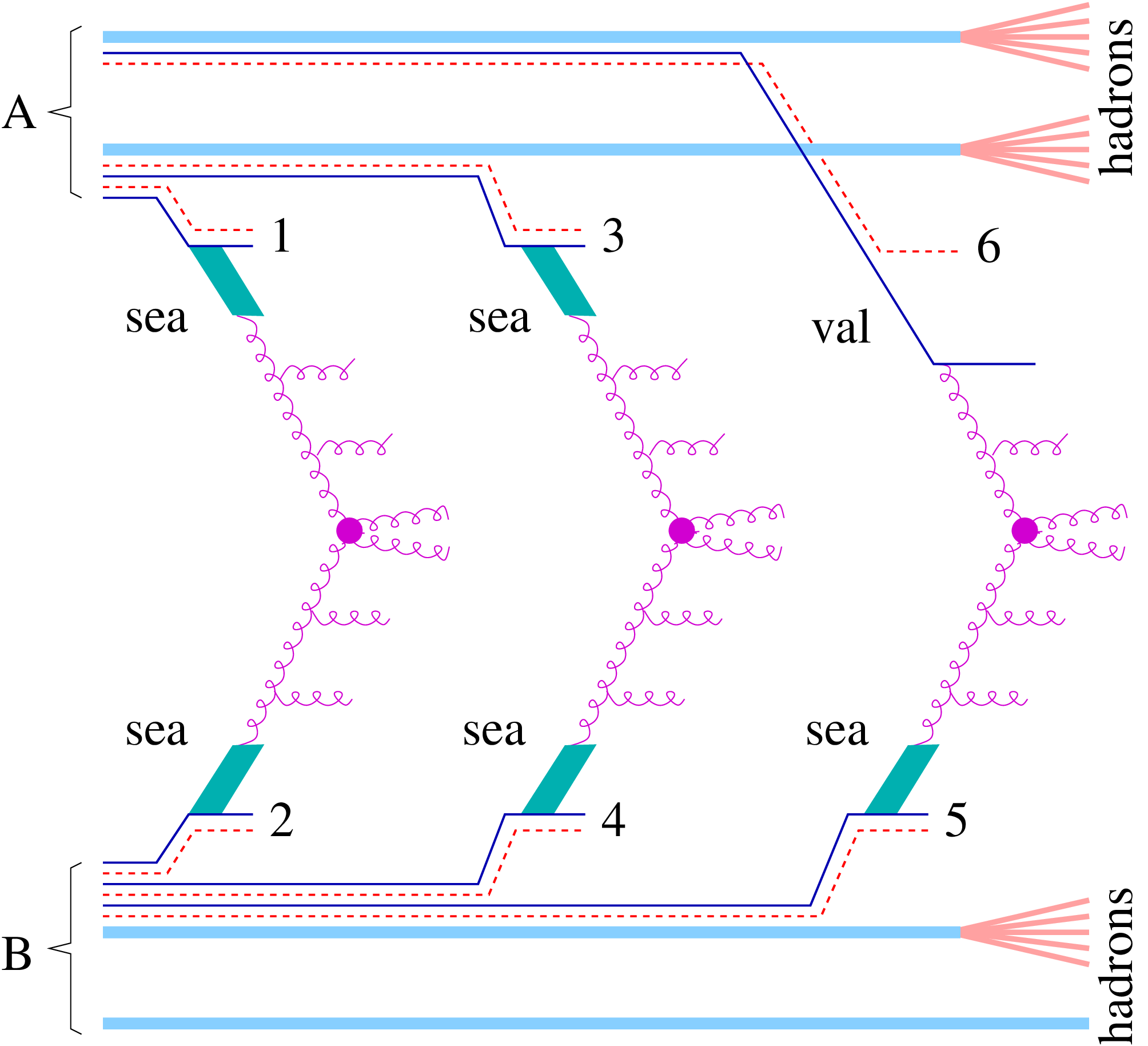}\caption{Partonic configuration of two colliding nuclei $A$ and $B$, each
one composed of two nucleons, with three scatterings (from three cut
Pomerons). Dark blue lines mark active quarks, red dashed lines active
antiquarks, and light blue thick lines projectile and target remnants.
One of the target nucleons is just a spectator.\label{partonic-configuration}}
\end{figure}

In order to be as general as possible, let us consider a collision
of two nuclei, as shown in Fig. \ref{partonic-configuration}, where
we consider the partonic configuration of two colliding nuclei $A$
and $B$, each one composed of two nucleons. Dark blue lines mark
active quarks, red dashed lines active antiquarks, and light blue
thick lines projectile and target remnants (nucleons minus the active
(anti)quarks). We have two scatterings of ``sea-sea'' type, and
one of ``val-sea'' type. We consider each incident nucleon as a
reservoir of three valence quarks plus quark-antiquark pairs. The
``objects'' which represent the ``external legs'' of the individual
scatterings are colorwise ``white'': quark-antiquark pairs in most
cases as shown in the figure, but one may as well have quark-diquark
pairs or even antiquark-antidiquark pairs \textendash{} in any case,
a $3$ and a $\bar{3}$ color representation.
Let us for simplicity consider the quark-antiquark option, and first
of all look at the ``sea'' cases (on the projectile or the target side).
In each case, a quark-antiquark pair is emitted as final-state
timelike (TL) parton pair (marked 1,2,3,4,5) and a spacelike (SL) ``soft
Pomeron'' (indicated by a thick cyan line), which is meant to be
similar to the QCD evolution, but emitting only soft gluons, which
we do not treat explicitly. Then emerging from this soft Pomeron,
we see a first perturbative SL gluon (another possibility is the emission
of a quark, to be discussed later), which initiates the partonic cascade.
In the case of ``val'', we also have a quark-antiquark pair as external
leg, but here first an antiquark is emitted as TL final particle (marked
6), plus an SL quark starting the partonic cascade.

In the case of multiple scattering as in Fig. \ref{partonic-configuration},
the projectile and target remnants remain colorwise white, but they
may change their flavor content during the multiple collision process. The
quark-antiquark pair ``taken out'' for a collision (the ``external
legs'' for the individual collisions), may be $u-\bar{s}$, then
the remnant for an incident proton has flavor $uds$. In addition,
the remnants get massive, much more than simply resonance excitation.
We may have remnants with masses of $10\,\mathrm{GeV/c^{2}}$ or more,
which contribute significantly to particle production (at large rapidities). 

\begin{figure}[h]
\noindent \centering{}\includegraphics[scale=0.25]{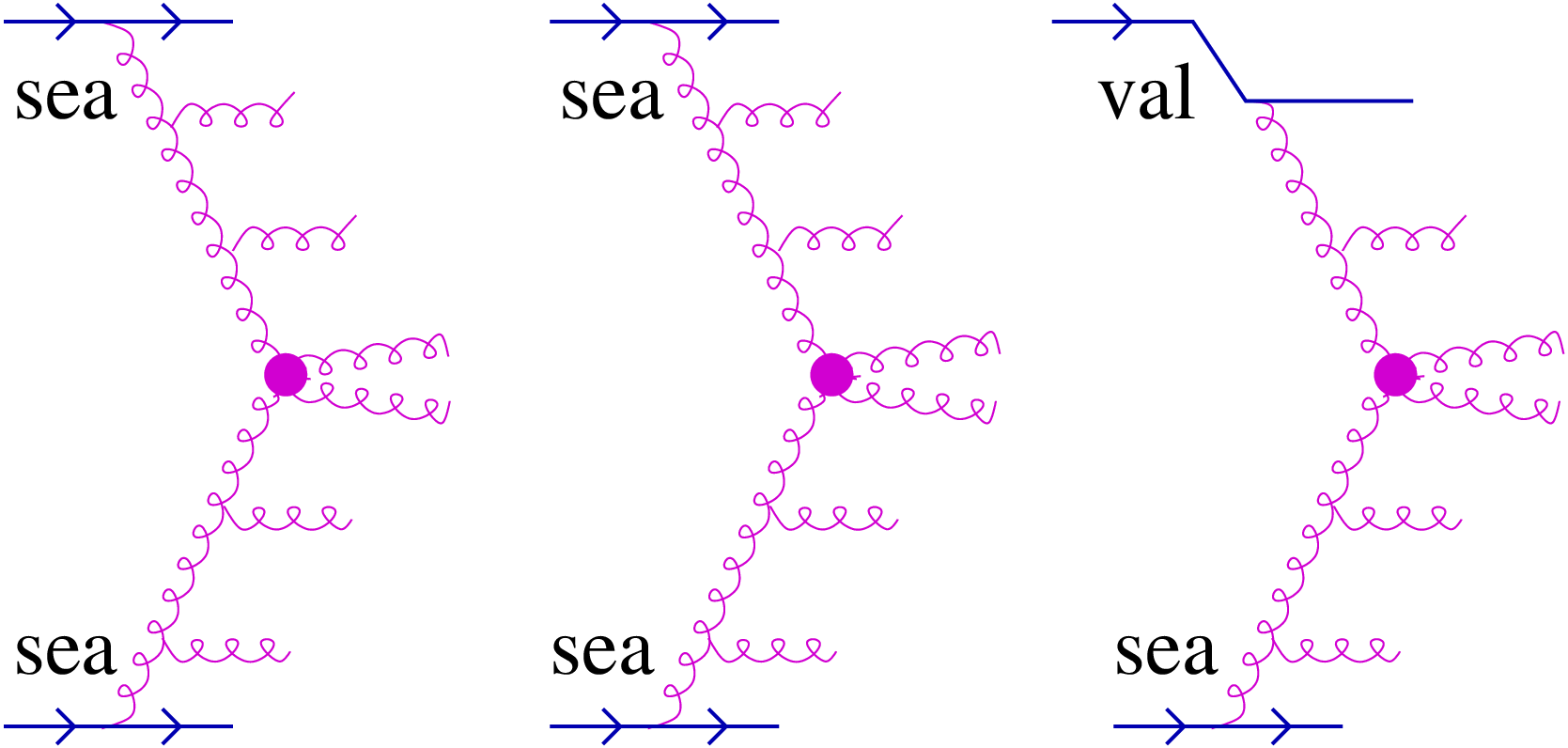}\caption{Configuration colorwise equivalent to the one of Fig. \ref{partonic-configuration}.
The outgoing antiquarks are drawn as incoming quarks (arrows towards
vertices). \label{partonic-configuration-1}}
\end{figure}

\begin{figure}[h]
\noindent \centering{}\includegraphics[scale=0.25]{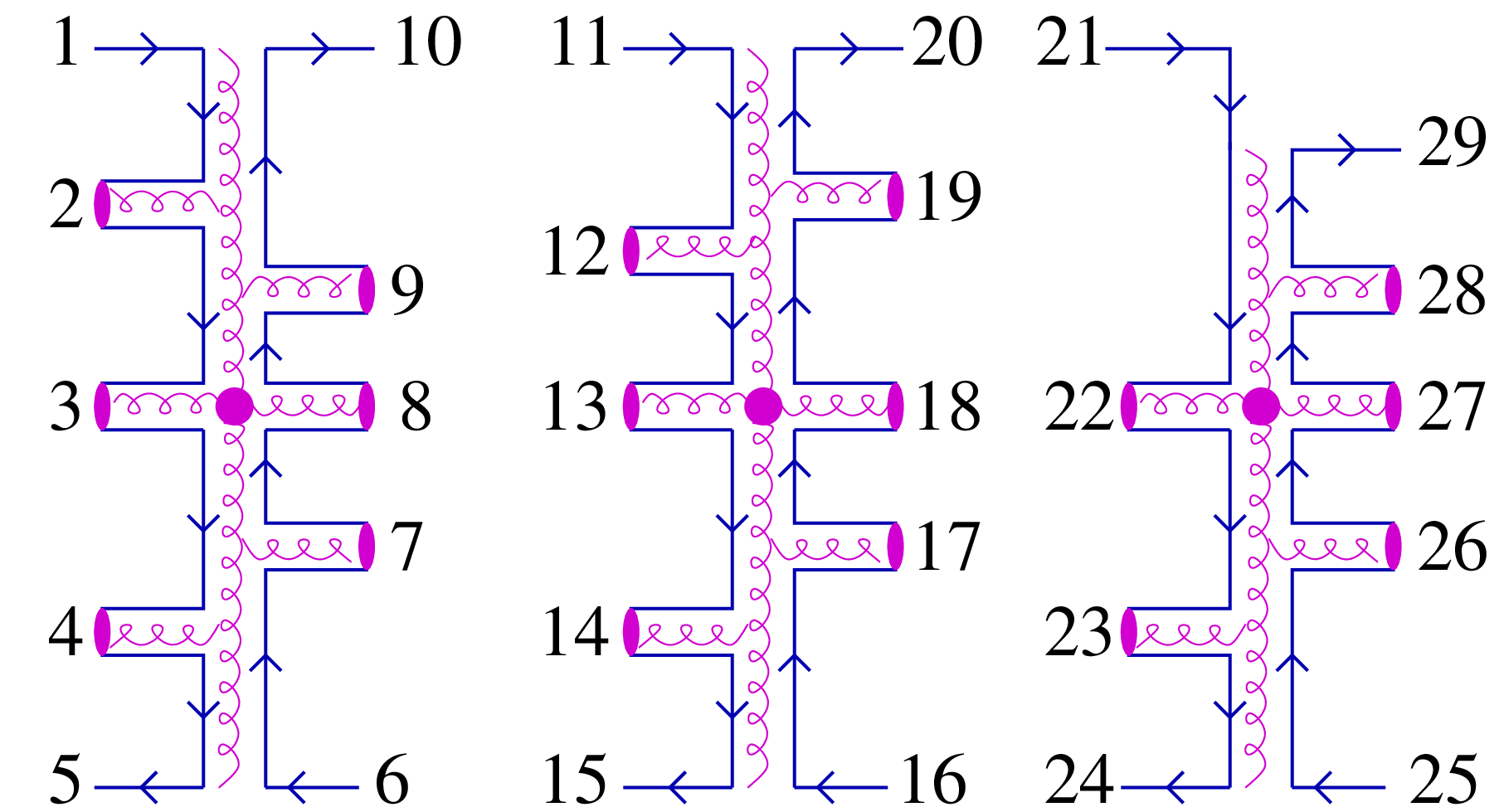}\caption{A possible color flow diagram for the three scatterings of Fig. \ref{partonic-configuration-1}.
\label{partonic-configuration-1-1}}
\end{figure}

\begin{figure}[h]
\noindent \centering{}\includegraphics[scale=0.25]{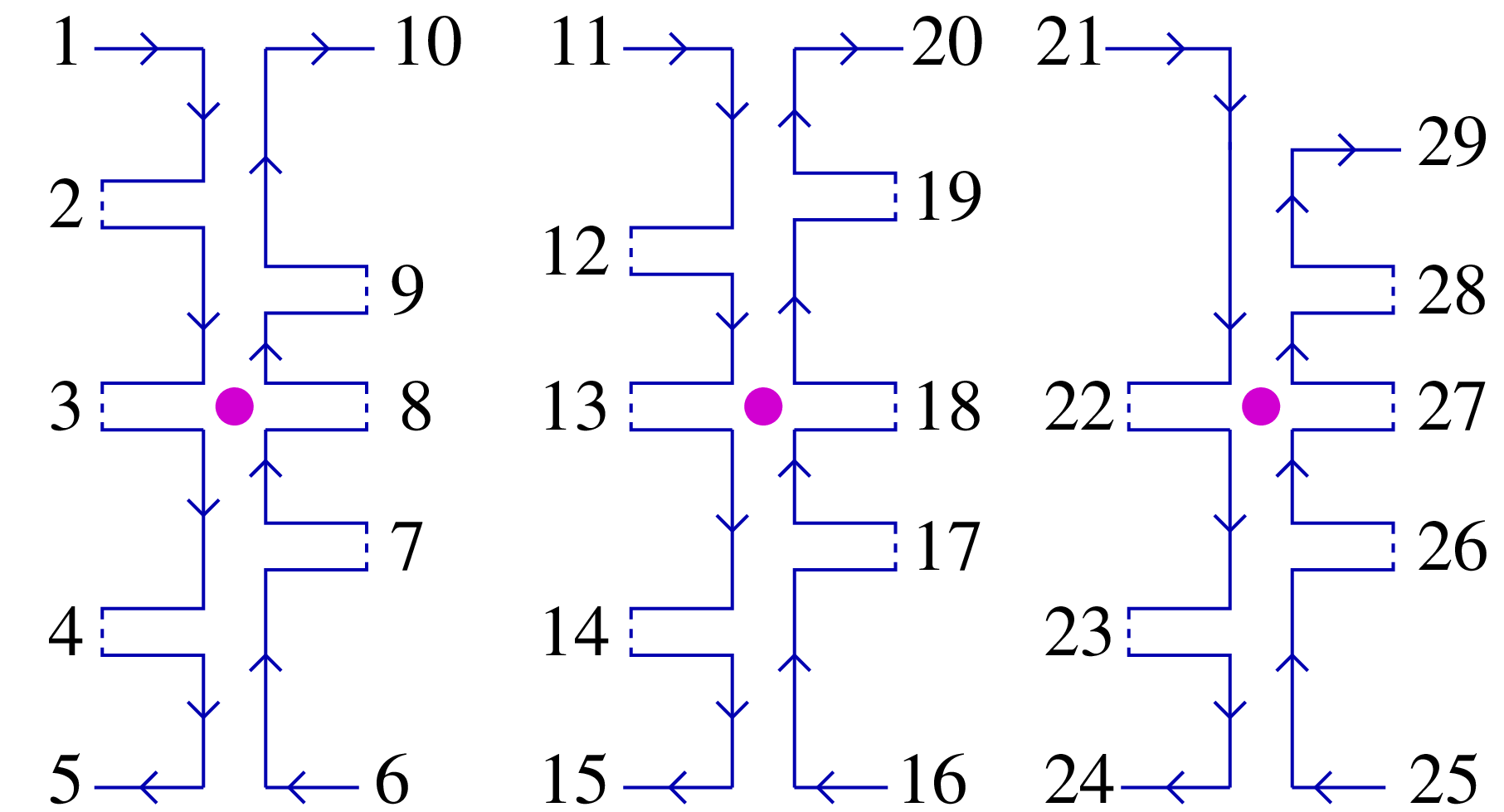}\caption{Simplified color flow diagram for the three scatterings of Fig. \ref{partonic-configuration-1-1}.
The dot indicates the Born process.\label{partonic-configuration-1-1-1}}
\end{figure}

In the following, we discuss the color flow for a given configuration,
as for example the one in Fig. \ref{partonic-configuration}. Since
the remnants are by construction white, we do not need to worry about
them, we just consider the rest of the diagram. In addition, colorwise,
the ``soft Pomeron'' part behaves as a gluon. Finally, we use the
following convention for the SL partons which are immediately emitted:
We use for the quarks an array away from the vertex, and for the antiquarks
an array towards the vertex. The diagram equivalent to Fig. \ref{partonic-configuration}
is then the one shown in Fig. \ref{partonic-configuration-1}.
Based on Fig. \ref{partonic-configuration-1}, considering the fact
that in the parton evolution and the Born process, the gluons are
emitted randomly to the right or to the left, we show in Fig. \ref{partonic-configuration-1-1}
a possible color flow diagram for the three scatterings. Horizontal
lines refer to TL partons, which later undergo a timelike cascade, while
the vertical lines refer to spacelike intermediate partons. We added 
integers
just to mark the different TL partons. For the leftmost scattering,
starting from one ``end'', say ``1'', we follow the color flow
to ``5'', and then starting from ``6'' to ``10'', so we
get two chains: 1-2-3-4-5 and 6-7-8-9-10. The end partons of each
chain are always quarks or antiquarks, and the inner partons are gluons.
Similar chains are obtained for the second scattering, 11-12-13-14-15
and 16-17-18-19-20, and for the third scattering, 21-22-23-24 and
25-26-27-28-29. These chains of partons will be mapped (in a unique
fashion) to kinky strings, where each parton corresponds to a kink,
and the parton four-momentum defines the kink properties, as already
done in earlier EPOS versions \cite{Drescher:2000ha}. In the following,
we simplify our graphs, instead of Fig. \ref{partonic-configuration-1-1}
we use a diagram as shown in Fig. \ref{partonic-configuration-1-1-1},
where we do not plot the gluon lines explicitly, since the double
lines with arrows in opposite direction allow us perfectly to identify
the gluons.

Considering these examples, it is useful to define ``initial  partons'',
being those which are immediately emitted as TL partons before the
parton evolution starts. In the case of the type ``sea'', a quark and
an antiquark (or diquark) are emitted, in Fig. \ref{partonic-configuration-1-1}
the partons 1,10 and 5,6 and 11,20 and 15,16 and 24,25. The partons
like 2,3,4 or 7,8,9 are emitted either in the spacelike partonic
cascade (like 2,4,7,9) or in the Born process (3,8). In case of ``val'',
a TL antiquark is emitted, namely the antiquark 21. Here, an SL quark
enters the partonic cascade (vertical line). Parton 29 is not
an initial parton, it is the first parton emitted in the SL cascade. 

\begin{figure}[h]
\noindent \centering{}(a)\hspace*{2.2cm}(b)~~~~~~~~~~~~~~~~\\
\includegraphics[scale=0.25]{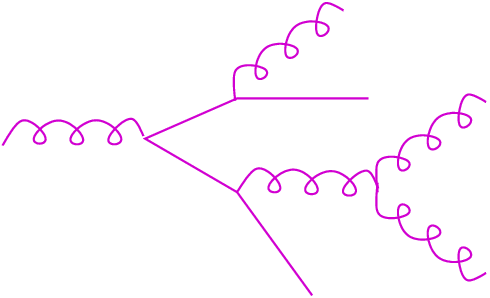}$\qquad$\includegraphics[scale=0.25]{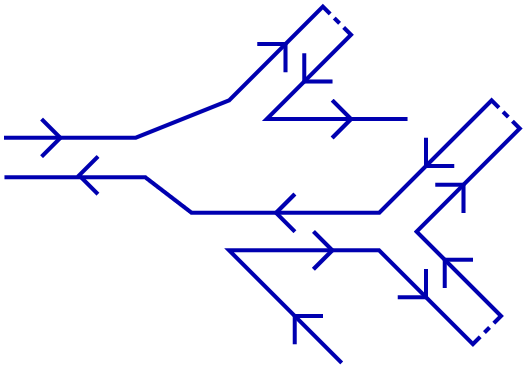}\caption{timelike cascade, with Feynman diagram (a) and the corresponding
color flow diagram (b). \label{timelike-cascade}}
\end{figure}

The above discussion demonstrates for an example, how to obtain ``chains
of partons'' (and then kinky strings), based on an analysis of the
color flow. But the picture is not yet complete. In general, each emitted
parton initiates a timelike cascade, as shown in Fig. \ref{timelike-cascade},
and this has to be taken into account when constructing the ``chains
of partons'' based on color flow. In the following, we discuss the
general rules, applicable for any multiple scattering configuration,
in pp or AA, including timelike cascades.

\subsection{The initial TL partons and the initial SL partons}

Even in the  case of multiple scatterings, the projectile and target remnants
remain color white and fragment independently, and correspondingly
the external legs of the individual scatterings are white. They are
quark-antiquark pairs (or less likely quark-diquark pairs). These
external legs emit immediately TL partons, referred to as ``initial
partons'', as compared to the partons emitted later during the parton
evolution. The number and the properties of the initial partons for
each individual scattering depend on the Pomeron type.

\begin{figure}[h]
\noindent \centering{}\includegraphics[scale=0.25]{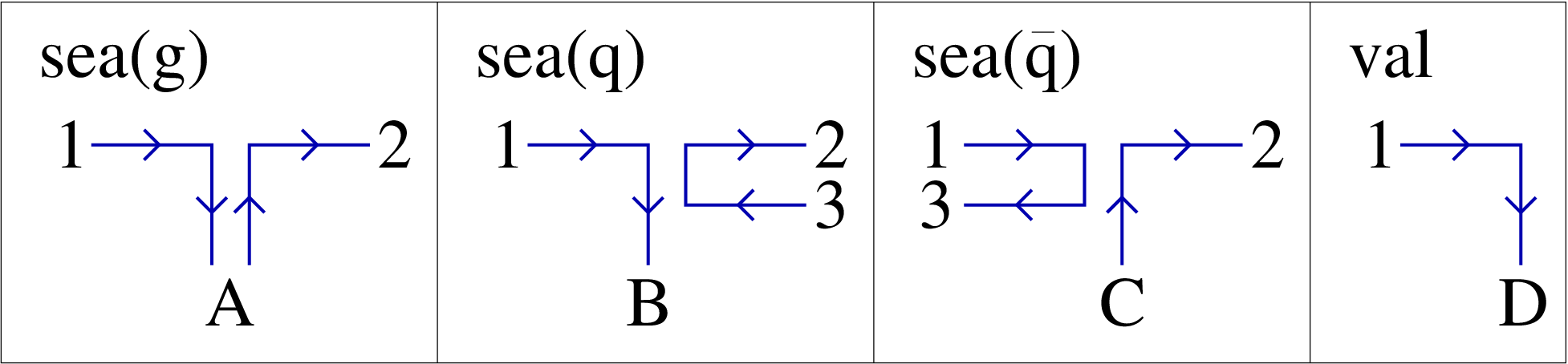}\caption{Initial TL partons 1,2,3, for the cases ``sea($g$)'' and ``sea($q$)'',
``sea($\bar{q}$)'', and ``val'', on the projectile side. The
``initial SL partons'' which initiate the SL cascade are marked
A, B, C, D. \label{timelike-cascade-1}}
\end{figure}

\begin{figure}[h]
\noindent \centering{}\includegraphics[scale=0.25]{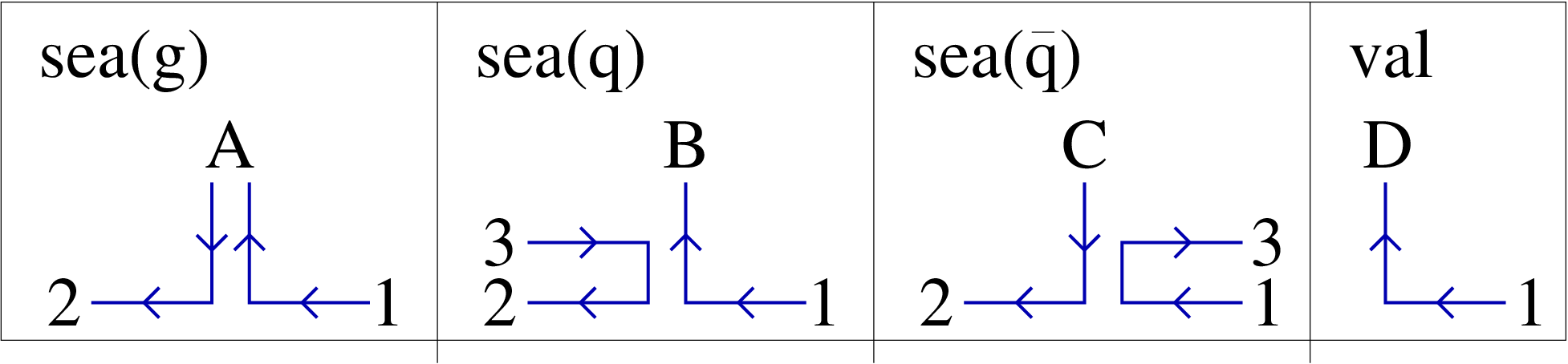}\caption{Initial TL partons 1,2,3, for the cases ``sea($g$)'' and ``sea($q$)'',
``sea($\bar{q}$)'', and ``val'', on the target side. The ``initial
SL partons'' which initiate the SL cascade are marked A, B, C, D. \label{timelike-cascade-1-1}}
\end{figure}

We have four scattering (or Pomeron) types: ``sea-sea'', ``val-sea'',
``sea-val'', and ``val-val'', where the first concerns the projectile,
the second the target. In the case of ``sea'' (on the projectile or
the target side), we have three cases: ``sea($g$)'' and ``sea($q$)'',
and ``sea($\bar{q}$)''. The notion of ``sea'' always means that
there is first a ``soft pre-evolution''. In case of ``sea($g$)''
the first parton entering the spacelike (SL) cascade, is a gluon,
in case of ``sea($q$)'' a quark, and in case of ``sea($\bar{q}$)''
an antiquark. These partons are referred to as ``initial SL partons''.
In Fig. \ref{timelike-cascade-1},
we show the four possibilities, for the projectile side. In case of
``sea($g$)'', the ``external leg'' is a quark-antiquark pair,
which emits a TL quark and a TL antiquark (2 and 1 in Fig. \ref{timelike-cascade-1})
and an SL gluon, which starts the SL cascade and is color connected
to 1 and 2. In the case of sea($q$), in addition to 1 and 2, the TL antiquark
3 is emitted, and an SL quark is initiating the cascade, color connected
to 1. The TL antiquark 3 is color connected to the TL quark 2 (so
2 and 3 constitute a ``little'' string, which will be given a small
amount of energy). In case of sea($\bar{q}$), in addition to 1 and
2, a TL quark is emitted (3), color connected to the TL antiquark
1. Here an SL antiquark is initiating the SL cascade, color connected
to 2. Finally, in the case of val, a TL antiquark is emitted (1), and
we have an SL quark initiating the SL cascade, color connected to antiquark
1. The discussion for the target side is identical, we just use a
somewhat different graphical representation, as shown in Fig. \ref{timelike-cascade-1-1}.
Here the arrows from 1 (emitted antiquark) to 2 (emitted quark) go
from right to left.

Knowing the color connections of the initial partons among each other
and with respect to the SL partons initiating the SL cascade, we need
as a next step to extend the construction of the color connections
to the SL cascade. 

\subsection{The spacelike (SL) parton evolution \label{-------spacelike-evolution-------}}

\begin{figure}[h]
\noindent \centering{}\includegraphics[scale=0.25]{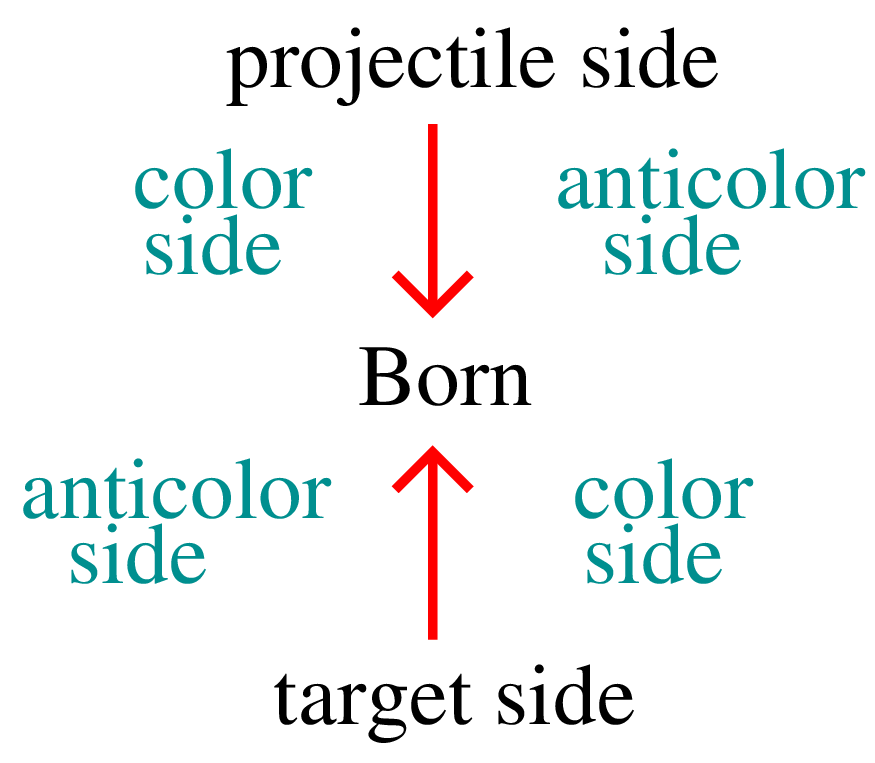}\caption{Defining color and anticolor side for the SL evolution. The red arrows
represent the directions of the evolutions, starting from the projectile
or the target side.\label{fig:Defining-color} }
\end{figure}

Whereas for technical reasons, we employ (for each individual scattering) 
a backward evolution method to generate the kinematical variables of the
partons, we will here consider the SL parton evolution
starting from the ``initial SL partons'' towards the Born process.
To have a unique definition of the color flow in the direction of
the evolution, we define (instead of ``right'' and ``left'') the
terms ``color side'' and ``anticolor side'', as shown in Fig.
\ref{fig:Defining-color}.
\begin{figure}[h]
\noindent \centering{}\includegraphics[scale=0.25]{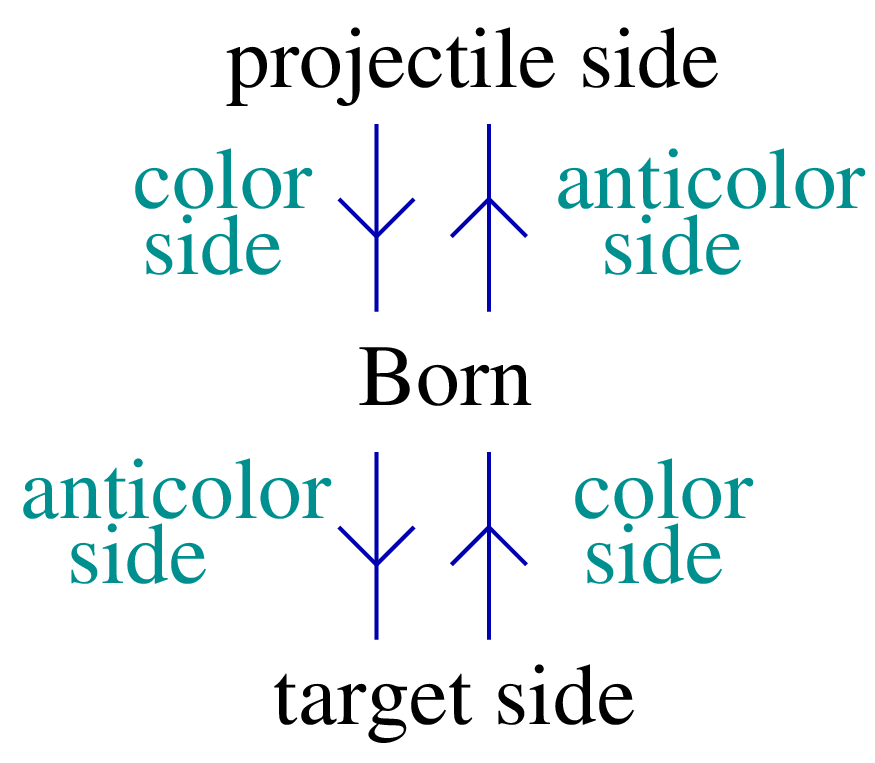}\caption{The color arrows for a gluon evolution, from the projectile or the
target side towards the Born process.\label{fig:Defining-color-1} }
\end{figure}
In this way, in the direction of the evolution, ``color side'' corresponds
to ``right'',  and ``anticolor side'' to ``left''. For the evolution
of an SL gluon, the two arrows representing the color flow are always
such that the forward arrow (in the direction of the evolution)
is on the ``color side'', and the backward arrow on the ``anticolor
side'', as shown in Fig. \ref{fig:Defining-color-1}. 

The aim is to construct the color flow, i.e., sequences of TL partons
$j_{1}-j_{2}-...-j_{n}$ (like the sequence $1-2-3-4-5$ shown in
in Fig. \ref{partonic-configuration-1-1-1}). To do so we number all
the timelike partons $j=1,2,3...$ (starting with the initial ones).
Referring to this parton numbering $j$, we define an integer array, 
the connection array $n_{\mathrm{CJ}}^{\eta}(i,j)$ for TL partons
$j$, 
defined such that
\begin{equation}
    j_{\mathrm{before}}=n_{\mathrm{CJ}}^{\eta}(1,j)
\end{equation}
defines the TL parton in the chain just before and 
\begin{equation}
    j_{\mathrm{after}}=n_{\mathrm{CJ}}^{\eta}(2,j)
\end{equation}
the TL parton just after the parton $j$. So for the example of the
chain $1-2-3-4-5$ shown in in Fig. \ref{partonic-configuration-1-1-1},
considering the subsequence $2-3-4$, for $j=3$, we have $j_{\mathrm{before}}=n_{\mathrm{CJ}}^{\eta}(1,j)=2$
and $j_{\mathrm{after}}=n_{\mathrm{CJ}}^{\eta}(2,j)=4$. When we talk
about sequences $j_{1}-j_{2}-...-j_{n}$, we have two options (which
we use both): following the color flow ($\to$) or the anticolor flow
($\leftarrow$). In the first case, we use $\eta=1$, in the second
one $\eta=2$. The index $\eta$ is referred to as ``color orientation''.
At the end of the emission procedures, this array will contain the
complete information about which partons are color connected. 

For the evolution process, it is useful to define the ``current SL
parton'' (say parton $a$), based on which the emission process is
considered ($a\to b+c$), with an SL parton $b$ and a TL $c$. In
the next step, parton $b$ is considered to be the current SL parton,
and so on. During the evolution procedure, we use for the current
SL parton some integer array,
the connection array $n_{\mathrm{CC}}(i,k)$ for the current SL parton,
providing information to which TL partons the current SL parton is
connected to. Here, 
\begin{equation}
    j_{\mathrm{proj}} =  n_{\mathrm{CC}}(i,1)
\end{equation}
refers to the connection on the projectile side and
\begin{equation}
    j_{\mathrm{targ}} =  n_{\mathrm{CC}}(i,2)
\end{equation}
to the connection on the target side. For gluons, the index $i=1$ refers 
to the color side, and $i=2$ to the anticolor side. In the case of quarks and antiquarks,
only $i=1$ will be used, and the value for $i=2$ is zero.

\begin{figure}[h]
\noindent \centering{}\includegraphics[scale=0.25]{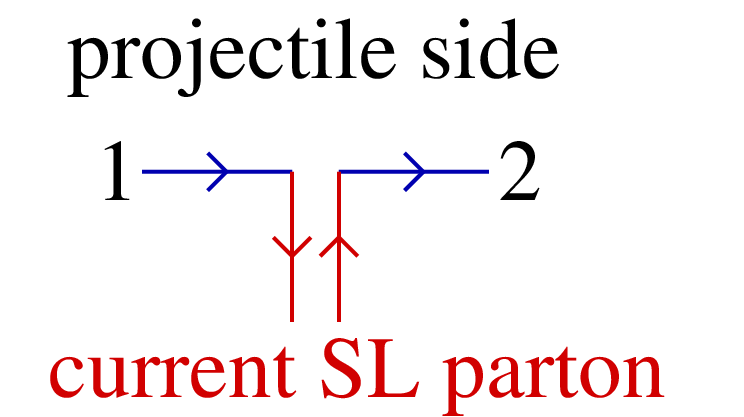}$\quad$\caption{gluon as first SL parton, connected to 1 (color side) and 2 (anticolor
side). \label{fig:First-emission}}
\end{figure}

Consider for example the first SL parton on the projectile side, as
shown in Fig. \ref{fig:First-emission}.
 The first ``current parton'' is a gluon. It is color connected
to 1 (color side) and 2 (anticolor side), so we have $n_{\mathrm{CC}}(1,k)=1$
and $n_{\mathrm{CC}}(2,k)=2$, here for $k=1$, since we consider
the projectile side. 

\begin{figure}[h]
\noindent \centering{}(a)\hspace*{4cm}(b)\hspace*{10cm}$\quad$\\
\includegraphics[scale=0.5]{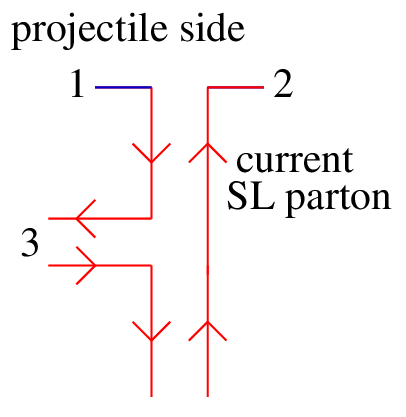}$\quad\qquad$\includegraphics[scale=0.5]{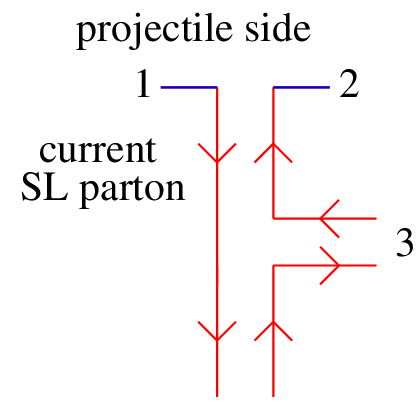}\caption{The two possibilities of color flow for a gluon emission: (a) emission
on the color side, corresponding to $\eta=1$ or (b) the anticolor
side, corresponding to $\eta=2$. \label{fig:First-emission-2}}
\end{figure}

In the next step, let us assume we have a gluon emission
(parton $3$). Concerning color flow, we have two possibilities, an
emission on the color side or the anticolor side, as shown in Fig.
\ref{fig:First-emission-2}.
 In the first case, we have a ``subchain'' $1\to3$, i.e., for the
new TL parton $j=3$, we have $n_{\mathrm{CJ}}^{\eta}(1,j)=1$, and
since we follow the color flow ($\to$), we use $\eta=1$. In the
second case, we have a ``subchain'' $2\to3$, i.e., for the new TL
parton $j=3$, we have $n_{\mathrm{CJ}}^{\eta}(1,j)=2$, and since
we follow the anticolor flow ($\to$), we use $\eta=2$. In other
words, the two choices correspond to the two choices $\eta=1$ and
$\eta=2$, and they are chosen randomly.%

\begin{figure}[h]
\noindent \centering{}\includegraphics[scale=0.5]{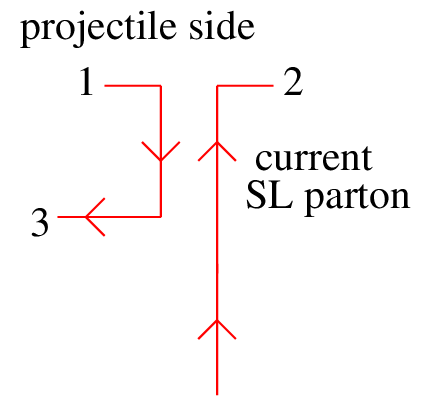}\caption{The two possibilities of color flow for the first emission. The l.h.s.
of the figure corresponds to the color orientation $\eta=1$, the r.h.s.
to $\eta=2$ \label{fig:First-emission-1}}
\end{figure}

Having discussed the case of a gluon being the first emitted TL parton
(marked 3), let us discuss the case of the first emitted TL parton
being a quark (also marked 3). Here, we have no choice, as can be
seen in Fig. \ref{fig:First-emission-1}.
Here, we have mandatory color orientation $\eta=1$, the quark is
emitted on the color side. And in the case of antiquark emission,
we have $\eta=2$. The color orientation determines which of the
partons, $n_{CC}(1,k)$ or $n_{CC}(2,k)$, is the one the new emitted
TL parton is connected to. In the first case (quark emission), we
have for the new parton $j=3$ the connection $n_{\mathrm{CJ}}^{\eta}(1,j)=n_{\mathrm{CC}}(\eta,k)=1$
(using $\eta=1$). In the second case (antiquark emission), we have
for the new parton $j=3$ the connection $n_{\mathrm{CJ}}^{\eta}(1,j)=n_{\mathrm{\mathrm{CC}}}(\eta,k)=2$
(using $\eta=2$) always for $k=1$ (projectile). %

So far we treated the first TL emission of the SL parton cascade, and
we explained how to use the two arrays $n_{\mathrm{CJ}}$ and $n_{\mathrm{\mathrm{CC}}}$.
In a similar way we can treat all subsequent emissions. However, an
emitted TL parton will, in general, develop a TL parton cascade, and
we first need to deal with that. We will discuss this in the next
section. 

\subsection{The timelike cascade}

Here we consider the situation, at some stage of the SL cascade, where
starting from some ``current SL parton'' we have an emitted TL parton
with (some unique) number $j$, and suppose we know the color orientation
$\eta$ of this TL emission and the color connection array $n_{\mathrm{CJ}}^{\eta}(1,j)$.
To simplify the notation (reduce the indices), we define for given
$\eta$ and given $j$ the integer $n_{\mathrm{CR}}(1)=n_{\mathrm{CJ}}^{\eta}(1,j)$.
There is also the variable $n_{\mathrm{CR}}(2)$, referring to the
connected parton on the other side, not yet known here, but important
in case of TL emissions in the Born process, to be discussed later. 

\begin{figure}[h]
\noindent \centering{}\includegraphics[scale=0.5]{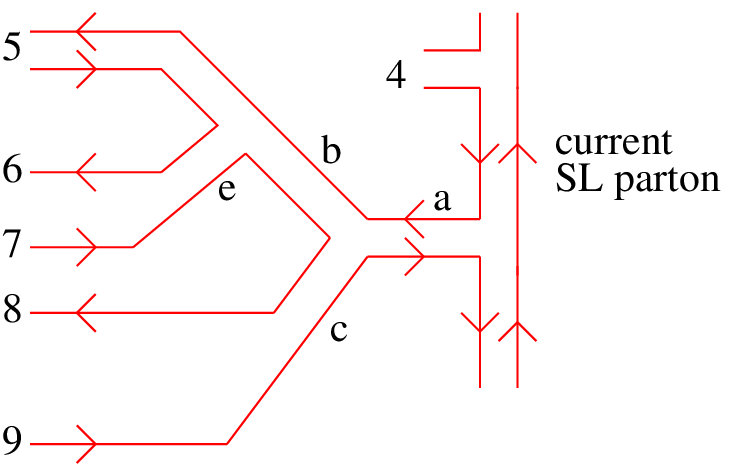}\caption{Color flow in TL cascade. \label{fig:TL-cascade-1}}
\end{figure}

We need to reconstruct the color flow of the TL cascade and connect
it to the rest of the diagram, knowing the connected parton $n_{\mathrm{CR}}(1)$,
which points to the previous TL parton in the color flow chain, and
the color orientation $\eta$. We assume that we know at this stage
the complete TL cascade, with the flavors and four-momenta of all
the partons. Let us consider a concrete case as shown in Fig. \ref{fig:TL-cascade-1},
with $\eta=1$ (emission of a gluon on the color side), and with
the previous TL parton in the color flow chain being $n_{\mathrm{CR}}(1)=4$
(assuming that the indices 1,2,3 have been used already before). We
also assume here that parton 4 is a final particle. We actually use
a rigorous unique numbering (1,2,3,...) only for the final partons,
for the intermediate partons in the plot we use symbols $a,\,b,\,c\ldots$.
The first emitted TL parton $a$ splits into $b$ and $c$, and these
daughter partons split again in partons, and at the end, we have final
partons 5, 6, 7, 8, 9. At this stage (after realizing the TL cascade,
as discussed earlier), the planar structure is already fixed, and each
vertex like $a\to b+c$ has a clearly defined ``first daughter''
D1 (upper) and a ``second daughter`` D2 (lower), when considering
the evolution from right to left. The fact that there is some freedom
concerning the exchange of D1 and D2 is taken care of during the evolution.
Here, we simply need to identify the final partons, associate a number
(the next free integer), and reconstruct the color connections. This
may be done as follows: 
\begin{enumerate}
\item Define the initial parton to be the current parton.
\item Starting from the current parton, loop over all the first daughters
until a final parton is reached, now being the new current parton.
\item Update connection information for the current parton itself, its parent,
and the second daughter.
\item If the second daughter is not final, consider it to be the current
parton, go back to 2, 
\item If the second daughter is final, define the parent to be the current
parton, go back to 3.
\end{enumerate}
In this way, we move through the whole diagram, from top to bottom
in the graph of Fig.~\ref{fig:TL-cascade-1}. For our concrete example,
we identify in this way the final partons 5, 6, 7, 8, and 9 and the color
connections 4-5-6, 7-8, 9-X, where ``X'' in the last expression
means that the corresponding parton is not yet known. 

\subsection{A complete example including quarks}

The above procedures concern one step in the SL cascade, so we need
to iterate all steps of the SL cascade on the projectile side and,
of course, do the same on the target side. Finally, we need to treat
the Born process, where we have two TL cascades, and for each one
we employ the above procedures. 

\begin{figure}[h]
\noindent \centering{}\includegraphics[scale=0.25]{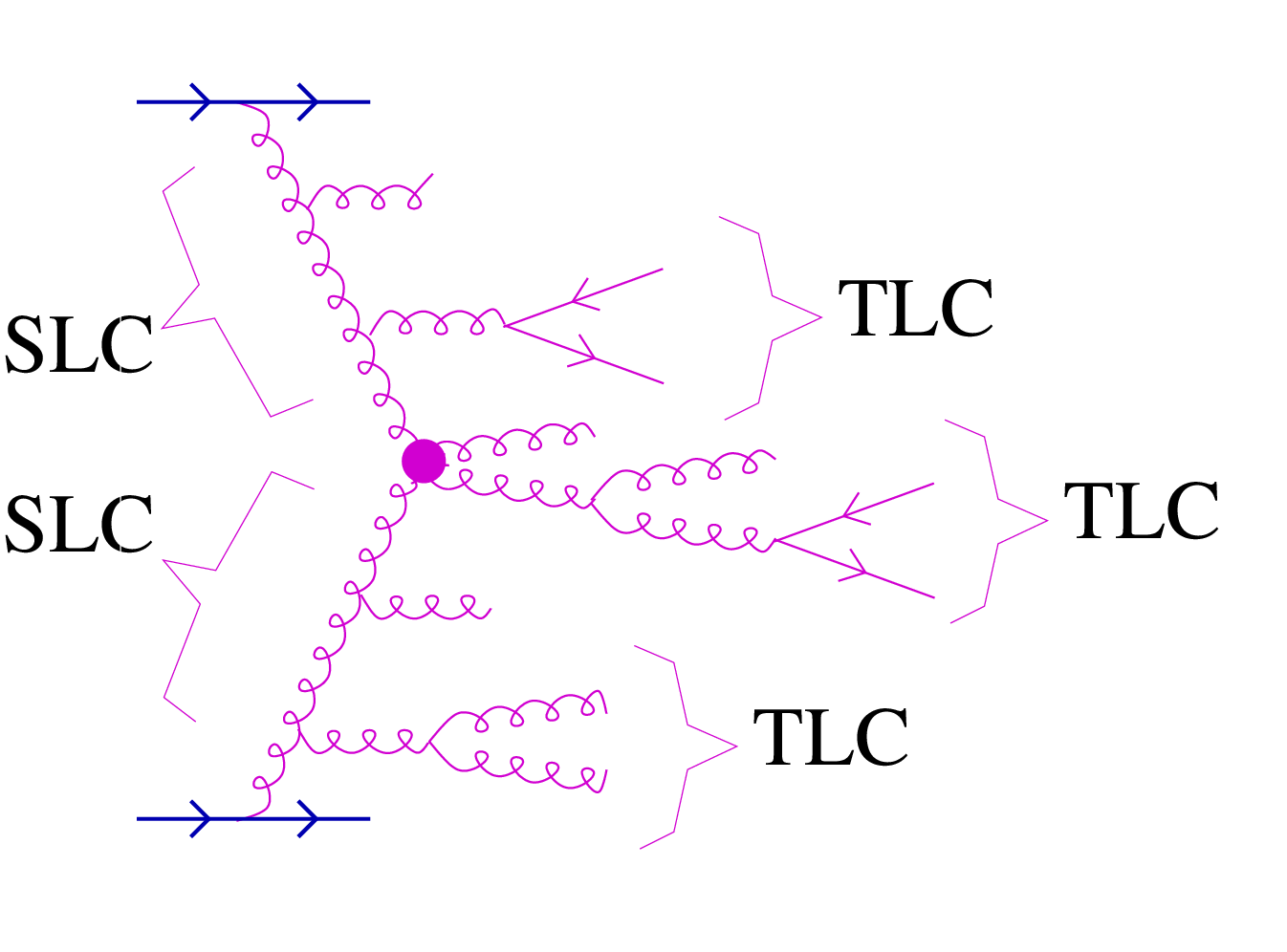}\caption{A single parton-parton scattering diagram, with SL and TL cascades,
including quarks. \label{complete-example}}
\end{figure}

\begin{figure}[h]
\noindent \centering{}\includegraphics[scale=0.25]{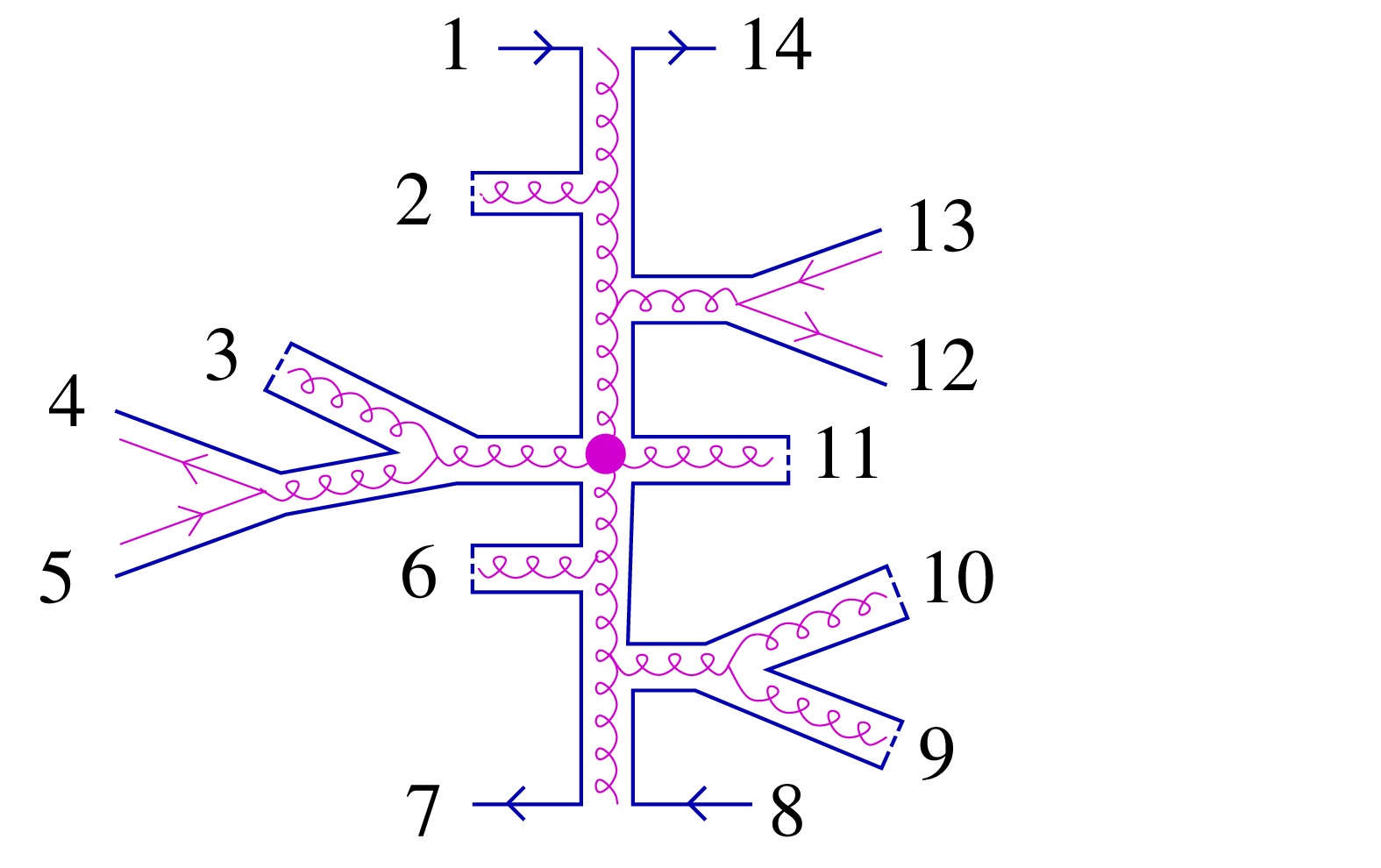}\caption{A possible color flow diagram corresponding to Fig. \ref{complete-example}.\label{complete-example-1}}
\end{figure}

Let us consider an example of a single scattering, with a diagram
with SL and TL cascades, including quarks, as shown in Fig. \ref{complete-example}.
In order to construct the color flow diagram, we remember that in
case of a $g\to g+g$ splitting, the emitted gluon may be emitted
(with equal probability) on the color side or the anticolor side
(see discussion in section \ref{-------spacelike-evolution-------}).
A possible choice is shown in Fig. \ref{complete-example-1}.
Following the color flow, always starting and ending with a single
arrow, we identify the following chains: 1-2-3-4, then 5-6-7, then
8-9-10-11-12, and finally 13-14.

One should note, as already mentioned several times, that even in
the case of multiple scatterings, the remnants are color-white, as are
the individual scatterings. In the multiple scattering procedure,
where we determine the momenta of all the Pomerons, we also fix the
flavors of the ``Pomeron ends''. Once these are known, the individual
scattering can be treated independently of each other. This is why
we discuss here individual scatterings.

\subsection{Strings}

\begin{figure}[h]
\noindent \centering{}\includegraphics[scale=0.25]{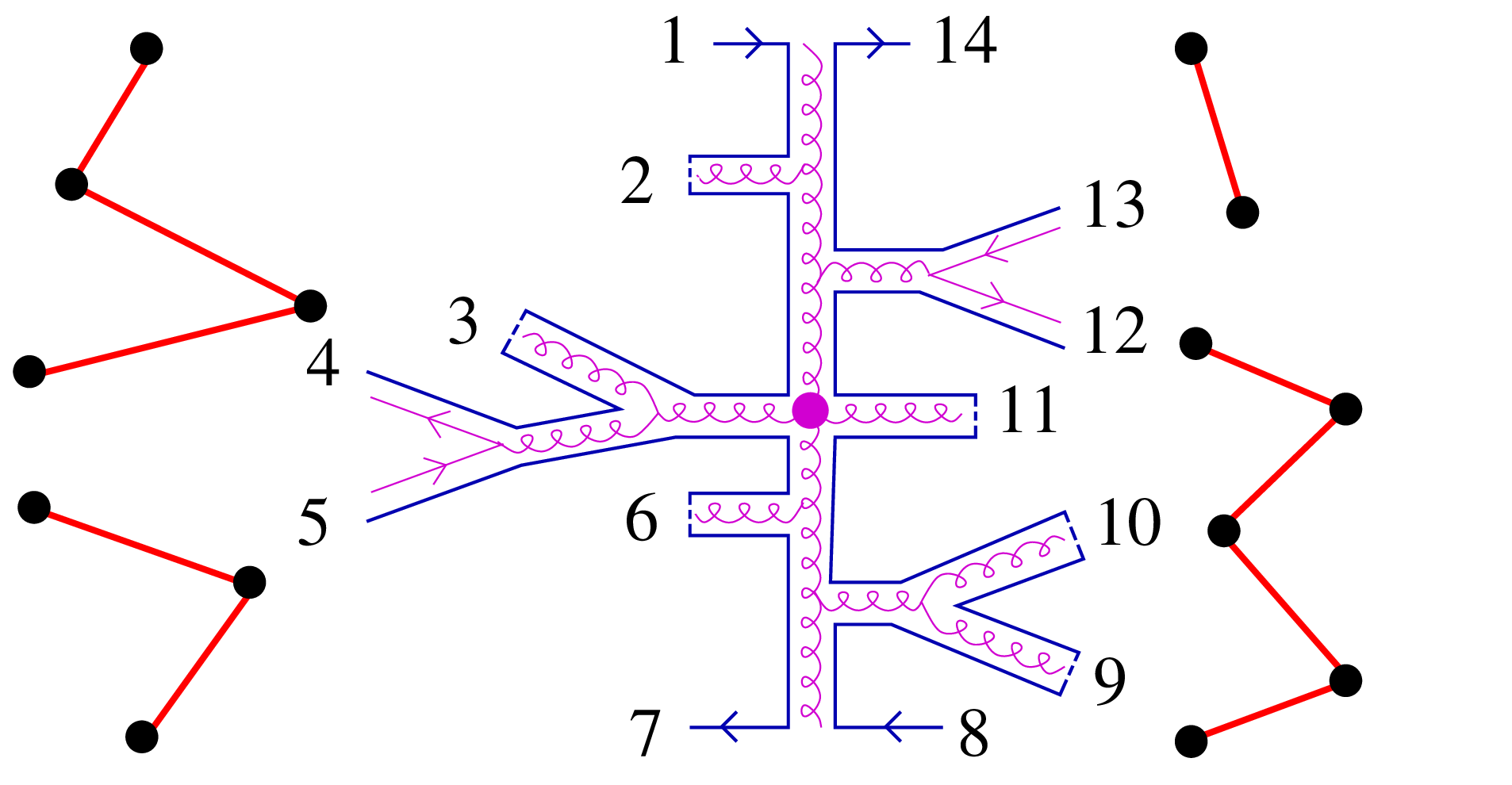}\caption{The four color flow chains 1-2-3-4, 5-6-7, 8-9-10-11-12, and 13-14
are mapped to kinky strings (red lines). The black points indicate
the kinks.\label{complete-example-1-1}}
\end{figure}

In the previous sections, we discussed how to construct chains of
partons, $j_{1}-j_{2}-...-j_{n}$, where the inner partons are gluons,
and the two outer partons, in general, a quark and an antiquark (in
any case a $3$ and $\bar{3}$ color representation). These chains
of partons are mapped (in a unique fashion) to kinky strings, where
each parton corresponds to a kink, and the parton four-momentum defines
the kink properties, as already done in earlier EPOS versions, for
a detailed discussion see \cite{Drescher:2000ha}. Let us consider
the example of Figs. \ref{complete-example} and \ref{complete-example-1-1}.
Following the color flow, we have identified four chains, namely 1-2-3-4,
5-6-7, 8-9-10-11-12, and 13-14. Each of these chains is mapped to
a kinky string, as indicated in Fig. \ref{complete-example-1-1}.
 Each parton corresponds to a kink. The inner kinks correspond to
the inner gluons in the chains. The kinks carry the momenta of the
corresponding partons. The chain 13-14 corresponds to a string with
no inner kinks, so it is a flat (yo-yo) string. %

\subsection{Heavy quark issues}

\begin{figure}[h]
\noindent \centering{}(a)\hspace*{3cm}(b)\hspace*{3cm}(c)\hspace*{6cm}$\qquad$\\
\includegraphics[scale=0.25]{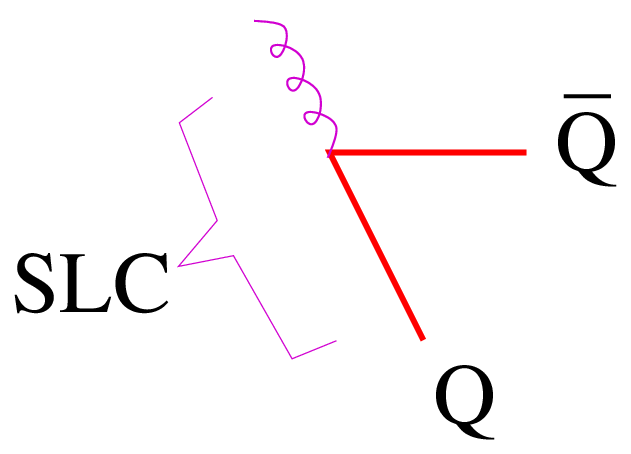}$\qquad$\includegraphics[scale=0.25]{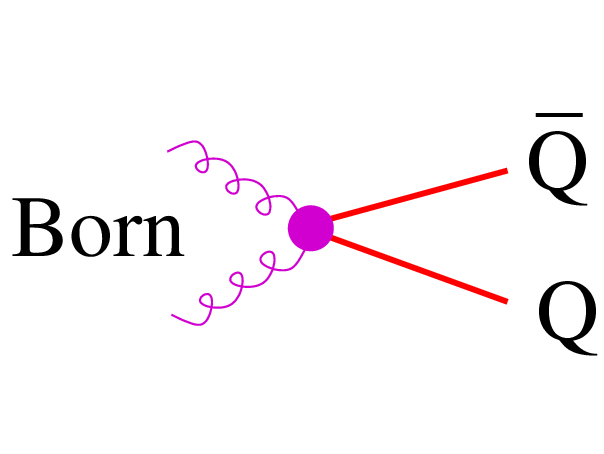}$\qquad$\includegraphics[scale=0.25]{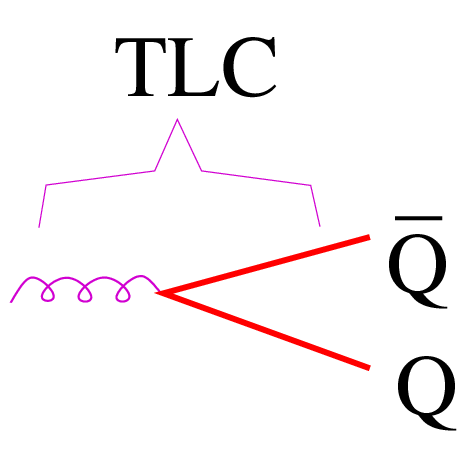}\caption{Different possibilities to create heavy flavor, (a) in the spacelike
cascade (SLC), (b) in the Born process, (c) in the timelike cascade
(TLC). \label{charm-3}}
\end{figure}

\begin{figure}[h]
\noindent \centering{}$\qquad$(a)\hspace*{4cm}(b)\hspace*{4cm}$\qquad$\\
\includegraphics[scale=0.25]{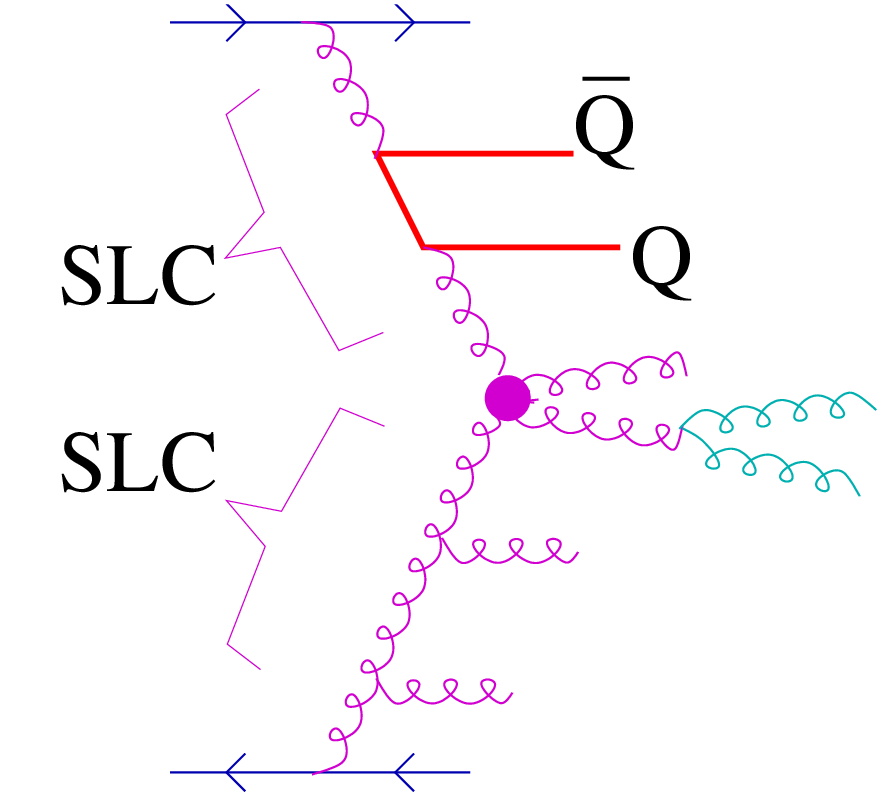}$\qquad$\includegraphics[scale=0.25]{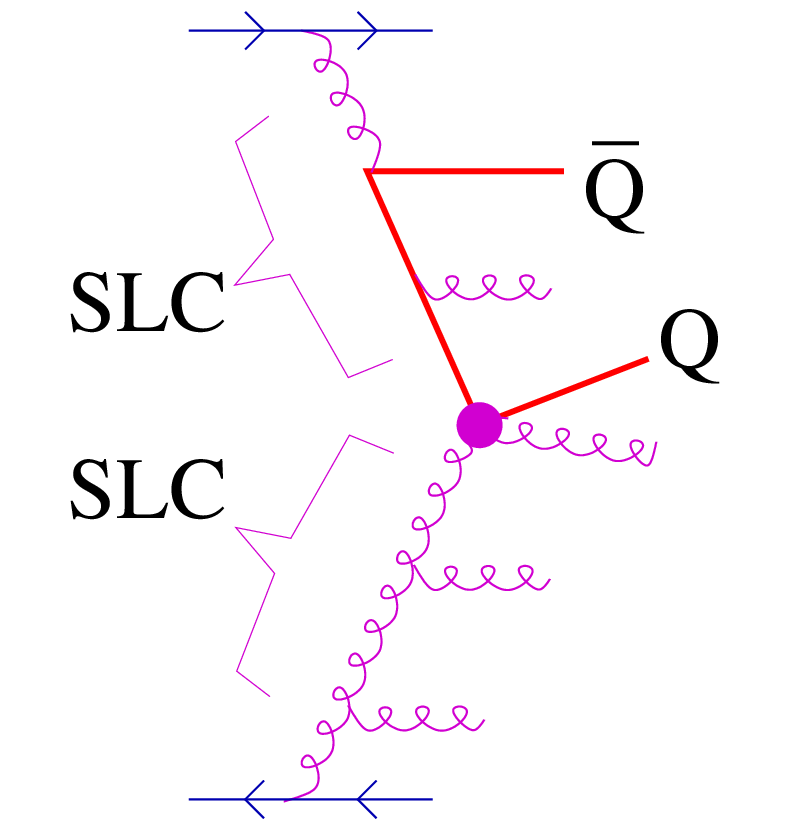}\\
$\qquad$\hspace*{2cm}(c)\hspace*{7cm}$\qquad$\\
$\qquad$\includegraphics[scale=0.25]{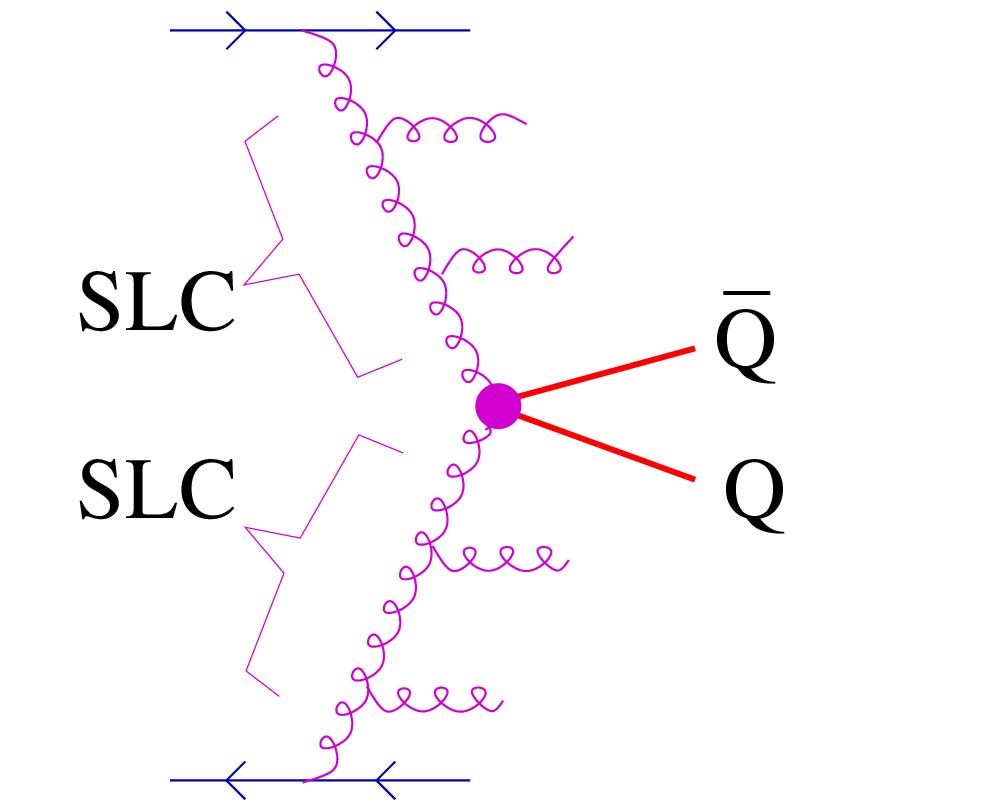}\caption{Heavy flavor production (a,b) in the SL cascade and (c) in the Born
process. The magenta point indicated the Born process. \label{charm-1}}
\end{figure}

\begin{figure}[h]
\noindent \centering{}$\qquad$(a)\hspace*{4cm}(b)\hspace*{4cm}$\qquad$\\
\includegraphics[scale=0.25]{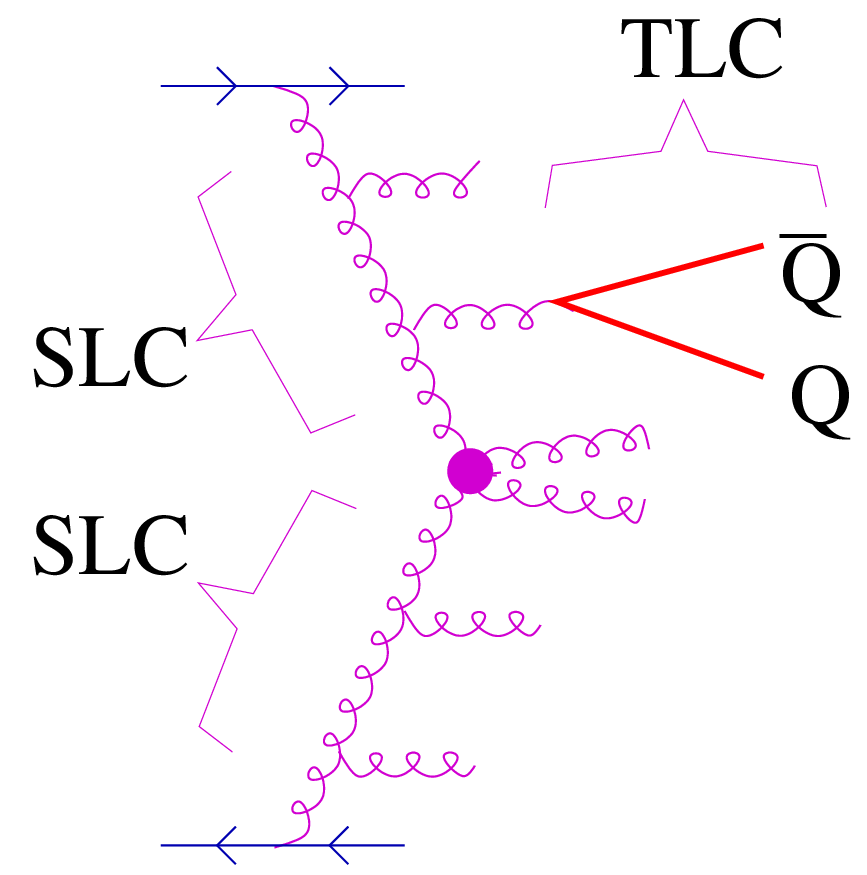}$\qquad$\includegraphics[scale=0.25]{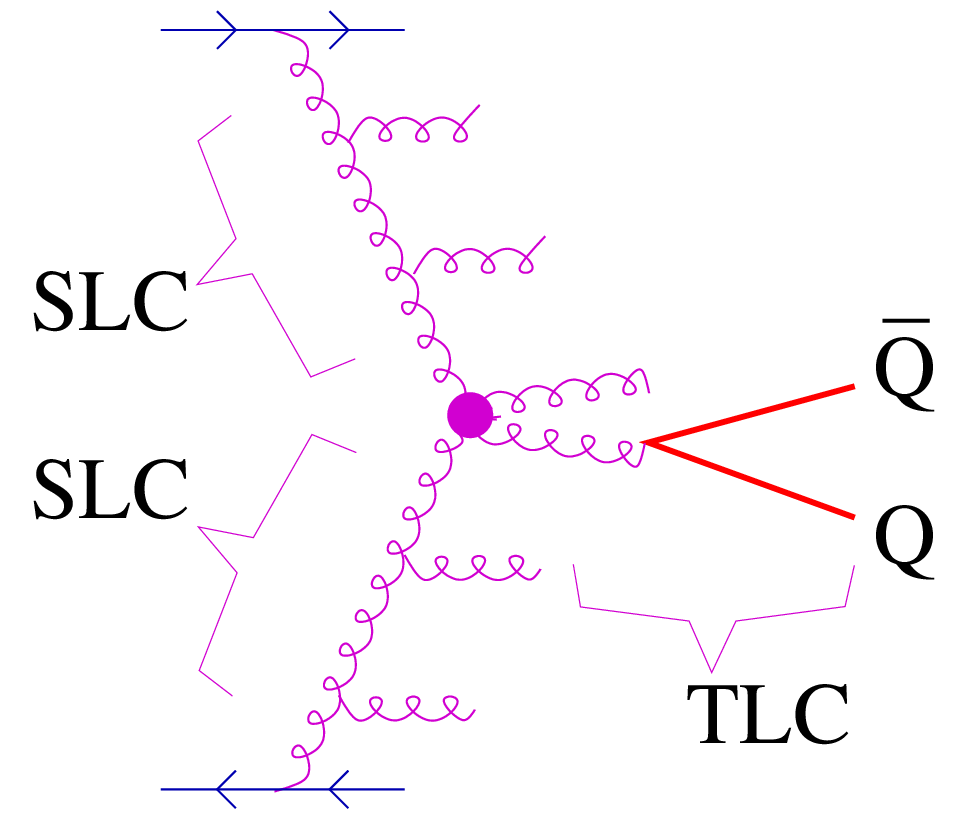}\caption{Heavy flavor production in the TL cascade. \label{charm-4-2}}
\end{figure}

At each step in the SL cascade, there is the possibility of quark-antiquark
production, and in the Born process as well. In the following, we
discuss in particular the case of heavy flavor quarks, with the general
notation $Q$ for quarks and $\bar{Q}$ for antiquarks. Heavy flavor
may be produced in different ways, as shown in Fig. \ref{charm-3}.
Starting from a gluon, a $Q-\bar{Q}$ pair may be produced in the
SL cascade, as shown in Fig. \ref{charm-3}(a), provided the virtuality
is large enough. The number of allowed flavors is considered to be
depending on the virtuality (variable flavor number scheme). It is
also possible to create a $Q-\bar{Q}$ in the Born process, via $g+g\to Q+\bar{Q}$
or $q+\bar{q}\to Q+\bar{Q}$ (for light flavor quarks $q$), as shown
in Fig. \ref{charm-3}(b), and finally $Q-\bar{Q}$ may be produced
in the TL cascade, via $g\to Q+\bar{Q}$, as shown in Fig. \ref{charm-3}(c).
Let us first consider the $Q-\bar{Q}$ production in the SLC. We may
have the situation as shown in Fig. \ref{charm-1}(a),
where a heavy flavor parton (here a $\bar{Q}$) is emitted, and the
corresponding antiparticle (here a $Q$) continues the SLC. But before
reaching the Born process, it is emitted, and a gluon continues the
SLC. The two heavy flavor partons have in general low transverse momenta.
Another possibility is shown in Fig. \ref{charm-1}(b), where a heavy
flavor parton produced in the SLC ``survives'' till the Born process,
and the latter has most likely the form $Q+l\to Q+l$, with $l$ being
a light flavor parton. Other than the production during the SLC, heavy
flavor may be produced in the Born process, via $g+g\to Q+\bar{Q}$
or $q+\bar{q}\to Q+\bar{Q}$ (for light flavor quarks $q$), as shown
in Fig. \ref{charm-1}(c). Finally, heavy flavor may be produced during
the timelike cascade, as shown in Fig. \ref{charm-4-2},
initiated either from a TL parton in the SL cascade [Fig. \ref{charm-4-2}(a)],
or from an outgoing parton of the Born process (Fig. \ref{charm-4-2}(b)).
In the first case, the transverse momenta are in general small.

\begin{figure}[h]
\noindent \centering{}\includegraphics[scale=0.25]{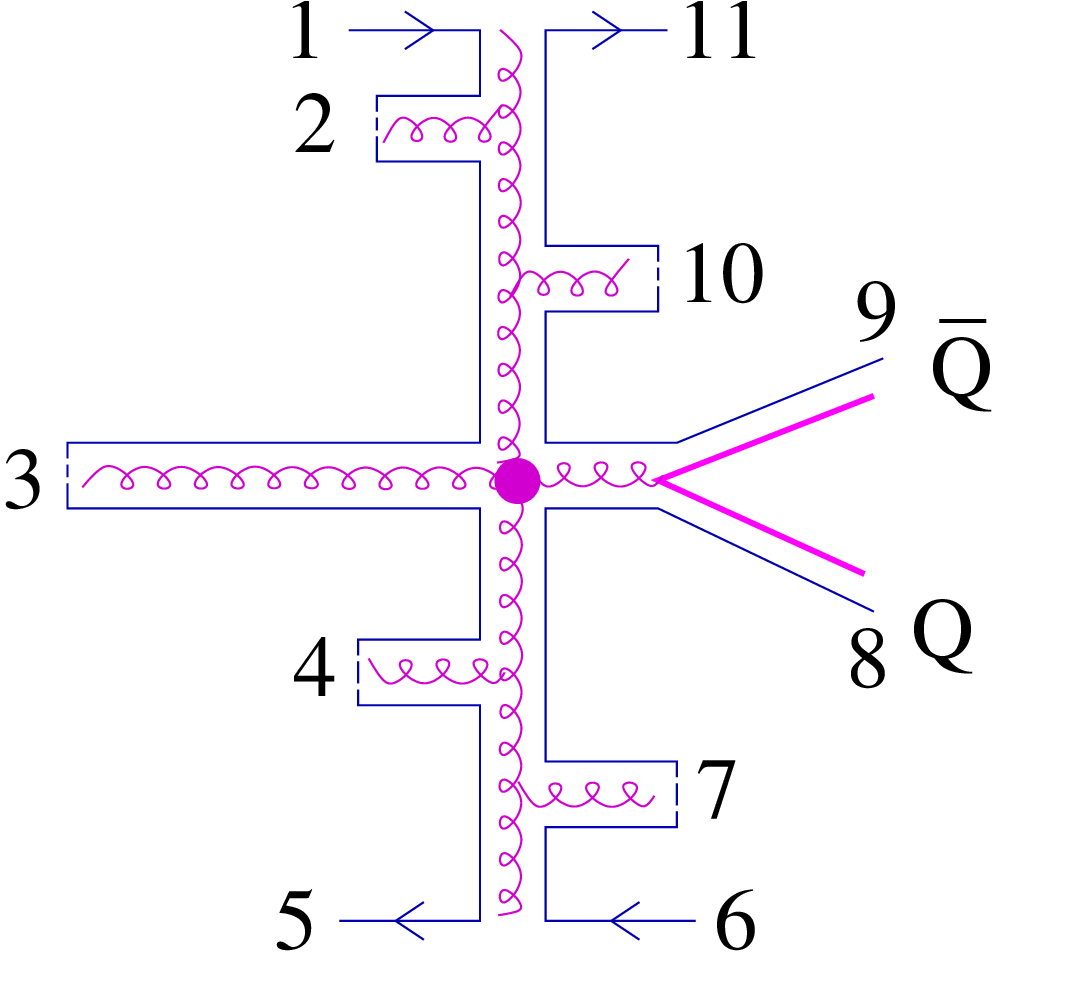}\caption{A possible color flow diagram corresponding to the graph of Fig. \ref{charm-4-2}(b)
\label{charm-4-1}}
\end{figure}

The next step will be, for a given Feynman diagram, to construct the
color flow diagram, as discussed in the previous sections. Let us
take the graph of Fig. \ref{charm-4-2}(b), i.e., heavy flavor production
during the TLC  of an outgoing Born parton. As usual, the gluons are
emitted to either side with equal probability, so a possible color
flow diagram is the one shown in Fig. \ref{charm-4-1}.
We identify three chains of partons: 1-2-3-4-5, 6-7-8, and 9-10-11.
The initial TL partons (the horizontal blue lines with arrows) or
most likely quarks and antiquarks (in any case $3$ and $\bar{3}$
color representations). Let us assume that 3 is a quark, and 6 an
antiquark (light flavor, both), then the two chains containing
heavy flavor are of the form $\bar{Q}-g-q$ and $\bar{q}-g-Q$, in
both cases, the heavy flavor partons are ``end partons'' in the
chains. 

\begin{figure}[h]
\noindent \centering{}\includegraphics[scale=0.25]{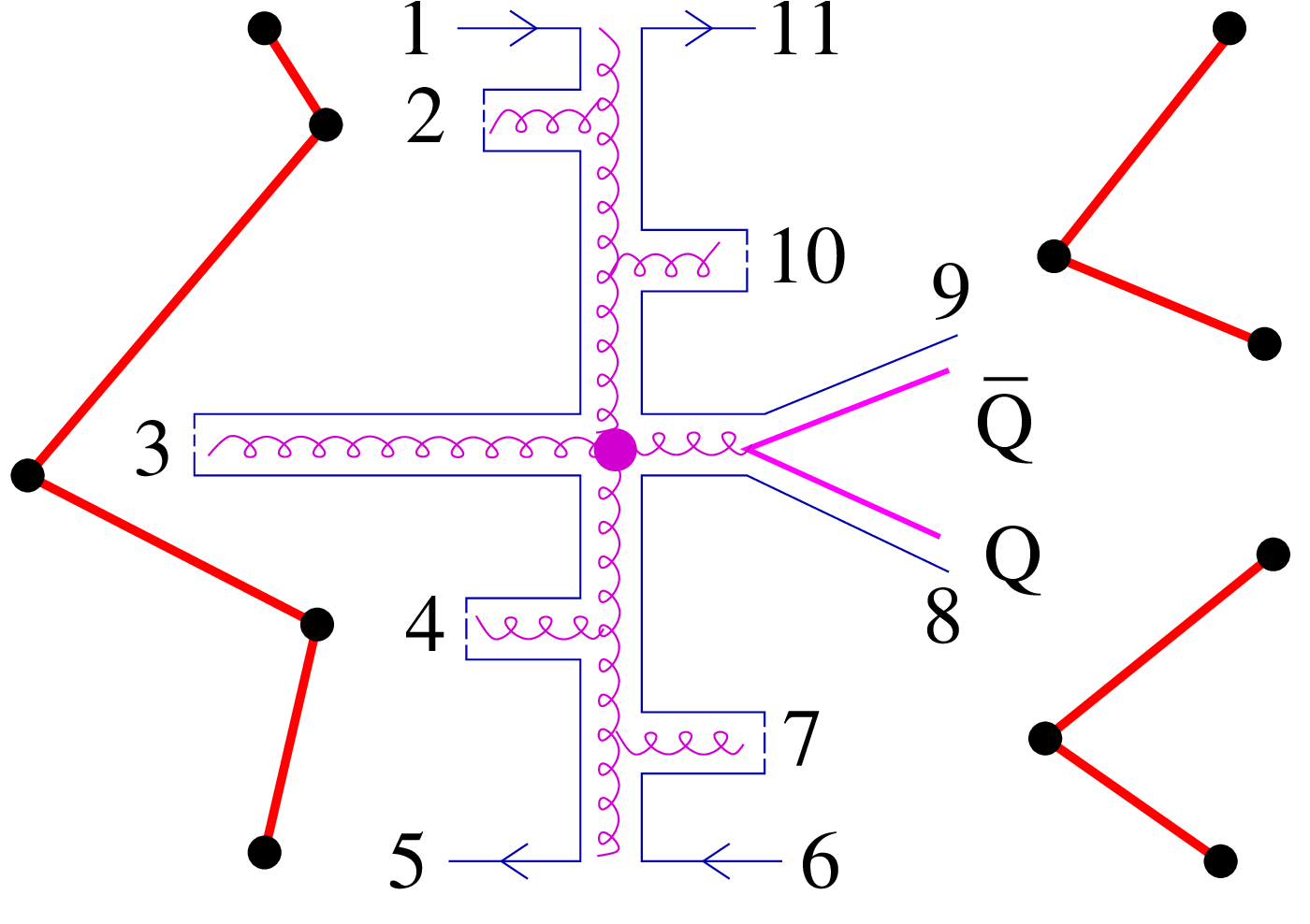}\caption{The chains 1-2-3-4-5, 6-7-8, and 9-10-11 are mapped to kinky strings
(red lines). The black points indicate the kinks, which carry the
parton momenta. \label{charm-4-1-1}}
\end{figure}

These chains of partons are finally mapped (in a unique fashion) to
kinky strings, where each parton corresponds to a kink, as shown in
Fig. \ref{charm-4-1-1}.

The general mapping procedure (chains of partons to kinky strings)
as well as the string decay procedures are described in detail in
\cite{Drescher:2000ha}.

\section{Summary}

\noindent We recently introduced new concepts \cite{werner:2023-epos4-overview}, 
implemented in EPOS4,
which allows to consistently accommodate factorization and saturation
in high-energy proton-proton and nucleus-nucleus collisions, in a rigorous
parallel scattering framework. 
EPOS4 has a modular structure: the multiple scattering contributions
to the total cross section in pp or AA scattering are expressed in
terms of (products of) cut Pomeron expressions $G$ (each one representing
a single scattering), and the latter ones are related to the ``real
QCD expressions'' $G_{\mathrm{QCD}}$ via some fundamental (new)
equation. In other words, $G_{\mathrm{QCD}}$ is the fundamental building
block of the multiple scattering framework of EPOS4. In this paper,
we provided detailed information about the precise structure
of $G_{\mathrm{QCD}}$ and  its calculation,
based on perturbative QCD, with special care concerning heavy flavor.
We discussed the implementation (for the first time in the EPOS framework)
of the ``backward parton evolution method'', which allows a much
better control of the hard processes and comparisons with the ``standard''
pQCD calculations based on factorization. But compared to those, there
are many technical subtleties in our approach, due to the fact that
we do the parton evolutions individually for each of the scatterings
in a multiple scattering configuration, and this has been discussed
in great detail. It is actually the occurrence of singularities for
cross section calculations as well as for parton generation which
needed to be taken care of. We also discussed the way to transform the 
partonic multiple scattering structure into strings.

\subsection*{Acknowledgements}

KW thanks Sergej Ostapchenko for many helpful discussions concerning
the technical aspects of this paper. KW also thanks Tanguy Pierog for
useful contributions related to the ``deformation function''.
BG acknowledges support from ANID PIA/APOYO AFB220004.

\vspace{3cm}

\appendix
\noindent \textbf{\huge{}APPENDIX}{\huge \par}

\section{Evolution function\label{=======appendix-evolution-function=======}}

\subsection{Differential equation for evolution function \label{-------Differential-equation-------}}

It will be useful to have a the differential equation, corresponding
to Eq. (\ref{evolution-equation-1}). We may write %
\begin{align}
 & E_{\mathrm{QCD}}^{ik}\left(x,t_{a},t\right)=\frac{\Delta^{k}(t)}{\Delta^{k}(t_{a})}E_{\mathrm{QCD}}^{ik}\left(x,t_{a},t_{a}\right)\\
 & \qquad+\sum_{j}\int_{t_{a}}^{t}\frac{dt'}{t'}\int\frac{dz}{z}\,E_{\mathrm{QCD}}^{ij}\left(\frac{x}{z},t_{a},t'\right)\bar{P}_{j}^{k}\!(t',z)\,\frac{\Delta^{k}(t)}{\Delta^{k}(t')},
\end{align}
with $\Delta^{k}(t)=\Delta^{k}(t_{\mathrm{ref}},t)$, with some arbitrary
reference value $t_{\mathrm{ref}}$. We divide by $\Delta^{k}(t)$,
to get 
\begin{align}
 & \frac{E_{\mathrm{QCD}}^{ik}\left(x,t_{a},t\right)}{\Delta^{k}(t)}=\frac{E_{\mathrm{QCD}}^{ik}\left(x,t_{a},t_{a}\right)}{\Delta^{k}(t_{a})}\\
 & \qquad+\sum_{j}\int_{t_{a}}^{t}\frac{dt'}{t'}\int\frac{dz}{z}\,\frac{E_{\mathrm{QCD}}^{ij}\left(\frac{x}{z},t_{a},t'\right)}{\Delta^{k}(t')}\bar{P}_{j}^{k}\!(t',z)\,,
\end{align}
then we compute the derivative $\frac{\partial}{\partial t}$ and
multiply by $t$, which gives%
\\
\smallskip{}
\noindent\fbox{\begin{minipage}[t]{1\columnwidth - 2\fboxsep - 2\fboxrule}%
\begin{equation}
t\frac{\partial}{\partial t}\left\{ \frac{E_{\mathrm{QCD}}^{ik}\left(x,t_{a},t\right)}{\Delta^{k}(t)}\right\} =\sum_{j}\int\frac{dz}{z}\,\frac{E_{\mathrm{QCD}}^{ij}\left(\frac{x}{z},t_{a},t\right)}{\Delta^{k}(t)}\bar{P}_{j}^{k}\!(t,z).\label{evolution-equation-3}
\end{equation}
\end{minipage}}\smallskip{}

\subsection{Evolution function theorem \label{-------evolution-function-theorem-------}}

For the proof of Eq.~(\ref{evolution-abc-1}) %
we start from Eq. (\ref{evolution-equation-3}). We first define a
matrix $\boldsymbol{E}_{\mathrm{QCD}}^{N}\left(t_{a},t\right)$ with
the components 
\begin{equation}
E_{\mathrm{QCD}}^{N,ki}\left(t_{a},t\right)=\int_{0}^{1}dx\,x^{N}\,E_{\mathrm{QCD}}^{ik}\left(x,t_{a},t\right),
\end{equation}
and we define a vector $\boldsymbol{g}_{N}^{i}(t_{a},t)$ to be the
$i^{th}$ column of the matrix $\boldsymbol{E}_{\mathrm{QCD}}^{N}\left(t_{a},t\right)$,
divided by $\Delta^{k}(t)$. We further define a matrix $\boldsymbol{\boldsymbol{\gamma}}_{N}(t)$
with elements
\begin{equation}
\boldsymbol{\gamma}_{N}^{kj}(t)=\int_{0}^{1}dx\,x^{N}\,\bar{P}_{j}^{k}\!(t,z).
\end{equation}
The Mellin transform of Eq. (\ref{evolution-equation-3}) is then
\begin{equation}
t\frac{\partial}{\partial t}\boldsymbol{g}_{N}^{i}\left(t_{a},t\right)=\boldsymbol{\boldsymbol{\gamma}}_{N}(t)\,\boldsymbol{g}_{N}^{i}\left(t_{a},t\right),\label{evolution-equation-4}
\end{equation}
which has the solution 
\begin{equation}
\boldsymbol{g}_{N}^{i}\left(t_{a},t\right)=\exp\left(\int_{t_{a}}^{t}\frac{dt'}{t'}\boldsymbol{\boldsymbol{\gamma}}_{N}\!(t)\right)\boldsymbol{g}_{N}^{i}\left(t_{a},t_{a}\right)\label{evolution-solution}
\end{equation}
(the exponential of a matrix $M$ is defined by its power series $\sum_{i=0}^{\infty}\frac{1}{i!}M^{i}$).
Using $E_{\mathrm{QCD}}^{ik}\left(x,t_{a},t_{a}\right)=\delta(1-x)\delta_{ik},$
we have %
$\boldsymbol{g}_{N}^{i}\left(t_{a},t_{a}\right)=\boldsymbol{e}_{i}/\Delta^{k}(t)$
(with $\boldsymbol{e}_{i}$ being a vector with the $i^{th}$ component
unity and all others zero), and so we get for the matrix $\boldsymbol{E}_{\mathrm{QCD}}^{N}$:
\begin{align}
\boldsymbol{E}_{\mathrm{QCD}}^{N}\left(t_{a},t\right) & =\left\{ \boldsymbol{g}_{N}^{1}\left(t_{a},t\right),\boldsymbol{g}_{N}^{2}\left(t_{a},t\right)...\right\} \Delta^{k}(t)\label{evolution-solution-1a}\\
 & =\exp\left(\int_{t_{a}}^{t}\frac{dt'}{t'}\boldsymbol{\boldsymbol{\gamma}}_{N}\!(t)\right).\label{evolution-solution-1}
\end{align}
Considering $t=t_{b}$, and splitting the integration interval, we
get 
\begin{equation}
\boldsymbol{E}_{\mathrm{QCD}}^{N}\left(t_{a},t_{b}\right)=\exp\left(\int_{t_{a}}^{t_{c}}\frac{dt'}{t'}\boldsymbol{\boldsymbol{\gamma}}_{N}\!(t)+\int_{t_{c}}^{t_{b}}\frac{dt'}{t'}\boldsymbol{\boldsymbol{\gamma}}_{N}\!(t)\right),
\end{equation}
which has the structure $e^{A+B}$ with two matrices $A$ and $B$.
The Baker-Campbell-Hausdorff formula states 
\begin{equation}
e^{A}e^{B}=e^{A+B+X},
\end{equation}
with $X=\frac{1}{2}[A,B]+...$, with ``...'' representing higher
commutators of $X$ and $Y$. In leading log accuracy, we can neglect
$X$, since it amounts to higher orders in $\alpha_{s}$, and so we
get%
{} 
\begin{equation}
\boldsymbol{E}_{\mathrm{QCD}}^{N}\left(t_{a},t_{b}\right)=\boldsymbol{E}_{\mathrm{QCD}}^{N}\left(t_{c},t_{b}\right)\,\boldsymbol{E}_{\mathrm{QCD}}^{N}\left(t_{a},t_{c}\right),\label{evolution-abc-matrix}
\end{equation}
or, with the corresponding matrix elements, 
\begin{equation}
E_{\mathrm{QCD}}^{N,ji}\left(t_{a},t_{b}\right)=\sum_{k}E_{\mathrm{QCD}}^{N,jk}\left(t_{c},t_{b}\right)\,E_{\mathrm{QCD}}^{N,ki}\left(t_{a},t_{c}\right).\label{evolution-abc-matrix-elements}
\end{equation}
The inverse Mellin transformation gives\\
\smallskip{}
\noindent\fbox{\begin{minipage}[t]{1\columnwidth - 2\fboxsep - 2\fboxrule}%
\begin{equation}
E_{\mathrm{QCD}}^{ij}\left(x,t_{a},t_{b}\right)=\sum_{k}\int\frac{dz}{z}E_{\mathrm{QCD}}^{ik}\left(\frac{x}{z},t_{a},t_{c}\right)E_{\mathrm{QCD}}^{kj}\left(z,t_{c},t_{b}\right)\,.\label{evolution-abc}
\end{equation}
\end{minipage}}\smallskip{}
QED.

\subsection{Sudakov factor \label{-------Sudakov-factor-------}}

In this section, we want to prove Eqs. (\ref{Sudakov-q},\ref{Sudakov-g}).
The usual form of the DGLAP equations is 
\begin{align}
 & \mu^{2}\frac{\partial q\left(x,\mu^{2}\right)}{\partial\mu^{2}}=\frac{\alpha_{s}\left(\mu^{2}\right)}{2\pi}\\
 & \quad\int_{x}^{1}\frac{d\xi}{\xi}\left[P_{qq}(\xi)q\left(x/\xi,\mu^{2}\right)+P_{qg}(\xi)g\left(x/\xi,\mu^{2}\right)\right],\nonumber 
\end{align}
\begin{align}
 & \mu^{2}\frac{\partial g\left(x,\mu^{2}\right)}{\partial\mu^{2}}=\frac{\alpha_{s}\left(\mu^{2}\right)}{2\pi}\\
 & \quad\int_{x}^{1}\frac{d\xi}{\xi}\left[\sum_{q,\bar{q}}P_{gq}(\xi)q\left(x/\xi,\mu^{2}\right)+P_{gg}(\xi)g\left(x/\xi,\mu^{2}\right)\right],\nonumber 
\end{align}

\noindent with LO splitting functions 
\begin{align}
 & P_{qq}(z)=C_{F}\left[\frac{1+z^{2}}{(1-z)_{+}}+\frac{3}{2}\delta(1-z)\right],\\
 & P_{qg}(z)=\frac{1}{2}\left[z^{2}+(1-z)^{2}\right],\\
 & P_{gq}(z)=C_{F}\frac{1+(1-z)^{2}}{z},\\
 & P_{gg}(z)=2C_{A}\left[\frac{z}{(1-z)_{+}}+\frac{1-z}{z}+z(1-z)\right]\nonumber \\
 & \qquad+\frac{11C_{A}-2N_{f}}{6}\delta(1-z),
\end{align}
with the usual ``+\textquotedbl{} prescription 
\begin{align}
 & f(x)_{+}=f(x),\quad0\leq x<1,\\
 & \int_{0}^{1}dxg(x)f(x)_{+}=\int_{0}^{1}dxf(x)(g(x)-g(1)).\label{def2+}
\end{align}
The ``+\textquotedbl{} and the $\delta$ functions comes from gluon
emissions. Without these terms, we obtain the real, or unregularized
splitting functions 
\begin{align}
 & \tilde{P}_{qq}(z)=C_{F}\frac{1+z^{2}}{1-z},\\
 & \tilde{P}_{gg}(z)=2C_{A}\left[\frac{z}{1-z}+\frac{1-z}{z}+z(1-z)\right],\\
 & \tilde{P}_{qg}(z)=P_{qg}(z);\quad\tilde{P}_{gq}(z)=P_{gq}(z).
\end{align}
We may use an alternative form of the diagonal splitting functions,
namely 
\begin{equation}
P_{qq}(z)=C_{F}\left[\frac{1+z^{2}}{(1-z)_{+}}+\frac{3}{2}\delta(1-z)\right]=C_{F}\left(\frac{1+z^{2}}{1-z}\right)_{+},
\end{equation}
\vspace{-0.5cm}
\begin{align}
P_{gg}(z) & =2C_{A}\left[\frac{z}{(1-z)_{+}}+\frac{1-z}{z}+z(1-z)\right]\qquad\qquad\nonumber \\
 & \qquad\qquad+\frac{11C_{A}-2N_{f}}{6}\delta(1-z),
\end{align}
\vspace{-0.5cm}
\begin{align}
 & =\left(2C_{A}\frac{z}{1-z}+C_{A}z(1-z)+\frac{N_{F}}{2}\left(z^{2}+(1-z)^{2}\right)\right)_{+}\nonumber \\
 & \quad+2C_{A}\frac{1-z}{z}+C_{A}z(1-z)-\frac{N_{F}}{2}\left(z^{2}+(1-z)^{2}\right).
\end{align}
Then we write 
\[
\int_{x}^{1}\frac{d\xi}{\xi}P_{qq}\left(\frac{x}{\xi}\right)q\left(\xi,\mu^{2}\right)=\int_{x}^{1}\frac{d\xi}{\xi}P_{qq}(\xi)q\left(x/\xi,\mu^{2}\right)
\]
\begin{equation}
=\int_{0}^{1}\frac{d\xi}{\xi}P_{qq}(\xi)q\left(x/\xi,\mu^{2}\right)-\int_{0}^{x}\frac{d\xi}{\xi}\tilde{P}_{qq}(\xi)q\left(x/\xi,\mu^{2}\right)
\end{equation}
\vspace{-0.5cm}
\begin{align}
 & =\int_{0}^{1}d\xi\tilde{P}_{qq}(\xi)\left[\frac{1}{\xi}q\left(x/\xi,\mu^{2}\right)-q\left(x,\mu^{2}\right)\right]\nonumber \\
 & \qquad-\int_{0}^{x}\frac{d\xi}{\xi}\tilde{P}_{qq}(\xi)q\left(x/\xi,\mu^{2}\right)
\end{align}
\vspace{-0.5cm}
\begin{align}
 & =\int_{1-\varepsilon}^{1}d\xi\tilde{P}_{qq}(\xi)\left[\frac{1}{\xi}q\left(x/\xi,\mu^{2}\right)-q\left(x,\mu^{2}\right)\right]\nonumber \\
 & +\int_{x}^{1-\varepsilon}\frac{d\xi}{\xi}\tilde{P}_{qq}(\xi)q\left(x/\xi,\mu^{2}\right)\nonumber \\
 & -q\left(x,\mu^{2}\right)\int_{0}^{1-\varepsilon}d\xi\tilde{P}_{qq}(\xi),
\end{align}
using that fact that in the interval $[0,x]$, we have $f(x)_{+}=f(x)$,
and the relation (\ref{def2+}). Finally, remarking that for $1-\varepsilon<\xi<1$
and $\varepsilon\to0$, we get 
\begin{equation}
\frac{1}{\xi}q\left(x/\xi,\mu^{2}\right)-q\left(x,\mu^{2}\right)\to0,
\end{equation}
we obtain 
\begin{align}
 & \int_{x}^{1}\frac{d\xi}{\xi}P_{qq}\left(\frac{x}{\xi}\right)q\left(\xi,\mu^{2}\right)\\
 & =\int_{x}^{1-\varepsilon}\frac{d\xi}{\xi}\tilde{P}_{qq}(\xi)q\left(x/\xi,\mu^{2}\right)-q\left(x,\mu^{2}\right)\int_{0}^{1-\varepsilon}d\xi\tilde{P}_{qq}(\xi).
\end{align}
The corresponding calculation for the gluon case is
\begin{align*}
 & \int_{x}^{1}\frac{d\xi}{\xi}P_{gg}\left(\frac{x}{\xi}\right)g\left(\xi,\mu^{2}\right)\\
 & =\int_{0}^{1}\frac{d\xi}{\xi}\left(\frac{2C_{A}\xi}{1-\xi}+C_{A}\xi(1-\xi)+\frac{N_{F}}{2}\left(\xi^{2}+(1-\xi)^{2}\right)\right)_{+}\\
 & \qquad\qquad g\left(x/\xi,\mu^{2}\right)
\end{align*}
\vspace{-0.5cm}
\begin{align}
 & -\int_{0}^{x}\frac{d\xi}{\xi}\left(\frac{2C_{A}\xi}{1-\xi}+C_{A}\xi(1-\xi)+\frac{N_{F}}{2}\left(\xi^{2}+(1-\xi)^{2}\right)\right)\nonumber \\
 & \qquad\qquad g\left(x/\xi,\mu^{2}\right)\nonumber \\
 & +\int_{x}^{1}\frac{d\xi}{\xi}\left[2C_{A}\frac{1-\xi}{\xi}+C_{A}\xi(1-\xi)-\frac{N_{F}}{2}\left(\xi^{2}+(1-\xi)^{2}\right)\right]\nonumber \\
 & \qquad\qquad g\left(x/\xi,\mu^{2}\right)\label{gg1}
\end{align}
\vspace{-0.5cm}
\begin{align}
 & \simeq\int_{x}^{1-\varepsilon}\frac{d\xi}{\xi}2C_{A}\left[\frac{\xi}{1-\xi}+\frac{1-\xi}{\xi}+\xi(1-\xi)\right]g\left(x/\xi,\mu^{2}\right)\nonumber \\
 & -g\left(x,\mu^{2}\right)\label{gg2}\\
 & \int_{\varepsilon}^{1-\varepsilon}d\xi\left(\frac{2C_{A}\xi}{1-\xi}+C_{A}\xi(1-\xi)+\frac{N_{F}}{2}\left(\xi^{2}+(1-\xi)^{2}\right)\right)\nonumber 
\end{align}
\begin{align}
 & =\int_{x}^{1-\varepsilon}\frac{d\xi}{\xi}\tilde{P}_{gg}(\xi)g\left(x/\xi,\mu^{2}\right)\\
 & -g\left(x,\mu^{2}\right)\int_{\varepsilon}^{1-\varepsilon}d\xi\left[\frac{1}{2}\tilde{P}_{gg}(\xi)+N_{F}\tilde{P}_{qg}(\xi)\right].\nonumber 
\end{align}
Going from Eq.~(\ref{gg1}) to (\ref{gg2}), the integrals 
\begin{align}
 & g\left(x,\mu^{2}\right)\int_{0}^{\varepsilon}d\xi\Bigg(\frac{2C_{A}\xi}{1-\xi}\\
 & +C_{A}\xi(1-\xi)+\frac{N_{F}}{2}\left(\xi^{2}+(1-\xi)^{2}\right)\Bigg)\nonumber 
\end{align}
and 
\begin{align}
 & \int_{1-\varepsilon}^{1}\frac{d\xi}{\xi}\Bigg[2C_{A}\frac{1-\xi}{\xi}\\
 & +C_{A}\xi(1-\xi)-\frac{N_{F}}{2}\left(\xi^{2}+(1-\xi)^{2}\right)\Bigg]g\left(x/\xi,\mu^{2}\right)\nonumber 
\end{align}
have been neglected. Then, we can write the DGLAP equation as
\begin{align}
 & \mu^{2}\frac{\partial q\left(x,\mu^{2}\right)}{\partial\mu^{2}}=\frac{\alpha_{s}\left(\mu^{2}\right)}{2\pi}\int_{x}^{1-\varepsilon}\frac{d\xi}{\xi}\nonumber \\
 & \qquad\Bigg[\tilde{P}_{qq}(\xi)q\left(x/\xi,\mu^{2}\right)+\tilde{P}_{qg}(\xi)g\left(x/\xi,\mu^{2}\right)\Bigg]\nonumber \\
 & \qquad-q\left(x,\mu^{2}\right)\left\{ \frac{\alpha_{s}\left(\mu^{2}\right)}{2\pi}\int_{0}^{1-\varepsilon}d\xi\tilde{P}_{qq}(\xi)\right\} \,,\label{DGq}
\end{align}
\begin{align}
 & \mu^{2}\frac{\partial g\left(x,\mu^{2}\right)}{\partial\mu^{2}}=\frac{\alpha_{s}\left(\mu^{2}\right)}{2\pi}\int_{x}^{1-\varepsilon}\frac{d\xi}{\xi}\nonumber \\
 & \qquad\left[\sum_{q,\bar{q}}\tilde{P}_{gq}(\xi)q\left(x/\xi,\mu^{2}\right)+\tilde{P}_{gg}(\xi)g\left(x/\xi,\mu^{2}\right)\right]\nonumber \\
 & \;-g\left(x,\mu^{2}\right)\left\{ \frac{\alpha_{s}\left(\mu^{2}\right)}{2\pi}\int_{\varepsilon}^{1-\varepsilon}d\xi\left[\frac{1}{2}\tilde{P}_{gg}(\xi)+N_{F}\tilde{P}_{qg}(\xi)\right]\right\} \,.\label{DGg}
\end{align}
Defining \\
\smallskip{}
\noindent\fbox{\begin{minipage}[t]{1\columnwidth - 2\fboxsep - 2\fboxrule}%
\noindent 
\begin{align}
A_{q} & =\frac{\alpha_{s}\left(\mu^{2}\right)}{2\pi}\int_{0}^{1-\varepsilon}d\xi\tilde{P}_{qq}(\xi),\label{Sudakov-2}\\
A_{g} & =\frac{\alpha_{s}\left(\mu^{2}\right)}{2\pi}\int_{\varepsilon}^{1-\varepsilon}d\xi\left[\frac{1}{2}\tilde{P}_{gg}(\xi)+N_{F}\tilde{P}_{qg}(\xi)\right],\label{Sudakov-3}
\end{align}
\end{minipage}}\smallskip{}

\noindent (representing the expressions in curly brackets in Eqs.
(\ref{DGq},\ref{DGg}), we get from Eqs. (\ref{DGq},\ref{DGg})
\begin{align}
 & \mu^{2}\frac{\partial q\left(x,\mu^{2}\right)}{\partial\mu^{2}}+q\left(x,\mu^{2}\right)A_{q}=\frac{\alpha_{s}\left(\mu^{2}\right)}{2\pi}\int_{x}^{1-\varepsilon}\frac{d\xi}{\xi}\nonumber \\
 & \qquad\Bigg[\tilde{P}_{qq}(\xi)q\left(x/\xi,\mu^{2}\right)+\tilde{P}_{qg}(\xi)g\left(x/\xi,\mu^{2}\right)\Bigg]\:,\label{DGq-1}
\end{align}
\begin{align}
 & \mu^{2}\frac{\partial g\left(x,\mu^{2}\right)}{\partial\mu^{2}}+g\left(x,\mu^{2}\right)A_{g}=\frac{\alpha_{s}\left(\mu^{2}\right)}{2\pi}\int_{x}^{1-\varepsilon}\frac{d\xi}{\xi}\nonumber \\
 & \qquad\left[\sum_{q,\bar{q}}\tilde{P}_{gq}(\xi)q\left(x/\xi,\mu^{2}\right)+\tilde{P}_{gg}(\xi)g\left(x/\xi,\mu^{2}\right)\right]\:.\label{DGg-1}
\end{align}

\noindent Defining Sudakov factors $\Delta_{g}$ and $\Delta_{q}$
as\\
\smallskip{}
\noindent\fbox{\begin{minipage}[t]{1\columnwidth - 2\fboxsep - 2\fboxrule}%
\noindent 
\begin{equation}
\Delta_{a}=\exp\left(-\int_{\mu_{0}^{2}}^{\mu^{2}}\frac{dq^{2}}{q^{2}}A_{a}\right),\label{Sudakov-1}
\end{equation}
\end{minipage}}\smallskip{}
 we find for $f_{q}=q$ and $f_{g}=g$:
\begin{equation}
\mu^{2}\frac{\partial}{\partial\mu^{2}}f_{a}(x,\mu^{2})+f_{a}A_{a}=\Delta_{a}\,\mu^{2}\frac{\partial}{\partial\mu^{2}}\left(\frac{f_{a}(x,\mu^{2})}{\Delta_{a}(\mu^{2})}\right).
\end{equation}
Reporting this into Eqs.~(\ref{DGq-1}) and (\ref{DGg-1}), we get\\
\smallskip{}
\noindent\fbox{\begin{minipage}[t]{1\columnwidth - 2\fboxsep - 2\fboxrule}%
\noindent 
\begin{align}
 & \Delta_{q}(\mu^{2})\mu^{2}\frac{\partial}{\partial\mu^{2}}\left\{ \frac{q(x,\mu^{2})}{\Delta_{q}(\mu^{2})}\right\} =\frac{\alpha_{s}\left(\mu^{2}\right)}{2\pi}\int_{x}^{1-\varepsilon}\frac{d\xi}{\xi}\\
 & \qquad\qquad\Bigg[\tilde{P}_{qq}(\xi)q\left(x/\xi,\mu^{2}\right)+\tilde{P}_{qg}(\xi)g\left(x/\xi,\mu^{2}\right)\Bigg]\nonumber \\
 & \Delta_{g}(\mu^{2})\mu^{2}\frac{\partial}{\partial\mu^{2}}\left\{ \frac{g\left(x,\mu^{2}\right)}{\Delta_{g}(\mu^{2})}\right\} =\frac{\alpha_{s}\left(\mu^{2}\right)}{2\pi}\int_{x}^{1-\varepsilon}\frac{d\xi}{\xi}\\
 & \qquad\qquad\left[\sum_{q,\bar{q}}\tilde{P}_{gq}(\xi)q\left(x/\xi,\mu^{2}\right)+\tilde{P}_{gg}(\xi)g\left(x/\xi,\mu^{2}\right)\right],\nonumber 
\end{align}
\end{minipage}}\smallskip{}
which is the DGLAP equation based on unregularized splitting functions,
as we use for our $E$ functions, see Eq. (\ref{evolution-equation-3}).
We use the convention $P_{i}^{j}=\tilde{P}_{ji}$. Eqs. (\ref{Sudakov-1},\ref{Sudakov-2},\ref{Sudakov-3})
agree with Eq. (\ref{Sudakov-q},\ref{Sudakov-g}).

\section{Parton-parton cross section formulas \label{=======appendix:xsection-formulas=======}}

Here we will show how to explicitly do some of the integrations,
in order to compute cross sections, starting from Eqs. (\ref{integrated-cross-section},\ref{differential-cross-section}).

\subsection{Rapidity integral}

Here we simplify the expression to get integrated cross sections as
rapidity integral. From Eqs. (\ref{integrated-cross-section},\ref{differential-cross-section}),
we get
\begin{align}
 & \sigma_{\mathrm{hard}}^{ij}=\sum_{klmn}\int dy_{3}dy_{4}dp_{3t}^{2}d^{2}p_{4t}\int dx_{1}dx_{2}\label{eq:jetcross-1-1-1}\\
 & \qquad E_{\mathrm{QCD}}^{ik}(x_{1},Q_{1}^{2},\mu_{\mathrm{F}}^{2})E_{\mathrm{QCD}}^{jl}(x_{2},Q_{2}^{2},\mu_{\mathrm{F}}^{2})\nonumber \\
 & \qquad\frac{1}{32s\pi^{2}}\bar{\sum}|\mathcal{M}^{kl\to mn}|^{2}\delta^{4}(p_{1}+p_{2}-p_{3}-p_{4})\frac{1}{1+\delta_{mn}}.\nonumber 
\end{align}
Integrating explicitly over $\vec{p}_{4t}$ , which removes 
\begin{equation}
\delta^{2}(\vec{p}_{1t}+\vec{p}_{2t}-\vec{p}_{3t}-\vec{p}_{4t}),
\end{equation}
and using $d^{2}p_{3t}=d^{2}p_{t}$$=\pi dp_{t}^{2}$, we get 
\begin{align}
 & \sigma_{\mathrm{hard}}^{ij}=\sum_{klmn}\int dy_{3}dy_{4}dp_{t}^{2}dx_{1}dx_{2}\nonumber \\
 & \qquad E_{\mathrm{QCD}}^{ik}(x_{1},Q_{1}^{2},\mu_{\mathrm{F}}^{2})E_{\mathrm{QCD}}^{jl}(x_{2},Q_{2}^{2},\mu_{\mathrm{F}}^{2})\:\frac{1}{32s\pi}\\
 & \qquad\bar{\sum}|\mathcal{M}^{kl\to mn}|^{2}\,\frac{1}{J}\,\delta(P_{\mathrm{in}}^{+}-P_{\mathrm{out}}^{+})\delta(P_{\mathrm{in}}^{-}-P_{\mathrm{out}}^{-})\frac{1}{1+\delta_{mn}},\nonumber \\
\end{align}
with $J=1/2.$ Here we used the identity 
\begin{equation}
\delta^{2}(\vec{x}-\vec{x}_{0})=\frac{1}{J}\delta^{2}(\vec{y}-\vec{y}_{0}),\;J=\left|\frac{\partial(x_{1},x_{2})}{\partial(y_{1},y_{2})}\right|,
\end{equation}
valid for any $\vec{x}(\vec{y})$, and in this case used for $x_{1}=E$,
$x_{2}=P_{\Vert}$, $y_{1}=P^{+},$ $y_{2}=P^{-},$ which gives $J=1/2$.
The LC momenta are given as $p_{1}^{+}=x_{1}\sqrt{s_{\mathrm{lad}}}$,
$p_{1}^{-}=0$, $p_{2}^{+}=0$, $p_{2}^{-}=x_{2}\sqrt{s_{\mathrm{lad}}}$,
and for the out-going partons as $p_{3}^{+}=p_{t}e^{y_{3}}$, $p_{3}^{-}=p_{t}e^{-y_{3}}$
and $p_{4}^{+}=p_{t}e^{y_{4}}$, $p_{4}^{-}=p_{t}e^{-y_{4}}$, so
we have 
\begin{equation}
\delta(P_{\mathrm{in}}^{+}-P_{\mathrm{out}}^{+})=\delta(x_{1}\sqrt{s_{lad}}-p_{t}e^{y_{3}}-p_{t}e^{y_{4}}),
\end{equation}
\begin{equation}
\delta(P_{\mathrm{in}}^{-}-P_{\mathrm{out}}^{-})=\delta(x_{2}\sqrt{s_{\mathrm{lad}}}-p_{t}e^{-y_{3}}-p_{t}e^{-y_{4}}),
\end{equation}
which gives 
\begin{equation}
\delta(P_{\mathrm{in}}^{+}-P_{\mathrm{out}}^{+})=\frac{1}{\sqrt{s_{\mathrm{lad}}}}\delta\left(x_{1}-\frac{1}{2}x_{t}\left(e^{y_{3}}+e^{y_{4}}\right)\right),
\end{equation}
\begin{equation}
\delta(P_{\mathrm{in}}^{-}-P_{\mathrm{out}}^{-})=\frac{1}{\sqrt{s_{\mathrm{lad}}}}\delta\left(x_{2}-\frac{1}{2}x_{t}\left(e^{-y_{3}}+e^{-y_{4}}\right)\right),
\end{equation}
with $x_{t}=2p_{t}/\sqrt{s_{\mathrm{lad}}}$, and so we get \\
\noindent\fbox{\begin{minipage}[t]{1\columnwidth - 2\fboxsep - 2\fboxrule}%
\vspace*{-0.3cm}
\begin{align}
 & \sigma_{\mathrm{hard}}^{ij}=\sum_{klmn}\int dy_{3}dy_{4}dp_{t}^{2}\nonumber \\
 & \qquad E_{\mathrm{QCD}}^{ik}(x_{1},Q_{1}^{2},\mu_{\mathrm{F}}^{2})E_{\mathrm{QCD}}^{jl}(x_{2},Q_{2}^{2},\mu_{\mathrm{F}}^{2})\nonumber \\
 & \qquad\frac{1}{16\pi s\,s_{\mathrm{lad}}}\bar{\sum}|\mathcal{M}^{kl\to mn}|^{2}\frac{1}{1+\delta_{mn}}\,,\label{xsection-rapidity-integral}
\end{align}
\vspace*{-0.3cm}%
\end{minipage}}\medskip{}
\\
with (from integrating over the delta functions)
\begin{equation}
x_{1}=\frac{p_{t}}{\sqrt{s_{\mathrm{lad}}}}\left(e^{y_{3}}+e^{y_{4}}\right),\:x_{2}=\frac{p_{t}}{\sqrt{s_{\mathrm{lad}}}}\left(e^{-y_{3}}+e^{-y_{4}}\right)\label{eq:x1x2y3y4-2}
\end{equation}
and $\delta_{mn}$ accounting for properly counting identical partons.%

\subsection{LC momentum fraction integral \label{subsec:LC-momentum-fraction}}

Here we want to show that $\sigma_{\mathrm{hard}}^{ij}$ in Eq. (\ref{xsection-rapidity-integral})
can be written as an integral over light-cone momentum fractions. 

It will be useful to consider quantities in the CMS of the two partons
entering the Born process (BornCMS). We may write (with $E=\sqrt{s}/2$)
\begin{align}
p_{1}=&(E,\:E\:,\,\,0\:,0),\quad p_{2}=(E,\:-E\:,\:0\,\,,0),\\
p_{3}=&(E,E\cos\theta,E\sin\theta,0), \\
p_{4}=&(E,-E\cos\theta,-E\sin\theta,0),
\end{align}
which allows computing 
\begin{equation}
t=-\frac{s}{2}(1-\cos\theta),\:u=-\frac{s}{2}(1+\cos\theta).\label{eq:tu}
\end{equation}
Using 
\begin{equation}
\cos\theta=\frac{(p_{3})_{z}}{(p_{3})_{0}}=\tanh\left(y_{3\,\mathrm{BornCMS}}\right)=\tanh\left(\frac{y_{3}-y_{4}}{2}\right),
\end{equation}
and Eq. (\ref{eq:x1x2y3y4-2}) with $a=1/\sqrt{s_{\mathrm{lad}}}$,
\begin{equation}
x_{1}=ap_{t}\left(e^{y_{3}}+e^{y_{4}}\right),\:x_{2}=ap_{t}\left(e^{-y_{3}}+e^{-y_{4}}\right),\label{eq:x1x2y3y4-1}
\end{equation}
we get the Jacobian matrix
\begin{align}
 & \frac{\partial(x_{1},x_{2},\cos\theta)}{\partial(y_{3},y_{4},p_{t}^{2})}=\\
 & \left(\begin{array}{ccc}
ap_{t}e^{y_{3}} & ap_{t}e^{y_{4}} & \frac{a}{2p_{t}}\left(e^{y_{3}}+e^{y_{4}}\right)\\
-ap_{t}e^{-y_{3}} & -ap_{t}e^{-y_{4}} & \frac{a}{2p_{t}}\left(e^{-y_{3}}+e^{-y_{4}}\right)\\
\frac{1/2}{\cosh^{2}(\frac{y_{3}-y_{4}}{2})} & \frac{-1/2}{\cosh^{2}(\frac{y_{3}-y_{4}}{2})} & 0
\end{array}\right),
\end{align}
and the Jacobian determinant
\begin{align}
 J=&\frac{a^{2}/4}{\cosh^{2}(\frac{y_{3}-y_{4}}{2})}\Bigg(\left|\begin{array}{cc}
e^{y_{4}} & \left(e^{y_{3}}+e^{y_{4}}\right)\\
-e^{-y_{4}} & \left(e^{-y_{3}}+e^{-y_{4}}\right)
\end{array}\right|\nonumber \\
 & +\left|\begin{array}{cc}
e^{y_{3}} & \left(e^{y_{3}}+e^{y_{4}}\right)\\
-e^{-y_{3}} & \left(e^{-y_{3}}+e^{-y_{4}}\right)
\end{array}\right|\Bigg),\\
  =&\frac{a^{2}/4}{\cosh^{2}(\frac{y_{3}-y_{4}}{2})}\Bigg(e^{y_{4}}\left(e^{-y_{3}}+e^{-y_{4}}\right)+e^{-y_{4}}\left(e^{y_{3}}+e^{y_{4}}\right)\nonumber \\
 & +e^{y_{3}}\left(e^{-y_{3}}+e^{-y_{4}}\right)+e^{-y_{3}}\left(e^{y_{3}}+e^{y_{4}}\right)\Bigg),\\
=&\frac{a^{2}/2}{\cosh^{2}(\frac{y_{3}-y_{4}}{2})}\left(e^{y_{3}-y_{4}}+e^{-y_{3}+y_{4}}+1+1\right),\\
=&\frac{a^{2}/2}{\cosh^{2}(\frac{y_{3}-y_{4}}{2})}\left(e^{\frac{y_{3}-y_{4}}{2}}+e^{-\frac{y_{3}-y_{4}}{2}}\right)^{2},\\
=&2a^{2}=2/s_{\mathrm{lad}}.
\end{align}
So we get
\begin{equation}
dy_{3}dy_{4}dp_{t}^{2}=\frac{1}{2}s_{\mathrm{lad}}\,dx_{1}dx_{2}d\cos\theta,
\end{equation}
and with [from Eq. (\ref{eq:tu})] 
\[
dt=\frac{1}{2}s\,d\cos\theta,
\]
we find

\begin{equation}
dy_{3}dy_{4}dp_{t}^{2}=\frac{1}{x_{1}x_{2}}dx_{1}dx_{2}dt,
\end{equation}
and so, from Eq. (\ref{xsection-rapidity-integral}) 

\[
\sigma_{\mathrm{hard}}^{ij}=\sum_{klmn}\int dx_{1}dx_{2}dt\,E_{\mathrm{QCD}}^{ik}(x_{1},Q_{1}^{2},\mu_{\mathrm{F}}^{2})E_{\mathrm{QCD}}^{jl}(x_{2},Q_{2}^{2},\mu_{\mathrm{F}}^{2})
\]
\begin{equation}
\times\frac{1}{16\pi s^{2}}\bar{\sum}|\mathcal{M}^{kl\to mn}|^{2}\frac{1}{1+\delta_{mn}},
\end{equation}
which we may write as\\
\noindent\fbox{\begin{minipage}[t]{1\columnwidth - 2\fboxsep - 2\fboxrule}%
\vspace*{-0.3cm}
\[
\sigma_{\mathrm{hard}}^{ij}\!=\!\!\sum_{klmn}\int\!\!dx_{1}\,dx_{2}\,dt\,E_{\mathrm{QCD}}^{ik}(x_{1},Q_{1}^{2},\mu_{\mathrm{F}}^{2})E_{\mathrm{QCD}}^{jl}(x_{2},Q_{2}^{2},\mu_{\mathrm{F}}^{2})
\]
\begin{equation}
\times\frac{\pi\alpha_{s}^{2}}{s^{2}}\left\{ \frac{1}{g^{4}}\bar{\sum}|\mathcal{M}^{kl\to mn}|^{2}\right\} \frac{1}{1+\delta_{mn}},\label{xsection-LC-momentum-integral}
\end{equation}
\vspace*{-0.3cm}%
\end{minipage}}\medskip{}
\\
with $\left\{ ...\right\} $being of the form in which the squared matrix elements
are usually tabulated, with $\alpha_{s}=g^{2}/4\pi$. The corresponding
differential cross section is 
\[
\frac{d\sigma_{\mathrm{hard}}^{ij}}{dx_{1}\,dx_{2}\,dt}\!=\!\!\sum_{klmn}E_{\mathrm{QCD}}^{ik}(x_{1},Q_{1}^{2},\mu_{\mathrm{F}}^{2})E_{\mathrm{QCD}}^{jl}(x_{2},Q_{2}^{2},\mu_{\mathrm{F}}^{2})
\]
\begin{equation}
\times\frac{\pi\alpha_{s}^{2}}{s^{2}}\left\{ \frac{1}{g^{4}}\bar{\sum}|\mathcal{M}^{kl\to mn}|^{2}\right\} \frac{1}{1+\delta_{mn}}.\label{eq:LC-momentum-integral-1}
\end{equation}

\subsection{Single jet parton-parton cross section\label{Appendix: differential parton-parton}}

Here we show that $d^{3}\sigma_{\mathrm{hard}}^{ij}/dy\,d^{2}p_{t}$
can be written as a single integral $\int dx...$, where $y$ and
$p_{t}$ refer to one outgoing parton (jet).

Again starting from Eq. (\ref{differential-cross-section}), one may
integrate over $\vec{p}_{4}$, to get
\begin{align}
 & \frac{d^{3}\sigma_{\mathrm{hard}}^{ij}}{dyd^{2}p_{t}}=\sum_{klmn}\int\!\int\!\!dx_{1}dx_{2}\label{eq:differential-2-1}\\
 & \qquad\,E_{\mathrm{QCD}}^{ik}(x_{1},Q_{1}^{2},\mu_{\mathrm{F}}^{2})E_{\mathrm{QCD}}^{jl}(x_{2},Q_{2}^{2},\mu_{\mathrm{F}}^{2})\nonumber \\
 & \qquad\frac{1}{32s\pi^{2}}\bar{\sum}|\mathcal{M}^{kl\to mn}|^{2}\frac{1}{1+\delta_{mn}}\frac{1}{E_{4}}\delta(E_{1}+E_{2}-E_{3}-E_{4}),\nonumber 
\end{align}
where $y$ and $p_{t}$ refer to particle 3. In the BornCMS, we have
(with $E_{3}=E_{4}=E$ and $E_{1}=E_{2}=\sqrt{s}/2$, not yet imposing
energy conservation) 
\begin{equation}
t=-E\sqrt{s}(1-\cos\theta),\:u=-E\sqrt{s}(1+\cos\theta),\label{eq:tu-1}
\end{equation}
and so
\begin{align}
 & \frac{1}{E_{4}}\delta(E_{1}+E_{2}-E_{3}-E_{4})\nonumber \\
 & \:=\delta\Big(E(\sqrt{s}-2E)\Big)=\delta\Big(\frac{\sqrt{s}}{2}(\sqrt{s}-2E)\Big)\nonumber \\
 & \:=2\delta\Big(s-2\sqrt{s}E)\Big)=2\delta(s+t+u),
\end{align}
which leads to
\begin{align}
 & \frac{d^{2}\sigma_{\mathrm{hard}}^{ij}}{dydp_{t}^{2}}\!=\!\sum_{klmn}\int\!dx_{1}dx_{2}\,E_{\mathrm{QCD}}^{ik}(x_{1},Q_{1}^{2},\mu_{\mathrm{F}}^{2})E_{\mathrm{QCD}}^{jl}(x_{2},Q_{2}^{2},\mu_{\mathrm{F}}^{2})\nonumber \\
 & \:\times\frac{1}{16\pi s^{2}}\bar{\sum}|\mathcal{M}^{kl\to mn}|^{2}\frac{1}{1+\delta_{mn}}\,s\,\delta(s+t+u).
\end{align}
Using $p_{1}^{+}=x_{1}\sqrt{s_{\mathrm{lad}}}$, $p_{1}^{-}=0$, $p_{3}^{+}=p_{t}e^{y}$,
$p_{3}^{-}=p_{t}e^{-y}$, we get 
\begin{equation}
t=(p_{1}-p_{3})^{+}(p_{1}-p_{3})^{-}-p_{t}^{2}=-p_{t}x_{1}\sqrt{s_{\mathrm{lad}}}e^{-y}
\end{equation}
and with $p_{2}^{-}=x_{2}\sqrt{s_{\mathrm{lad}}}$, $p_{2}^{+}=0$,
we find 
\begin{equation}
u=(p_{2}-p_{3})^{+}(p_{2}-p_{3})^{-}-p_{t}^{2}=-p_{t}x_{2}\sqrt{s_{\mathrm{lad}}}e^{+y}
\end{equation}
and therefore (using $x_{t}=2p_{t}/\sqrt{s_{\mathrm{lad}}}$)
\begin{align}
 & s\,\delta(s+t+u)\\
 & =x_{1}x_{2}\delta\left(x_{1}x_{2}-x_{1}\begin{array}{c}
\frac{x_{t}}{2}\end{array}e^{-y}-x_{2}\begin{array}{c}
\frac{x_{t}}{2}\end{array}e^{y}\right)\\
 & =x_{1}x_{2}\delta\big(x(x_{2}-\frac{x_{1}}{x}\begin{array}{c}
\frac{x_{t}}{2}\end{array}e^{-y})\big)\\
 & =x_{1}x_{2}\frac{1}{x}\delta(x_{2}-\frac{x_{1}}{x}\begin{array}{c}
\frac{x_{t}}{2}\end{array}e^{-y})
\end{align}
with $x=x_{1}-\begin{array}{c}
\frac{x_{t}}{2}\end{array}e^{y}$. So we have finally
\begin{align}
 & \frac{d^{2}\sigma_{\mathrm{hard}}^{ij}}{dydp_{t}^{2}}=\sum_{klmn}\int dx\,E_{\mathrm{QCD}}^{ik}(x_{1},Q_{1}^{2},\mu_{\mathrm{F}}^{2})E_{\mathrm{QCD}}^{jl}(x_{2},Q_{2}^{2},\mu_{\mathrm{F}}^{2})\nonumber \\
 & \qquad\times\frac{1}{16\pi s^{2}}\,\bar{\sum}|\mathcal{M}^{kl\to mn}|^{2}\frac{1}{1+\delta_{mn}}\,x_{1}x_{2}\frac{1}{x}\\
 & \mathrm{\qquad\quad with}\;x_{1}=x+\begin{array}{c}
\frac{x_{t}}{2}\end{array}e^{y},\,x_{2}=\begin{array}{c}
\frac{x_{1}}{x}\,\frac{x_{t}}{2}\end{array}e^{-y}.\nonumber 
\end{align}
or\\
\noindent\fbox{\begin{minipage}[t]{1\columnwidth - 2\fboxsep - 2\fboxrule}%
\vspace*{-0.3cm}
\begin{align}
 & \frac{d^{2}\sigma_{\mathrm{hard}}^{ij}}{dydp_{t}^{2}}=\sum_{klmn}\int_{b}^{a}dx\,E_{\mathrm{QCD}}^{ik}(x_{1},Q_{1}^{2},\mu_{\mathrm{F}}^{2})E_{\mathrm{QCD}}^{jl}(x_{2},Q_{2}^{2},\mu_{\mathrm{F}}^{2})\nonumber \\
 & \qquad\times\frac{\pi\alpha_{s}^{2}}{s^{2}}\left\{ \begin{array}{c}
\frac{1}{g^{4}}\end{array}\bar{\sum}|\mathcal{M}^{kl\to mn}(s,t)|^{2}\right\} \frac{1}{1\!+\!\delta_{mn}}\,x_{1}x_{2}\frac{1}{x}\,,\label{jet-differential-xsection}
\end{align}
\vspace*{-0.3cm}%
\end{minipage}}\smallskip{}
\\
with\\
\noindent\fbox{\begin{minipage}[t]{1\columnwidth - 2\fboxsep - 2\fboxrule}%
\vspace*{-0.3cm}
\begin{align}
 & x_{1}=x+\begin{array}{c}
\frac{p_{t}}{\sqrt{s_{\mathrm{lad}}}}\end{array}e^{y},\,x_{2}=\begin{array}{c}
\frac{x_{1}}{x}\,\frac{p_{t}}{\sqrt{s_{\mathrm{lad}}}}\end{array}e^{-y},\\
 & s=x_{1}x_{2}\sqrt{s_{\mathrm{lad}}},\,t=-p_{t}x_{1}\sqrt{s_{\mathrm{lad}}}e^{-y}\,,\\
 & a=\begin{array}{c}
\frac{2p_{t}^{2}}{s_{\mathrm{lad}}}\end{array}/(2-\!\!\begin{array}{c}
\frac{2p_{t}}{\sqrt{s_{\mathrm{lad}}}}\end{array}\!e^{-y}),\:b=1-\!\!\begin{array}{c}
\frac{p_{t}}{\sqrt{s_{\mathrm{lad}}}}\end{array}\!e^{y},
\end{align}
\vspace*{-0.3cm}%
\end{minipage}}\medskip{}
\\
with $\left\{ ...\right\} $being the form the squared matrix elements
are usually tabulated, with $\alpha_{s}=g^{2}/4\pi$, and where the
integration limits $a$ and $b$ are obtained from $x_{2}\le1$ and
$x_{1}\le1$. This formula is useful to compute (without Monte Carlo)
transverse momentum spectra of produced partons for tests and comparisons.

In the EPOS framework, we have evolution functions with an initial
condition $\delta(1-x)$, which forces us to distinguish four cases,
namely both-sided, one-sided(lower), one-sided(upper), or no emissions.
In the case of both-sided emissions, we use Eq. (\ref{jet-differential-xsection}),
with $\tilde{E}_{\mathrm{QCD}}^{ab}$ instead of $E_{\mathrm{QCD}}^{ab}$.
In case of no emissions on either side, we have $E_{\mathrm{QCD}}^{ab}(x_{i},Q_{i}^{2},\mu_{\mathrm{F}}^{2})=\Delta_{\mathrm{QCD}}^{a}(Q_{i}^{2},\mu_{\mathrm{F}}^{2})\delta(1-x_{i})\delta_{ab}$
on both sides, so $\int\!dx_{1}dx_{2}...$ is trivial, and using $\delta(s+t+u)=\delta(f(y))$
$=\delta\big((y-y_{0})f'(y_{0})\big)$ $+\delta\big((y+y_{0})f'(-y_{0})\big)$,
with 
\begin{align}
f(y) & =s+t+u,\\
 & =s-x_{1}p_{t}\sqrt{s_{\mathrm{lad}}}e^{-y}-x_{2}p_{t}\sqrt{s_{\mathrm{lad}}}e^{y},\\
 & =s_{\mathrm{lad}}-p_{t}\sqrt{s_{\mathrm{lad}}}e^{-y}-p_{t}\sqrt{s_{\mathrm{lad}}}e^{y},\\
 & =s_{\mathrm{lad}}(1-x_{t}\cosh(y)),
\end{align}
and $f'(y)=-t+u$, and with $y_{0}$ being the (unique) positive root
of $f(y)$, we get (after integrating over $y$) 
\begin{align}
 & \frac{d^{2}\sigma_{\mathrm{hard}}^{ij}}{dp_{t}^{2}}\!=\!\sum_{mn}\Delta_{\mathrm{QCD}}^{i}(Q_{1}^{2},\mu_{\mathrm{F}}^{2})\Delta_{\mathrm{QCD}}^{j}(Q_{2}^{2},\mu_{\mathrm{F}}^{2})\nonumber \\
 & \:\times\frac{\pi\alpha_{s}^{2}}{s^{2}}\left\{ \begin{array}{c}
\frac{1}{g^{4}}\end{array}\bar{\sum}|\mathcal{M}^{ij\to mn}|^{2}\right\} \frac{2}{1+\delta_{mn}}\,s_{\mathrm{lad}}\,/|f'(y_{0})|.
\end{align}
with $|f'(y_{0})|=|t-u|=s-2|t|\approx s_{\mathrm{lad}}$. In case
of no emissions on the upper side ($x_{1}=1$), we have t+s+t
\begin{align}
 & \frac{d^{2}\sigma_{\mathrm{hard}}^{ij}}{dydp_{t}^{2}}=\sum_{lmn}\Delta_{\mathrm{QCD}}^{i}(Q_{1}^{2},\mu_{\mathrm{F}}^{2})E_{\mathrm{QCD}}^{jl}(x_{2},Q_{2}^{2},\mu_{\mathrm{F}}^{2})\nonumber \\
 & \qquad\times\frac{\pi\alpha_{s}^{2}}{s^{2}}\left\{ \begin{array}{c}
\frac{1}{g^{4}}\end{array}\bar{\sum}|\mathcal{M}^{il\to mn}|^{2}\right\} \frac{1}{1\!+\!\delta_{mn}}\,x_{2}\frac{1}{x}\\
 & \qquad\quad\mathrm{with}\:x=1-ae^{y},\,x_{2}=\frac{1}{x}ae^{-y}.\nonumber 
\end{align}
In case of no emissions on the lower side ($x_{2}=1$), we have
\begin{align}
 & \frac{d^{2}\sigma_{\mathrm{hard}}^{ij}}{dydp_{t}^{2}}=\sum_{kmn}\int dx\,E_{\mathrm{QCD}}^{ik}(x_{1},Q_{1}^{2},\mu_{\mathrm{F}}^{2})\Delta_{\mathrm{QCD}}^{j}(Q_{2}^{2},\mu_{\mathrm{F}}^{2})\nonumber \\
 & \qquad\times\frac{\pi\alpha_{s}^{2}}{s^{2}}\left\{ \begin{array}{c}
\frac{1}{g^{4}}\end{array}\bar{\sum}|\mathcal{M}^{kj\to mn}|^{2}\right\} \frac{1}{1\!+\!\delta_{mn}}\,x_{1}\frac{1}{x}\\
 & \qquad\quad\mathrm{with}\:x=1-ae^{-y},\,x_{1}=\frac{1}{x}ae^{y}.\nonumber 
\end{align}
We finally mention that the relation 
\begin{equation}
x_{1}x_{2}-x_{1}ae^{-y}-x_{2}ae^{y}=0
\end{equation}
(from $s+t+u=0)$ not only allows to express $x_{2}$ in terms of
$x_{1}$ and $y$ or $x_{1}$ in terms of $x_{2}$ and $y$ (for given
$a=\sqrt{p_{t}^{2}/s_{\mathrm{lad}}}$), as
\begin{equation}
x_{2}=\frac{x_{1}ae^{-y}}{x_{1}-ae^{y}},\;x_{1}=\frac{x_{2}ae^{-y}}{x_{2}-ae^{y}},
\end{equation}
but also the other way round $y$ in terms of $x_{1}$ and $x_{2}$,
as
\begin{equation}
e^{-y}=\frac{2a/x_{1}}{1\pm\sqrt{1-\frac{4a^{2}}{x_{1}x_{2}}}},
\end{equation}
or
\begin{equation}
e^{y}=\frac{2a/x_{2}}{1\pm\sqrt{1-\frac{4a^{2}}{x_{1}x_{2}}}},
\end{equation}
useful formulas for later applications.

\section{Born scattering kinematics: Explicit formulas \label{=======appendix:born-kinematics=======}}

Here we will provide explicit formulas for different kinematic quantities
and relations related to the Born process, as being used in the numerical
procedures. 

We consider two partons entering the Born process (let us note them
$1$ and $2$), producing two partons (let us note them $3$ and $4$),
so altogether we have the elementary process $1+2\to3+4$. All kinematic
formulas depend on the four masses, $m_{1}$, $m_{2}$, $m_{3}$,
and $m_{4}$. In order to present explicit formulas for the different
cases, we will use the notation 
\begin{equation}
case\;m_{1}m_{2}m_{3}m_{4},
\end{equation}
so for example ``$case\;0000$'' represents the case of massless
incoming and outgoing partons, like $g+g\to g+g$ or $u+\bar{u}\to g+g$. 

\subsection{The W variable}

Be $E$ and $E'$ the modules of the momenta of particles 1 and 3,
respectively. We then define $W=4E^{2}$ and $W'=4E'^{2}$. We have
shown [see Eqs. (\ref{eq:variableW1},\ref{eq:variableW2})]
\begin{eqnarray}
 & W=s-2\left(m_{1}^{2}+m_{2}^{2}\right)+\frac{1}{s}\left(m_{1}^{2}-m_{2}^{2}\right)^{2},\\
 & W'=s-2\left(m_{3}^{2}+m_{4}^{2}\right)+\frac{1}{s}\left(m_{3}^{2}-m_{4}^{2}\right)^{2}.
\end{eqnarray}
We get for the ``interesting cases'' explicitly
\begin{eqnarray}
 &  & case\;0000\,:W=W'=s\:,\nonumber \\
 &  & case\;m0m0\,:W=W'=\left(s-m^{2}\right)^{2}/s,\nonumber \\
 &  & case\;00mm\,:W=s,\,W'=s-4m^{2},\nonumber \\
 &  & case\;mm00\,:W=s-4m^{2},\,W'=s\,,\label{cases-1}\\
 &  & case\;mm\tilde{m}\tilde{m}\,:W=s-4m^{2},\,W'=s-4\tilde{m}^{2}\nonumber \\
 &  & case\;m\tilde{m}m\tilde{m}\,:W=W'=s-2(m^{2}+\tilde{m}^{2})\nonumber \\
 &  & \qquad\qquad\qquad\qquad+(m^{2}-\tilde{m}^{2})^{2}/s\,.\nonumber 
\end{eqnarray}
Here $m$ refers to the charm and $\tilde{m}$ to the bottom quark
mass, and $u$, $d$, and $s$ are considered to be massless. These are
the formulas used in the code. Essentially all important kinematic
relations are expressed in terms of $W$ and $W'$.

\subsection{Energy limits}

The minimum value of the Mandelstam $s$ of the Borm process is [see
Eq.~(\ref{smin})] 
\begin{equation}
s_{\min}=d\Big(1+\sqrt{1-\left((m_{3}^{2}-m_{4}^{2})/d\right)^{2}}\Big)\;\label{smin-2}
\end{equation}
\[
\mathrm{with}\quad d=m_{3}^{2}+m_{4}^{2}+2\,p_{t\,\mathrm{min}}^{2}.
\]
For the ``interesting cases'', we get for $s_{\min}$ : 
\begin{eqnarray}
 &  & 0000\,:\;\quad4\,p_{t\,\mathrm{min}}^{2},\\
 &  & m0m0\,:\quad d\left(1+\sqrt{1-m^{4}/d^{2}}\right),\\
 &  & \qquad\qquad\qquad\mathrm{with\:d=m^{2}+2\,p_{t\,\mathrm{min}}^{2}},\nonumber \\
 &  & 00mm\,:\quad4\Big(m^{2}+p_{t\,\mathrm{min}}^{2}\Big)\,,\\
 &  & mm00\,:\quad4\,p_{t\,\mathrm{min}}^{2}\,,\\
 &  & mm\tilde{m}\tilde{m}\,:\quad4(\tilde{m}^{2}+p_{t\,\mathrm{min}}^{2})\,,\\
 &  & m\tilde{m}m\tilde{m}\,:\,d\Big(1+\sqrt{1-\left((m^{2}-\tilde{m}^{2})/d\right)^{2}}\Big)\,,\\
 &  & \qquad\qquad\qquad\mathrm{with\:d=m^{2}+\tilde{m}^{2}+2\,p_{t\,\mathrm{min}}^{2}\,.}\nonumber 
\end{eqnarray}

\subsection{The variable \emph{t} }

The module of the Mandelstam variable $t$ is given as [see Eq. (\ref{pt2t})]
\begin{eqnarray}
 & |t|=\frac{1}{2}\,\Big(\,\sqrt{W+4m_{1}^{2}}\sqrt{W'+4m_{3}^{2}}-2(m_{1}^{2}+m_{3}^{2})\qquad\nonumber \\
 & \qquad\qquad\qquad\mp\sqrt{W}\sqrt{W'-4p_{t}^{2}}\,\Big),\label{pt2t-1}
\end{eqnarray}
the ``$\mp$'' referring to respectively $p_{z}\ge0$ and $p_{z}\le0$.
For the different cases, we have 
\begin{eqnarray}
 &  & m0m0\,:\;|t|=\frac{W}{2}\!\Big(\!1\mp\sqrt{1-\frac{4p_{t}^{2}}{W}}\Big),\\
 &  & 00mm\,:\;|t|=\frac{W}{2}\!\Big(\!1\mp\sqrt{1-4\frac{p_{t}^{2}+m^{2}}{W}}\Big)\!-\!m^{2},\\
 &  & mm00\,:\;|t|=\frac{W}{2}\!\Big(\!1\mp\sqrt{1-4\frac{p_{t}^{2}-m^{2}}{W}}\Big)\!+\!m^{2},\\
 &  & mm\tilde{m}\tilde{m}\,:\,|t|=\frac{W}{2}\Big(1\mp\sqrt{1-4\frac{p_{t}^{2}-m^{2}+\tilde{m}^{2}}{W}}\,\Big)\nonumber \\
 &  & \qquad\qquad\qquad\qquad\qquad\qquad\qquad+m^{2}-\tilde{m}^{2}\\
 &  & m\tilde{m}m\tilde{m}\,:\,|t|=\frac{W}{2}\!\Big(\!1\mp\sqrt{1-\frac{4p_{t}^{2}}{W}}\,\Big).
\end{eqnarray}
 From Eq. (\ref{pt2t-1}), we get
\[
|t|_{\mathrm{min/max}}\!=\!\frac{1}{2}\!\Big(\!\!\sqrt{W+4m_{1}^{2}}\sqrt{W'+4m_{3}^{2}}-2(m_{1}^{2}+m_{3}^{2})
\]
\vspace{-0.4cm}
\begin{equation}
\qquad\mp\sqrt{W}\sqrt{W'-4p_{t\,\min}^{2}}\,\Big).\label{tminmax-1}
\end{equation}
Considering the different cases, we get for the quantities $|t|_{\mathrm{min/max}}$
:
\begin{eqnarray}
 &  & m0m0\,:\;\frac{W}{2}\,\Bigg(1\mp\sqrt{1-4p_{t\,\mathrm{min}}^{2}/W}\,\Bigg),\\
 &  & 00mm\,:\;\frac{W}{2}\Big(1\mp\sqrt{1-4(p_{t\,\mathrm{min}}^{2}+m^{2})/W}\,\Big)-m^{2},\qquad\\
 &  & mm00\,:\;\frac{W}{2}\Big(1\mp\sqrt{1-4(p_{t\,\min}^{2}-m^{2})/W}\,\Big)+m^{2},\\
 &  & mm\tilde{m}\tilde{m}\,:\frac{W}{2}\Big(1\mp\sqrt{1-4\frac{p_{t\,\min}^{2}-m^{2}+\tilde{m}^{2}}{W}}\,\Big)\nonumber \\
 &  & \qquad\qquad\qquad\qquad\qquad\qquad\qquad\qquad+m^{2}-\tilde{m}^{2},\\
 &  & m\tilde{m}m\tilde{m}\,:\frac{W}{2}\,\Big(1\mp\sqrt{1-4p_{t\,\min}^{2}/W}\,\Big)\,.
\end{eqnarray}
Using the identity 
\begin{eqnarray}
 &  & \Big(1-\sqrt{1-C}\,\Big),\\
 &  & =\frac{C\Big(1-\sqrt{1-C}\,\Big)}{\Big(1+\sqrt{1-C}\,\Big)\Big(1-\sqrt{1-C}\,\Big)},\\
 &  & =\frac{C}{\Big(1+\sqrt{1-C}\,\Big)},
\end{eqnarray}
the latter expression being better for numerical purposes when \textbf{$C\ll1$},
we get for $|t|_{\mathrm{min}}$ : 
\begin{eqnarray}
 &  & m0m0:\,\frac{2p_{t\,\mathrm{min}}^{2}}{1+\sqrt{1-4p_{t\,\mathrm{min}}^{2}/W}},\\
 &  & 00mm:\,\frac{2(p_{t\,\mathrm{min}}^{2}+m^{2})}{1+\sqrt{1-4(p_{t\,\mathrm{min}}^{2}+m^{2})/W}}-m^{2},\quad\\
 &  & mm00:\,\frac{2(p_{t\,\mathrm{min}}^{2}-m^{2})}{1+\sqrt{1-4(p_{t\,\mathrm{min}}^{2}-m^{2})/W}}+m^{2},\quad\\
 &  & mm\tilde{m}\tilde{m}\,:\frac{2(p_{t\,\min}^{2}-m^{2}+\tilde{m}^{2})}{1+\sqrt{1-4(p_{t\,\min}^{2}-m^{2}+\tilde{m}^{2})/W}}\nonumber \\
 &  & \qquad\qquad\qquad\qquad\qquad\qquad\qquad+m^{2}-\tilde{m}^{2},\\
 &  & m\tilde{m}m\tilde{m}\,:\frac{2p_{t\,\mathrm{min}}^{2}}{1+\sqrt{1-4p_{t\,\mathrm{min}}^{2}/W}}.
\end{eqnarray}

\subsection{The limit t\_max+}

We define the maximum value $|t|_{\mathrm{max+}}$ of $|t|$ with
$p_{z}\ge0$ [see Eq. (\ref{eq:tmaxplus})], i.e.,
\begin{equation}
|t|_{\mathrm{max+}}=\frac{1}{2}\,\Big(\,\sqrt{W+4m_{1}^{2}}\sqrt{W'+4m_{3}^{2}}-2(m_{1}^{2}+m_{3}^{2})\Big).
\end{equation}
For the different cases, we get
\begin{eqnarray}
 &  & m0m0:\:|t|_{\mathrm{max+}}=\frac{W}{2}=\frac{\left(s-m\text{\texttwosuperior}\right)^{2}}{2\,s},\\
 &  & 00mm:\:|t|_{\mathrm{max+}}=\frac{s}{2}-m^{2},\\
 &  & mm00:\:|t|_{\mathrm{max+}}=\frac{s}{2}-m^{2},\\
 &  & mm\tilde{m}\tilde{m}\,:|t|_{\mathrm{max+}}=\frac{s}{2}-m^{2}-\tilde{m}^{2},\\
 &  & m\tilde{m}m\tilde{m}\,:|t|_{\mathrm{max+}}=\frac{W}{2}\,.
\end{eqnarray}

\subsection{Transverse momentum for given \emph{t}}

The general formula is [see Eq. (\ref{eq:t2pt})]
\begin{eqnarray}
 & p_{t}^{2}=\frac{W'}{4}-\frac{1}{W}\,\Big(\,|t|-\frac{1}{2}\sqrt{W+4m_{1}^{2}}\sqrt{W'+4m_{3}^{2}}\nonumber \\
 & \qquad\qquad\qquad+m_{1}^{2}+m_{3}^{2}\,\Big)^{2}.
\end{eqnarray}
For the different cases, we get 
\begin{eqnarray}
 &  & m0m0:\:p_{t}^{2}=|t|\Big(1-\frac{|t|}{W}\Big)\,,\\
 &  & 00mm:\:p_{t}^{2}=\Big(|t|+m^{2}\Big)\Big(1-\frac{|t|+m^{2}}{W}\Big)-m^{2},\\
 &  & mm00:\:p_{t}^{2}=\Big(|t|-m^{2}\Big)\Big(1-\frac{|t|-m^{2}}{W}\Big)+m^{2},\\
 &  & mm\tilde{m}\tilde{m}\,:p_{t}^{2}=\Big(|t|-m^{2}+\tilde{m}^{2}\Big)\Big(1-\frac{|t|-m^{2}+\tilde{m}^{2}}{W}\Big)\nonumber \\
 &  & \qquad\qquad\qquad\qquad\qquad\qquad\qquad\qquad+m^{2}-\tilde{m}^{2},\\
 &  & m\tilde{m}m\tilde{m}\,:p_{t}^{2}=|t|\Big(1-\frac{|t|}{W}\Big)\,.
\end{eqnarray}


\begin{thebibliography}{10}

\bibitem{Collins:1989}
J.~Collins, D.~Soper, and G.~Sterman,
\newblock in Perturbative Quantum Chromodynamics, edited by A.H. Mueller, World
  Scientific, Singapore  (1989).

\bibitem{Ellis:1996}
R.~Ellis, W.~Stirling, and B.~Webber,
\newblock in QCD and Collider Physics, Cambridge Monographs on Particle
  Physics, Nuclear Physics and Cosmology, Cambridge University Press, Cambridge  (1996).

\bibitem{Gross:1973}
D.~Gross and F.~Wilczek,
\newblock Physical Review Letters {\bf 30}, 1343 (1973).

\bibitem{Politzer:1973}
H.~Politzer,
\newblock Physical Review Letters {\bf 30}, 1346 (1973).

\bibitem{CMS:2010ifv}
CMS, V.~Khachatryan {\em et~al.},
\newblock JHEP {\bf 09}, 091 (2010), 1009.4122.

\bibitem{werner:2023-epos4-overview}
K.~Werner,
\newblock http://arxiv.org/abs/2301.12517.

\bibitem{Gribov:1967vfb}
V.~N. Gribov,
\newblock Zh. Eksp. Teor. Fiz. {\bf 53}, 654 (1967).

\bibitem{Gribov:1968jf}
V.~N. Gribov,
\newblock Sov. Phys. JETP {\bf 29}, 483 (1969).

\bibitem{GribovLipatov:1972}
V.~N. Gribov and L.~N. Lipatov,
\newblock Sov. J. Nucl. Phys. {\bf 15}, 438 (1972).

\bibitem{Abramovsky:1973fm}
V.~A. Abramovsky, V.~N. Gribov, and O.~V. Kancheli,
\newblock Yad. Fiz. {\bf 18}, 595 (1973).

\bibitem{Drescher:2000ha}
H.~J. Drescher, M.~Hladik, S.~Ostapchenko, T.~Pierog, and K.~Werner,
\newblock Phys. Rept. {\bf 350}, 93 (2001), hep-ph/0007198.

\bibitem{AltarelliParisi:1977}
G.~Altarelli and G.~Parisi,
\newblock Nuclear Physics B. {\bf 126}, 298 (1977).

\bibitem{Dokshitzer:1977}
Y.~L. Dokshitzer,
\newblock Sov. Phys. JETP {\bf 46}, 641 (1977).

\bibitem{MPI:2018}
Multiple Parton Interactions at the LHC, edited
by Paolo Bartalini and Jonathan Richard Gaunt, World Scientific, 
ISBN: 978-981-3227-75-0 (2018).

\bibitem{Guiot:2018kfy} B.~Guiot, Heavy-quark production with $k_{t}$-factorization:
The importance of the sea-quark distribution, Phys. Rev. D \textbf{99}
(2019) no.7, 074006. 

\bibitem{Guiot:2021vnp} B.~Guiot and A.~van Hameren, D and B-meson
production using kt-factorization calculations in a variable-flavor-number
scheme, Phys. Rev. D \textbf{104} (2021) no.9, 094038 

\bibitem{Dulat_2016-CTEQ-PDF}
S.~Dulat {\em et~al.},
\newblock Physical Review D {\bf 93} (2016).

\bibitem{ZEUS96}
ZEUS, M.~Derrick {\em et~al.},
\newblock Z. Phys. {\bf C72}, 399 (1996), hep-ex/9607002.

\bibitem{H1-94}
{H1 Collaboration, Abt et al.},
\newblock Z. Phys. {\bf C63}, 377 (1994).

\bibitem{H1-96a}
{H1 Collaboration, Aid et al.},
\newblock Nucl. Phys. {\bf B470}, 3 (1996).

\bibitem{H1-96b}
{H1 Collaboration, Adlhoff et al.},
\newblock Nucl. Phys. {\bf B485}, 3 (1996).

\bibitem{ATLAS:2017ble}
ATLAS, M.~Aaboud {\em et~al.},
\newblock JHEP {\bf 05}, 195 (2018), 1711.02692.

\bibitem{Cacciari:2012ny}
M.~Cacciari {\em et~al.},
\newblock JHEP {\bf 10}, 137 (2012), 1205.6344.

\bibitem{ALICE:2012-pp-charm-spectra}
ALICE, B.~Abelev {\em et~al.},
\newblock JHEP {\bf 01}, 128 (2012), 1111.1553.

\end{thebibliography}
\end{document}